\documentclass[12pt,preprint]{aastex}
  
\slugcomment{Resubmitted to ApJS}

\shorttitle{Warm Gas in Luminous Infrared Galaxies}
\shortauthors{Wilson et al.}

\begin{document}


\title{Luminous Infrared Galaxies with the Submillimeter Array: \\ I. Survey Overview and the Central Gas to Dust Ratio}

\author{Christine D. Wilson\altaffilmark{1,2},
Glen R. Petitpas\altaffilmark{2}, Daisuke Iono\altaffilmark{3,2},
Andrew J. Baker\altaffilmark{4},
Alison B. Peck\altaffilmark{2,12}, Melanie Krips\altaffilmark{2},
Bradley Warren\altaffilmark{1}, 
Jennifer Golding\altaffilmark{1}, Adam Atkinson\altaffilmark{1}, 
Lee Armus\altaffilmark{11}, T. J. Cox\altaffilmark{2},
Paul Ho\altaffilmark{9,2}, Mika
Juvela\altaffilmark{10}, Satoki Matsushita\altaffilmark{9}, 
J. Christopher Mihos\altaffilmark{8}, Ylva Pihlstrom\altaffilmark{7}, 
Min S. Yun\altaffilmark{6}
}

\altaffiltext{1}{Department of Physics \& Astronomy, McMaster University,
Hamilton, Ontario L8S 4M1 Canada; wilson@physics.mcmaster.ca,
bwarren@physics.mcmaster.ca, goldingj@physics.mcmaster.ca, atkinsa@muss.cis.mcmaster.ca}
\altaffiltext{2}{Harvard-Smithsonian Center for Astrophysics,
  Cambridge, MA 02138; gpetitpa@cfa.harvard.edu,
  mkrips@cfa.harvard.edu, tcox@cfa.harvard.edu} 
\altaffiltext{3}{National Astronomical Observatory of Japan, 2-21-1
  Osawa, Mitaka, Tokyo 181-0015, Japan; d.iono@nao.ac.jp}
\altaffiltext{4}{Department of Physics and Astronomy,
Rutgers, the State University of New Jersey,
136 Frelinghuysen Road,
Piscataway, NJ 08854-8019  U.S.A.;
ajbaker@physics.rutgers.edu
}
\altaffiltext{6}{Department of Astronomy, University of Massachusetts,
  Amherst, MA 01003;myun@astro.umass.edu} 
\altaffiltext{7}{Department of Physics and Astronomy, University of
  New Mexico, Albuquerque, NM 87131; ylva@unm.edu}
\altaffiltext{8}{Department of Astronomy, Case Western Reserve
  University, 10900 Euclid Avenue, Cleveland, OH 44106; mihos@case.edu}
\altaffiltext{9}{Academia Sinica Institute of Astronomy and
  Astrophysics, Taipei 106, Taiwan; 
  pho@asiaa.sinica.edu.tw, satoki@asiaa.sinica.edu.tw} 
\altaffiltext{10}{University of Helsinki Observatory, Finland; mika.juvela@helsinki.fi}
\altaffiltext{11}{Spitzer Science Center, California Institute of
  Technology, Pasadena, CA 91125; lee@ipac.caltech.edu}
\altaffiltext{12}{Joint ALMA Office, Avda El Golf 40, piso 18,
  Santiago, Chile 7550108; apeck@alma.cl}

\begin{abstract}
We present
new data obtained with the Submillimeter Array 
for a sample of fourteen nearby luminous and ultraluminous infrared
galaxies. The galaxies were
selected to have distances $D_L < 200$ Mpc and
far-infrared luminosities $\log L_{\rm FIR} > 11.4$.
The galaxies were observed with spatial resolutions of order 1 kpc
in the CO J=3-2, CO J=2-1, $^{13}$CO J=2-1,
and HCO$^+$ J=4-3 lines as well as the continuum at
880 $\mu$m and 1.3 mm.
We have combined our CO and continuum data to measure
an average gas-to-dust mass
ratio of $120 \pm 28$ (rms deviation 109) in the central regions of
these galaxies, very similar to the 
value of 150 determined for the Milky Way. 
This similarity is interesting given the more
intense heating from the starburst and possibly accretion activity
in the luminous infrared galaxies compared to the Milky Way.
We find that the peak H$_2$ surface density correlates with the
far-infrared luminosity, which
suggests that galaxies with higher gas surface
densities inside the central kiloparsec have a higher star formation rate.
The lack of 
a significant correlation between total H$_2$ mass and
far-infrared luminosity in our sample suggests that the increased
star formation rate is due to the increased availability of
molecular gas as fuel for star formation in the central regions.
In contrast to previous analyses by other authors, we do not find a
significant correlation between
central gas surface density and the star formation efficiency,
as trace by the ratio of far-infrared luminosity
to nuclear gas mass. Our data show that
it is the star formation rate, not the star formation efficiency, that
increases with increasing central gas surface density in these galaxies.
\end{abstract}

\keywords{galaxies: infrared --- galaxies: individual (Arp 55, Arp
  193, Arp 299,
IRAS 10565+2448, IRAS 17208-0014, Mrk 231, Mrk 273, 
NGC 1614, NGC 2623, NGC 5331, NGC 5257, NGC 5258, NGC 6240,
UGC 5101, VV 114)}


\section{Introduction}

Ultra-luminous infrared galaxies (ULIRGs) contain the regions of most intense
star formation in the local universe. Although their high
rates of star formation and accretion 
appear to be triggered by the merger of two
gas-rich galaxies \citep{s88a,v02},
the detailed physical connection between galaxy
mergers and star formation and, in particular, the time evolution of
this process, is not well understood.
Relating numerical hydrodynamical models \citep{miho96,cox04}
to observations is complicated by the difficulty in identifying
the precise stage of the merger \citep{murp01}. In addition,
while high resolution imaging has found that most ULIRGs have 
nuclear separations from $<$0.3 kpc to 48 kpc \citep{murp96}, other
strongly interacting galaxies with these nuclear separations which are
{\it not} 
ULIRGs have also been found \citep{brai04}. These
observations suggest that the onset of the intense star formation
and accretion
which produces a ULIRG is not a simple function of the age of the
merger and leaves open the question of whether all luminous infrared
galaxies (LIRGs\footnote{$L_{\rm FIR} = 4
\pi D_L^2 F_{\rm FIR}$ $L_\odot$, where $F_{\rm FIR} = 1.26 \times 10^{-14}
(2.58 f_{60} + f_{100})$ erg cm$^{-2}$ s$^{-1}$ and $f_{60}$ and $f_{100}$ are
the IRAS fluxes in Jy at 60 and 100 $\mu$m \citep{sanders96}.}; $11
\le \log (L_{\rm FIR}/L_\odot) < 12$) will pass through a
ULIRG phase ($\log (L_{\rm FIR}/L_\odot) \ge 12$) 
at some point in their evolution. 

Local ULIRGs are also important
as the closest analogs to the high-redshift submillimeter galaxies
 \citep[SMGs;][]{blai02}: both populations have high infrared
luminosities, large amounts of molecular gas 
\citep{fray98,fray99,neri03,grev05,t06},
 and morphological evidence of recent or ongoing
mergers \citep{v02,cons03}. 
Since galaxy
merger rates are substantially higher in the early universe 
\citep{lefe00,gott01},
understanding the physical and dynamical properties of nearby
ULIRGs is also important for understanding
the processes in the early universe which give rise to the
very luminous submillimeter galaxy population.

Because molecular gas is the fuel for current and future star
formation, the physical properties and distribution of the warm,
dense molecular gas are crucial for understanding the
processes and timescales controlling star formation in galaxy mergers.
Previous high-resolution studies of molecular gas in luminous
infrared galaxies have used primarily the ground-state rotational
transition of CO, which is sensitive to gas as cold as 10~K
\citep{s91,ds98,bs99}, with a few galaxies observed in the
CO J=2-1 line \citep{bs96,ds98,s99,t99}. 
However, since the CO J=3-2 line traces the warmer and
denser gas, it is more likely to be directly associated with the
starburst activity 
and/or fueling of the active galactic nuclei (AGN) in these galaxies. 
Indeed, observations of the CO J=3-2
emission in two luminous infrared
galaxies,  VV 114 \citep{iono04} and NGC 6090 \citep{wang04},
reveal that the large-scale distribution and kinematics of
the CO J=3-2 line can be significantly different from those of the CO
J=1-0 line.
 
In this paper, we
present new data obtained with the Submillimeter Array (SMA)
for a sample of fourteen luminous and ultraluminous infrared
galaxies in the CO J=3-2, CO J=2-1, $^{13}$CO J=2-1, and HCO$^+$ J=4-3
lines. In addition, we 
present new high-resolution observations of continuum emission
at 880 $\mu$m and 1.3 mm, which allow us to study the dust properties
in the central kiloparsec of the galaxies.
This SMA legacy survey aims to address five broad scientific
questions:

\begin{enumerate} 

\item {\it What are the distributions, kinematics, and
physical conditions of dense molecular gas in U/LIRGs?} The high
resolution CO J=3-2 data cubes trace the distribution of the
warm and dense gas that feeds the starburst (and any accretion) activity
in these luminous galaxies.
The new SMA CO J=3-2, J=2-1, and $^{13}$CO J=2-1 
data can be combined with
published CO J=1-0 and J=2-1 data for a 
detailed investigation of the physical properties of the molecular
gas using Large Velocity Gradient \citep[LVG,][]{ss74,gk74} 
and Monte Carlo \citep{juve97} models
as our primary diagnostic tools.
The CO J=3-2 kinematics  allow us to study the detailed gas
dynamics in the inner few hundred parsecs, yielding measurements of
the total enclosed mass and of the local linewidth that is a parameter
in models of disk turbulence.
The combination of morphology and kinematics offers clues to the
geometry of each merger via comparison of the separation and orientation of
the galaxy nuclei with the results from numerical simulations
\citep[see, e.g.,][for an analysis of Arp 220]{m01} .
 
\item {\it What is the distribution of the dust in U/LIRGs?}
The 880 $\mu$m  continuum images trace the spatial distribution of
the cold and warm (10--70 K) dust, which reflects both the local rate of star
formation activity and the available mass of gas. The submillimeter dust
emission is often significantly more compact than that of CO
\citep[e.g.,][]{s99,mt01,s06} 
arising exclusively from deep in the
gravitational potential wells of the galactic nuclei. 
High-resolution continuum images from the SMA can be combined with
spectra from 
the {\it Spitzer Space Telescope}  
\citep[e.g.,][]{a07} 
to estimate the dust temperature via the
mid-infrared to submillimeter spectral energy distribution (SED), the
dust mass %
\citep[including both small and large grains, e.g.,][]{m07}
and, indirectly,
the gas mass based on 3D radiative transfer modeling \citep{j03}.
In addition, these spatially resolved SEDs of local U/LIRGS
will improve our interpretation of the templates used
for determining photometric redshifts of high-redshift
submillimeter galaxies \citep{yun02,a03,a05}.

\item {\it Do the gas properties change as the interaction
progresses?} Our sample of fourteen U/LIRGs covers a range 
from mid to late merger stages and should be sufficiently large that we can
establish a merger sequence through comparison of global
morphologies with numerical models \citep{miho96,cox04}.
This data set allows us to determine how the distribution and
kinematics of the gas change as a function of physical conditions such
as density and temperature, or vice versa, 
and to correlate those changes with the
stage of the merger.
This type of detailed gas
physics on small scales still poses challenges for numerical simulations.
Thus, on large scales, where the dynamics of the system are well
described by the models, the numerical simulations can help with the
interpretation of the data, while on smaller scales, the data can
drive the development of more accurate descriptions for the gas physics in the
simulations. 
 
\item {\it How do the properties of the dense gas in local
U/LIRGs compare to those of the gas in high-redshift submillimeter galaxies?}
Armed with a robust local sample of fourteen U/LIRGs, we can make
a rigorous comparison of the properties of the gas with those in
higher redshift galaxies \citep{grev05,t06}.
Changes in gas characteristics over the age of the universe
will reveal important information about the process of star formation
and ultimately the formation and evolution of galaxies at early times.

\item {\it  What is the origin of nuclear OH megamasers?}
Bright 1667 MHz OH megamaser emission is observed in the nuclei of
some luminous infrared galaxies, including five galaxies in our
sample (IRAS 17208-0014, Mrk 231, Mrk 273, UGC 5101 and Arp 299). 
These extremely
bright masers are promising tracers of dust-obscured star formation
and mergers at high redshifts, and could ultimately be used
to estimate the merger rate as a function of redshift \citep{darl02}.
However, in order to use OH megamasers as high-redshift probes, we
need to understand whether there is a specific 
type or stage of merger that leads to OH maser emission. 
Whether maser emission occurs is likely governed
by the physical, chemical, and kinematic conditions in the molecular
gas in the nuclear
regions of the mergers. 
For example, using global CO and HCN luminosities, \citet{darling07}
concludes that OH megamasers are associated with high mean molecular gas
densities and high dense gas fractions.
Our sample, which contains galaxies with and
without megamasers, is well suited for identifying any unique nuclear
conditions that produce OH megamasers in luminous infrared galaxies.
\end{enumerate}
 
In this paper, we describe the sample selection, 
observations, and reduction
of the survey data (\S~\ref{sec-obs}). We also examine the gas to dust mass
ratio in the central kiloparsec (\S~\ref{sec-gasdust}) and correlations
between the central gas mass, gas surface density, infrared luminosity,
dust temperature, and CO J=3-2/2-1 line ratio (\S~\ref{sec-corr}).
A companion paper presents a detailed analysis of the galaxy VV 114
\citep{p08}; a detailed analysis of NGC 6240 is given in
\citet{i07}. Future papers  will 
compare the results from
this survey with similar observations of high-redshift
submillimeter galaxies \citep{i08}; examine
the physical properties of the molecular gas by combining molecular
line observations with radiative transfer models; use the molecular
gas data to place constraints on the origin of the OH megamaser activity
seen in some of the galaxies in our sample; 
compare the properties of the cold gas and dust
as seen with the SMA with the properties of the warm dust derived
from {\it Spitzer} data; and compare the
molecular gas and dust properties with the predictions of numerical
simulations to place the galaxies into a merger sequence.

\section{Observations and Data Reduction\label{sec-obs}}

\subsection{Sample selection}

\begin{figure}
\includegraphics[angle=0,scale=.8]{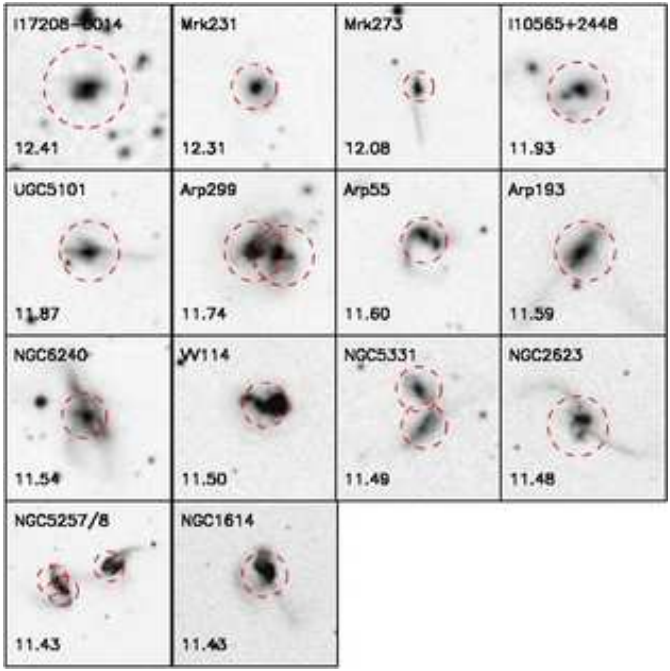}
\caption[ulirgs_beam.eps]{Optical images from the Second Digital
Sky Survey for the fourteen galaxies in our sample
with the approximate half-power beam-width
field of view of the SMA at 345 GHz overlaid. 
Each plot is labeled with 
$\log L_{\rm FIR}/{\rm L_\odot}$. 
In this figure and in the tables, the galaxies are
presented in order of decreasing $L_{\rm FIR}$.
\label{fig-dss_beam}}
\end{figure}

For this survey,
we selected a 
sample of luminous and ultraluminous
infrared galaxies with redshifts $z < 0.045$ (distances $D_L < 200$
Mpc, adopting $H_o = 70$ km s$^{-1}$ Mpc$^{-1}$, $\Omega_M =
  0.3$,
$\Omega_\Lambda = 0.7$) and
far-infrared luminosities $\log L_{\rm FIR} > 11.4$. The distance limit was 
chosen to allow spatial resolutions of order 1 kpc
or better in all the target galaxies. The luminosity cutoff was chosen
to allow us to span a wide range of merger properties and luminosities
while still concentrating on the most infrared-luminous nearby galaxies.
Out of a total of 39 galaxies above declination -20$^o$ \citep{s03} 
which meet these two
criteria, we selected for this survey 
14 galaxies with previous interferometric 
observations in the CO J=1-0 transition.
The galaxies observed in this survey are listed in Table~\ref{tbl-sample}
and shown in Figure~\ref{fig-dss_beam}. The sample includes five
systems in which the two progenitor galaxies are still distinct
(Arp 299, Arp 55, VV 114, NGC 5331, and NGC 5257/8); the remaining
 nine galaxies show a single central concentration.

Compared to the full sample of 39 galaxies, 
the sample observed with
the SMA has a much higher fraction of
galaxies with a large far-infrared flux (86\% with
$F_{\rm FIR} > 45$~Jy compared to 44\% for the full sample).
The SMA sample contains a higher proportion of nearby galaxies (29\%
with $z < 0.021$ 
compared to 15\% for the full sample), although the mean distance of
the SMA sample is only 4\% smaller than the mean distance of the full
sample. The SMA sample also 
has a higher proportion of more luminous galaxies (29\%
with $\log L_{\rm FIR} > 11.9$ compared to 13\% for the full sample), and 
has a mean far-infrared luminosity that
is 35\% larger (0.13 in $\log L_{\rm FIR}$) than the full sample. 
Thus, the
sample of galaxies discussed in this paper is slightly biased towards
more luminous galaxies than the full luminosity and distance limited
sample.

\subsection{Data from the Submillimeter Array}

Observations with the Submillimeter Array \citep[SMA:][]{h04} were obtained between
2005 May 16 and 2007 May 2. 
The correlator was configured to have a spectral resolution of 0.8125 MHz
(typically $\sim 0.7$ km s$^{-1}$ for CO J=3-2 and $\sim 1.1$ km s$^{-1}$ 
for CO J=2-1)
and a bandwidth of 2 GHz, which was covered by 24 ``chunks'' overlapping
slightly in
frequency in each of the upper and lower sidebands.
Each galaxy 
was observed with the CO J=3-2 transition in the lower sideband
of the receiver and continuum in the upper sideband,
10 GHz away. The HCO$^+$ J=4-3 line lies in the high-frequency end of
the upper sideband window and was detected in seven of the galaxies in
our sample. 
In addition, nine galaxies were observed in the CO and $^{13}$CO
J=2-1 lines and in 1.3 mm continuum in a single tuning. It was not
possible to also include the C$^{18}$O J=2-1 line in our passband due
to the broad lines of these galaxies.

Each galaxy was observed in a single configuration of the SMA,
either the extended array (unprojected baselines of 50 to 182 m) 
or the compact array (unprojected baselines of 16 to 69 m). 
The largest angular scale to which these data are sensitive, as
calculated from the minimum $uv$ distance that the data sample well,
ranges from 12-16$^{\prime\prime}$ (19-25$^{\prime\prime}$) for the
compact array CO J=3-2 (J=2-1) data to
7-10$^{\prime\prime}$  (10-15$^{\prime\prime}$)
for the extended
array CO J=3-2 (J=2-1) data.
The configuration
used in each case was chosen  to yield a spatial resolution of 0.7-1 kpc in the
CO J=3-2 line; the typical angular resolution at this wavelength 
is 2.5$^{\prime\prime}$ for the compact configuration and
0.9$^{\prime\prime}$ for the extended configuration. 
The exceptions were NGC 2623 in CO J=2-1, and NGC 6240
and IRAS 17208-0014 in CO J=3-2, which were observed in both the extended and
compact arrays, and NGC 5257/8, which was observed in both the compact
and sub-compact arrays.
For a given galaxy,
the same configuration was used for both the CO J=2-1 and the CO J=3-2
observations. The field of view of the SMA is $\sim 60^{\prime\prime}$
for the 
CO J=2-1 data and $\sim 40^{\prime\prime}$ for the CO J=3-2 data.
Three of the galaxies have sufficiently extended CO 
emission that a small mosaic of two (Arp 299, NGC 5331) or three (NGC 5257/58)
pointings was used.

The data for each line-configuration combination in each galaxy were 
obtained in a single night's observing except for 
NGC 2623 (CO J=2-1 extended
array data), NGC 5257/8 (CO J=3-2 data),
and VV 114 (CO J=2-1 data). Between 6 and 8 antennas were
used in the observations and the total on-source integration time ranged
from 3.0 to 13.0 hr.
The typical double sideband system temperatures at transit ranged
from 240 to 800 K for the CO J=3-2 data, except for IRAS 17208-0014, for
which the values ranged from 650 to 1900 K,
and from 140 to 550 K for the CO J=2-1 data.
The observing dates, on-source integration times, and
sensitivities  obtained with robust weighting
are given in Table~\ref{tbl-obs}.

The initial data calibration was done using the MIR software package.
Observations of Uranus, Neptune, Ganymede, or Callisto were used to
determine the current flux of the gain 
calibrator, which was a nearby quasar with a flux of at least 0.6 Jy.
The amplitude and phase gain variations with
time were calibrated using this gain calibrator.
Antenna-based gain solutions were used throughout except for
the extended array observations of NGC 2623 which required
baseline-based gain solutions. 
Bandpass calibration was determined using a strong quasar such as 3C273,
3C279, or 3C111.
We estimate the absolute flux calibration accuracy of these data
to be 20\%.
However, some of the galaxies are at redshifts such that ozone
lines affect the atmospheric transmission in the CO J=3-2 line (Arp 55,
UGC 5101) or 880 $\mu$m
continuum (Arp 299, Mrk 273, and, to a lesser extent, Arp 55, Mrk 231, 
NGC 5331, and UGC 5101, for which only 200 MHz of the continuum
bandwidth might be affected); thus,  the absolute calibration for these
galaxies is more uncertain.
After calibration in MIR, the data were exported to MIRIAD format 
\citep{s95} for
further editing and imaging.

The data were first flagged to remove the six beginning and ending channels
of each of the 24 chunks of the correlator. The data were also flagged
to remove high amplitude data points, with typically $<$ 1\% of the data
removed in this step. For a few data sets, additional flagging was used
to remove data at the beginning or end of the track when deteriorating 
weather conditions or high system temperatures at lower elevations
had caused a larger amplitude scatter. Data cubes were made
using velocity resolutions of 10, 20, and 40 km s$^{-1}$ for the CO
J=3-2 and J=2-1 lines,  20, 40, and 100 km s$^{-1}$ for the $^{13}$CO
J=2-1 line, and 40 and 100 km s$^{-1}$ for the HCO$^+$ J=4-3 line.
For the $^{13}$CO J=2-1 and HCO$^+$ data sets where significant 
continuum emission was detected (see below), the continuum emission was
subtracted in the $uv$ plane using line-free channels before imaging. 
All data cubes were inverted with  weighting by the system temperature
and moderate robustness, which gives 
the optimal tradeoff between sensitivity and resolution.

Each data cube was cleaned down to two times the rms noise,
except for NGC 6240, which was cleaned down to the rms noise. For most sources,
the beam was sufficiently clean and the emission sufficiently compact that
cleaning the inner quarter of the image produced good results. However,
four sources located near the equator (IRAS 17208-0014, NGC 5331,
NGC 5257, and NGC 6240)
had large sidelobes and/or extended emission and were cleaned using 
regions chosen by inspecting the dirty maps
to isolate the emission from the galaxy; the same fixed regions
were used for all velocity channels. The final data cubes
were corrected for the attenuation of the SMA primary beam 
or the analgous response for the combined mosaic before
we measured the integrated source fluxes.

Continuum maps  at 1.3 mm and 880 $\mu$m were made by imaging the 
individual line-free
regions of the spectra; there were four such regions at 1.3 mm (one
on either side of each of the CO and $^{13}$CO J=2-1 lines)
and three regions at 880 $\mu$m (one on either side of the CO J=3-2
line and a single broad region in the upper sideband that  excluded
the region of the spectrum  containing possible  emission from the
HCO$^+$ J=4-3 line).
These continuum images were then averaged together, weighting by
the inverse of the product of the bandwidth of each image and the
square of its noise,  
to obtain a single continuum image with the best sensitivity. 
The continuum images were not cleaned, as the signal-to-noise
ratio was low and the source(s) were generally compact and well separated
from any sidelobes of the main beam. The one exception is Mrk 231,
where the central source was sufficiently strong that cleaning was
useful.

\begin{figure}
\includegraphics[angle=0,scale=.7]{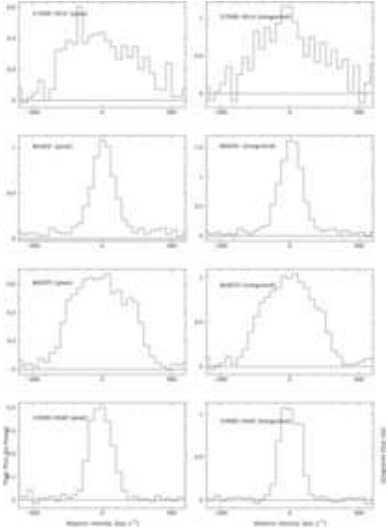}
\caption[12co32spectra_1.eps]{Peak and integrated spectra of the $^{12}$CO
J=3-2 emission for four U/LIRGs from our sample. The integrated
spectrum is the sum of the emission inside
a rectangular region whose size was chosen to encompass all
of the emission above $2\sigma$ in the moment 0 map. The velocity
scale is relative to the recession velocity for each galaxy given in
Table~\ref{tbl-sample}. 
\label{fig-co32_mrk_iras}}
\end{figure}

\begin{figure}
\includegraphics[angle=0,scale=.7]{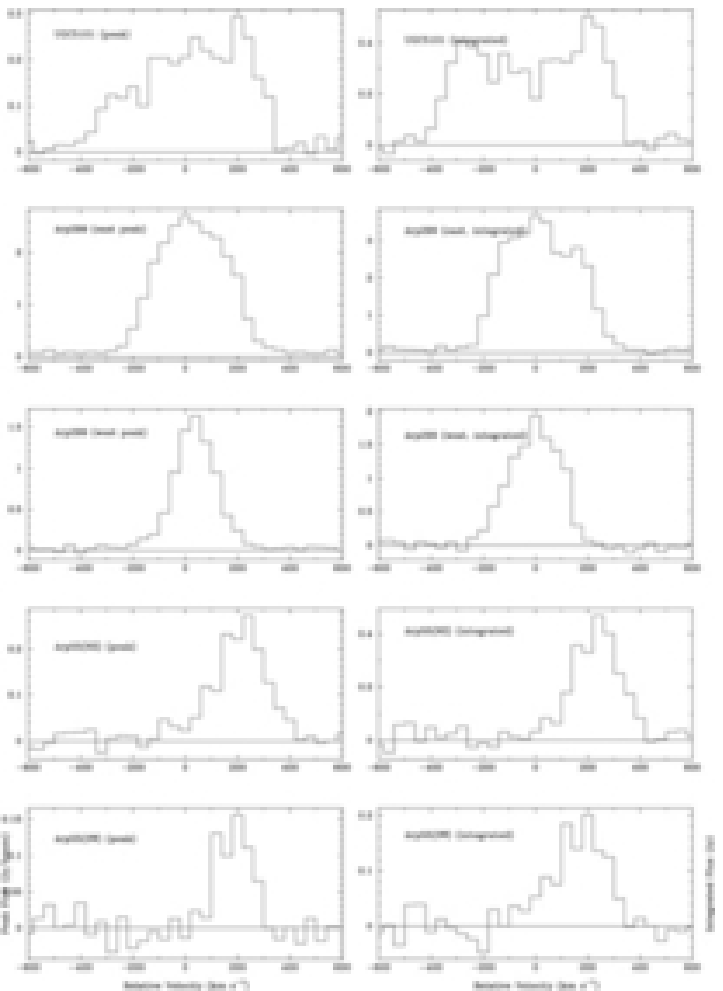}
\caption[12co32spectra_2.eps]{Peak and integrated spectra of the $^{12}$CO
J=3-2 emission for three U/LIRGs from our sample. 
See Figure~\ref{fig-co32_mrk_iras} for additional details. 
\label{fig-co32_arp55_arp299_u5101}}
\end{figure}

\begin{figure}
\includegraphics[angle=0,scale=.7]{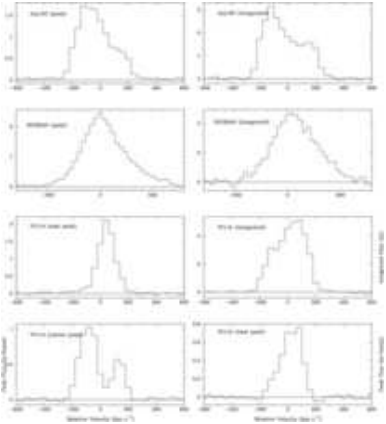}
\caption[12co32spectra_3.eps]{Peak and integrated spectra of the $^{12}$CO
J=3-2 emission for three U/LIRGs from our sample. 
Note the wider
velocity scale used for NGC 6240.
See Figure~\ref{fig-co32_mrk_iras} for additional details. 
\label{fig-co32_arp193_n6240_vv114}}
\end{figure}

\begin{figure}
\includegraphics[angle=0,scale=.7]{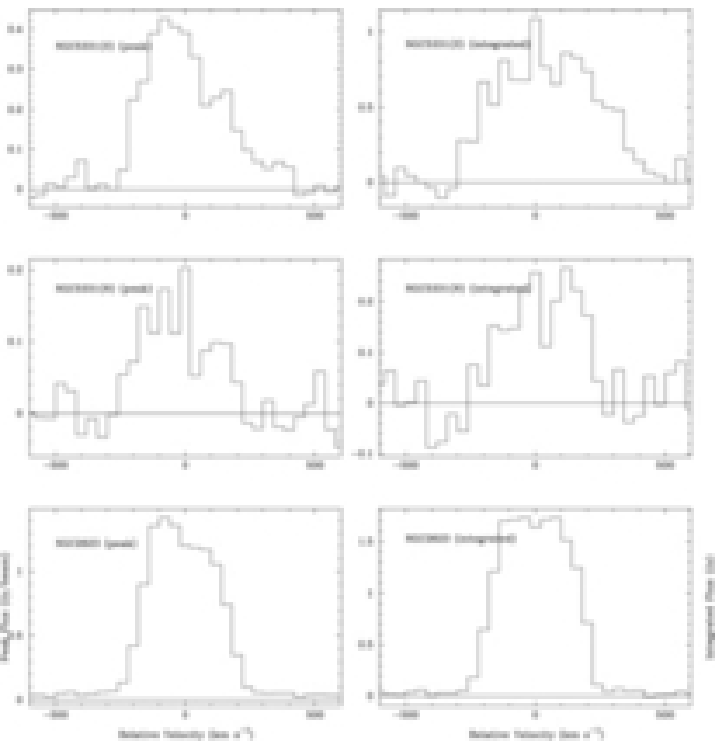}
\caption[12co32spectra_4.eps]{Peak and integrated spectra of the $^{12}$CO
J=3-2 emission for two U/LIRGs from our sample. 
See Figure~\ref{fig-co32_mrk_iras} for additional details. 
\label{fig-co32_n2623_n5257_n5331}}
\end{figure}

\begin{figure}
\includegraphics[angle=0,scale=.7]{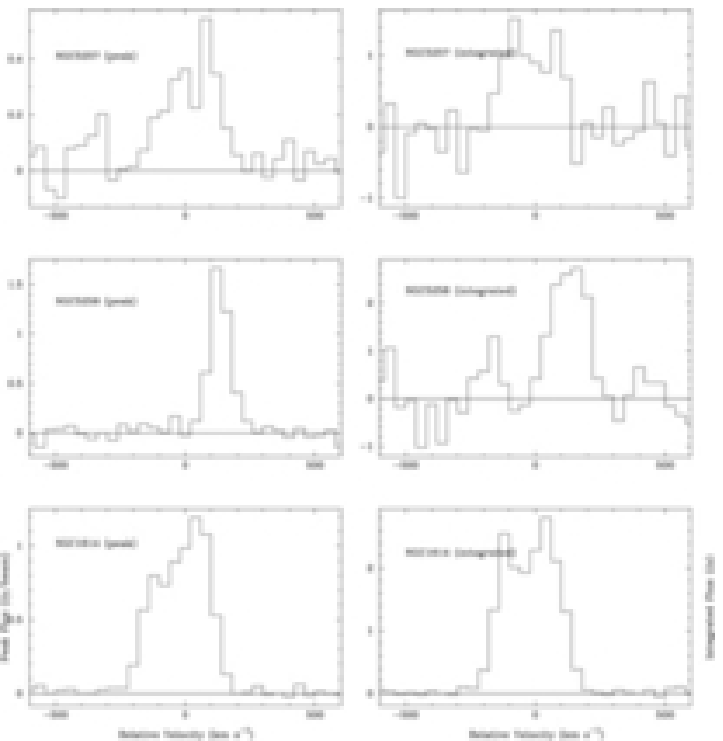}
\caption[12co32spectra_5.eps]{Peak and integrated spectra of the $^{12}$CO
J=3-2 emission for two U/LIRGs from our sample. 
See Figure~\ref{fig-co32_mrk_iras} for additional details. 
\label{fig-co32_n1614_n5258}}
\end{figure}

\begin{figure}
\includegraphics[angle=0,scale=.7]{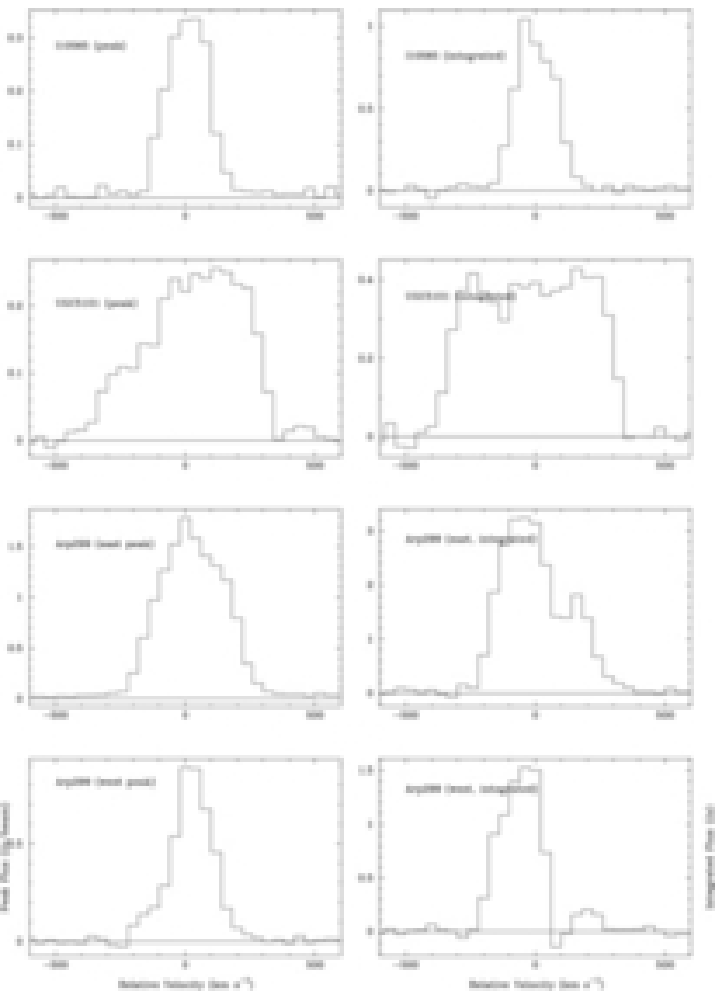}
\caption[12co21spectra_1.eps]{Peak and integrated spectra of the $^{12}$CO
J=2-1 emission for three U/LIRGs from our sample. 
See Figure~\ref{fig-co32_mrk_iras} for additional details. 
\label{fig-co21_i10565_u5101_arp299}}
\end{figure}

\begin{figure}
\includegraphics[angle=0,scale=.7]{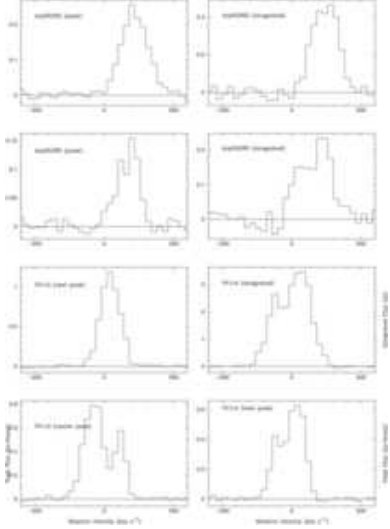}
\caption[12co21spectra_2.eps]{Peak and integrated spectra of the $^{12}$CO
J=2-1 emission for two U/LIRGs from our sample. 
The integrated spectrum for VV 114 encompasses the entire 
extended emission region.
See Figure~\ref{fig-co32_mrk_iras} for additional details. 
\label{fig-co21_arp55_vv114}}
\end{figure}

\begin{figure}
\includegraphics[angle=0,scale=.7]{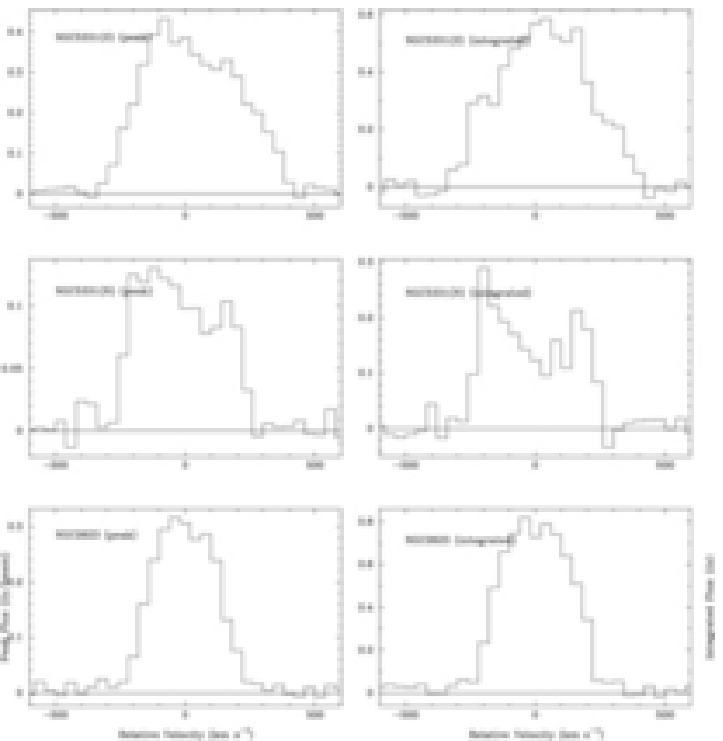}
\caption[12co21spectra_3.eps]{Peak and integrated spectra of the $^{12}$CO
J=2-1 emission for two U/LIRGs from our sample. 
See Figure~\ref{fig-co32_mrk_iras} for additional details. 
\label{fig-co21_vv114_n5331_n2623}}
\end{figure}

\begin{figure}
\includegraphics[angle=0,scale=.7]{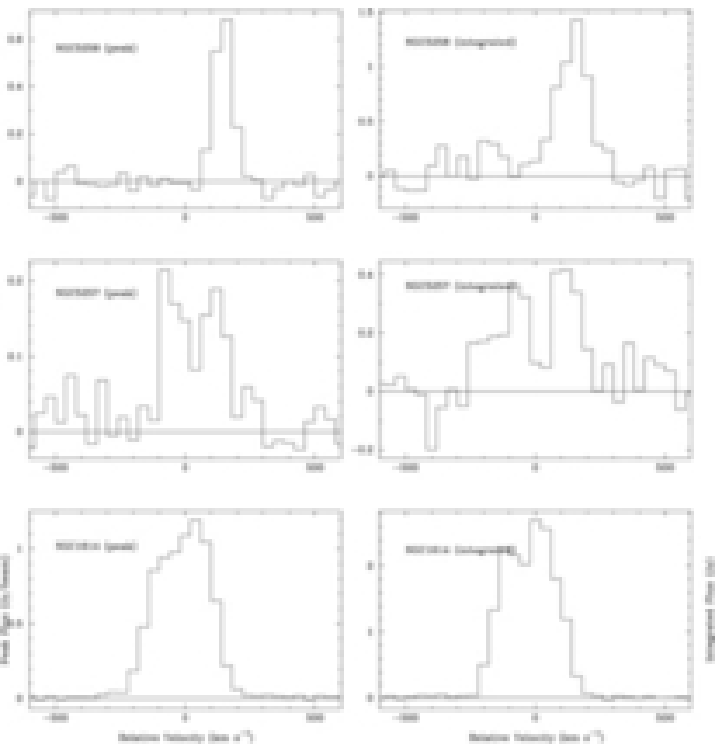}
\caption[12co21spectra_4.eps]{Peak and integrated spectra of the $^{12}$CO
J=2-1 emission for two U/LIRGs from our sample. 
See Figure~\ref{fig-co32_mrk_iras} for additional details. 
\label{fig-co21_n1614_n5257}}
\end{figure}

\begin{figure}
\includegraphics[angle=0,scale=.7]{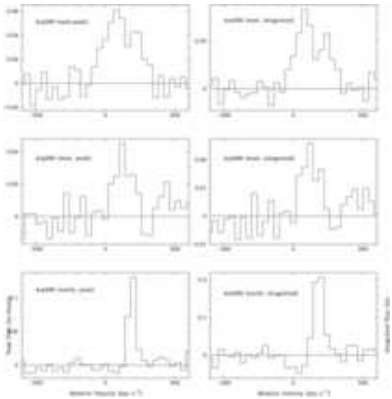}
\caption[13co21spectra_1.eps]{Peak and integrated spectra of the $^{13}$CO
J=2-1 emission for 3 regions in the interacting galaxy
Arp 299 
for which $^{13}$CO J=2-1 emission was detected. 
See Figure~\ref{fig-co32_mrk_iras} for additional details. 
\label{fig-13co21_arp299}}
\end{figure}

\begin{figure}
\includegraphics[angle=0,scale=.7]{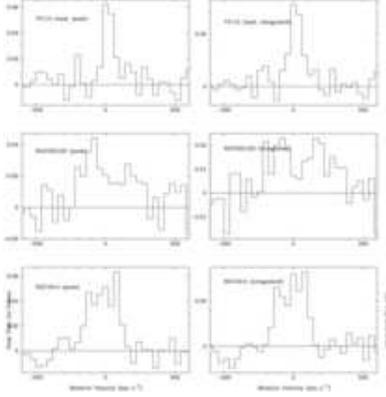}
\caption[13co21spectra_2.eps]{Peak and integrated spectra of the $^{13}$CO
J=2-1 emission for 3 U/LIRGs from our sample
for which $^{13}$CO J=2-1 emission was detected. 
See Figure~\ref{fig-co32_mrk_iras} for additional details. 
\label{fig-13co21_rest}}
\end{figure}

\begin{figure}
\includegraphics[angle=0,scale=.7]{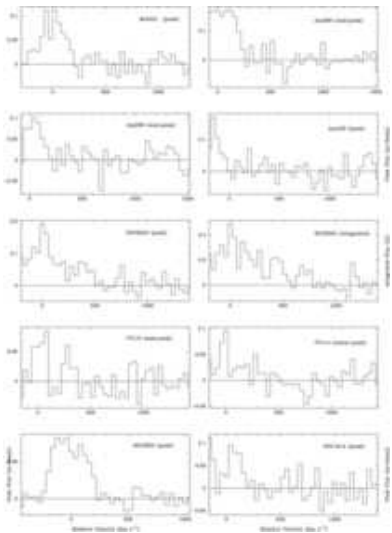}
\caption[HCOspectra.eps]{Peak spectra of the HCO$^+$
J=4-3 emission for nine galaxy components from our sample
for which HCO$^+$ J=4-3 emission was detected. The integrated spectrum
is also given for NGC6240.
See Figure~\ref{fig-co32_mrk_iras} for additional details. 
\label{fig-HCOp}}
\end{figure}

Table~\ref{tbl-fluxes} gives the continuum and integrated 
$^{12}$CO, $^{13}$CO, and HCO$^+$ fluxes measured
for the galaxies in our sample. The fluxes measured for the CO J=3-2
line and 880 $\mu$m continuum emission for NGC6240 differ somewhat
from the values given in \citet{i07} due to the different weighting
used.
Previously published data for the 
CO J=2-1 and J=1-0 lines are also given where available.
Figures~\ref{fig-I17208co32}-\ref{fig-NGC1614co21}
show the moment 0, 1, and 2 images
for the CO J=3-2 and CO J=2-1 data, the 880 $\mu$m and 1.3 mm
dirty maps where signal was detected from the galaxy
at the 3$\sigma$ level or better, and the $^{13}$CO J=2-1
and HCO$^+$ J=4-3
integrated intensity maps, where these lines were detected. 
Unless otherwise noted
in the figure captions, the moment 0 maps
have been made from the 40 km s$^{-1}$ resolution data cube using
only signal greater than $\pm 2\sigma$, while 
the moment 1 and 2 maps have been made from the 20 km s$^{-1}$ resolution data
cube using only positive signal greater than 4$\sigma$.
These images have not been corrected for the primary beam or mosaic
response.
Figures~\ref{fig-co32_mrk_iras}-\ref{fig-co32_n1614_n5258} show
the CO J=3-2 spectra,
Figures~\ref{fig-co21_i10565_u5101_arp299}-\ref{fig-co21_n1614_n5257} 
show the CO J=2-1 spectra, 
Figures~\ref{fig-13co21_arp299}-\ref{fig-13co21_rest}
show the $^{13}$CO J=2-1 spectra 
and
Figure~\ref{fig-HCOp} shows the HCO$^+$ J=4-3 spectra.
In all cases except HCO$^+$,
both the spectrum at the position of the emission peak and 
the spectrum integrated over the entire region
with emission are shown.

\subsubsection{Morphology and kinematics of the CO emission}

\begin{figure}
\includegraphics[angle=-90,scale=.35]{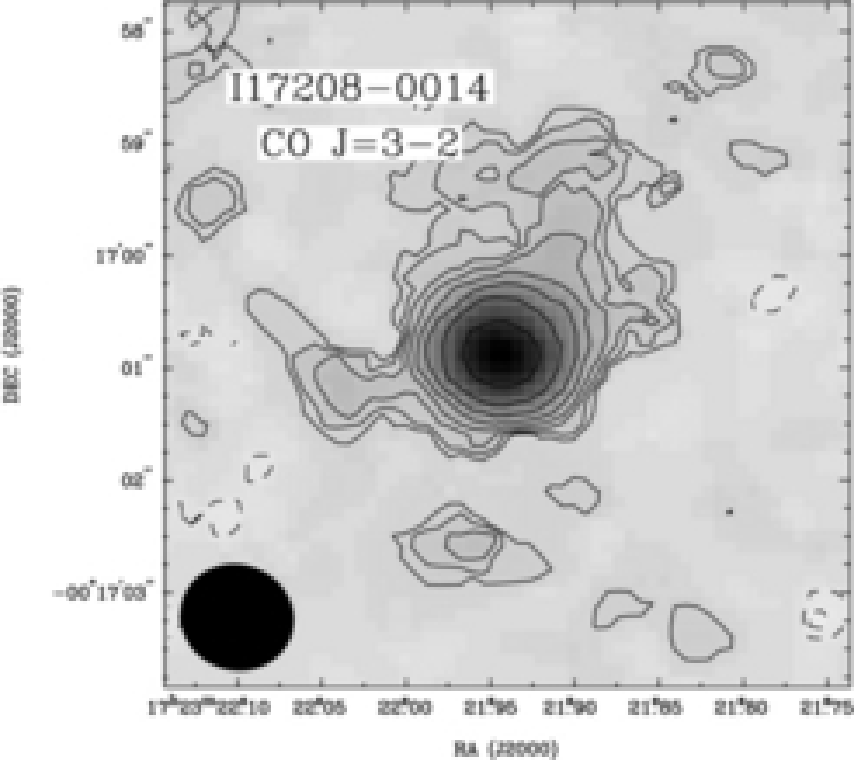}
\includegraphics[angle=-90,scale=.35]{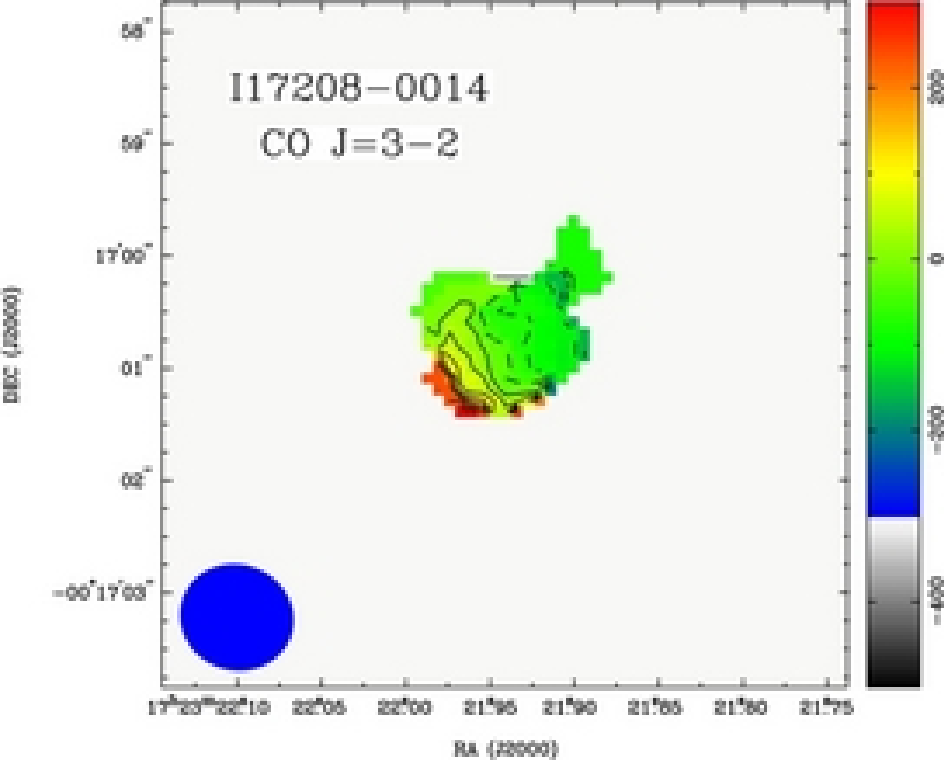}
\includegraphics[angle=-90,scale=.35]{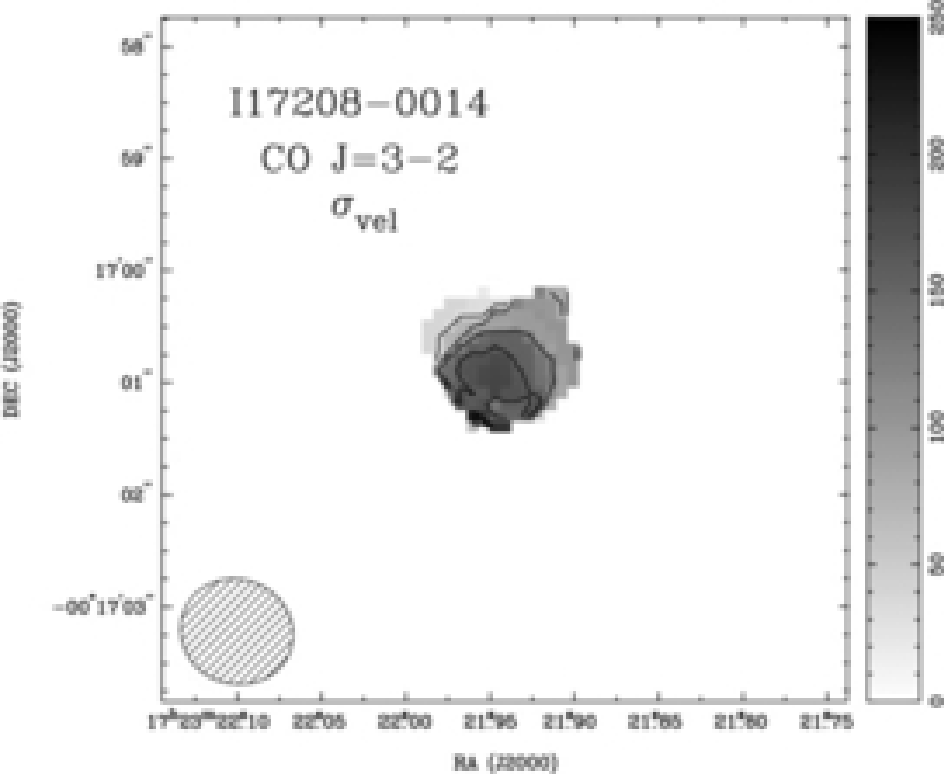}
\includegraphics[angle=-90,scale=.35]{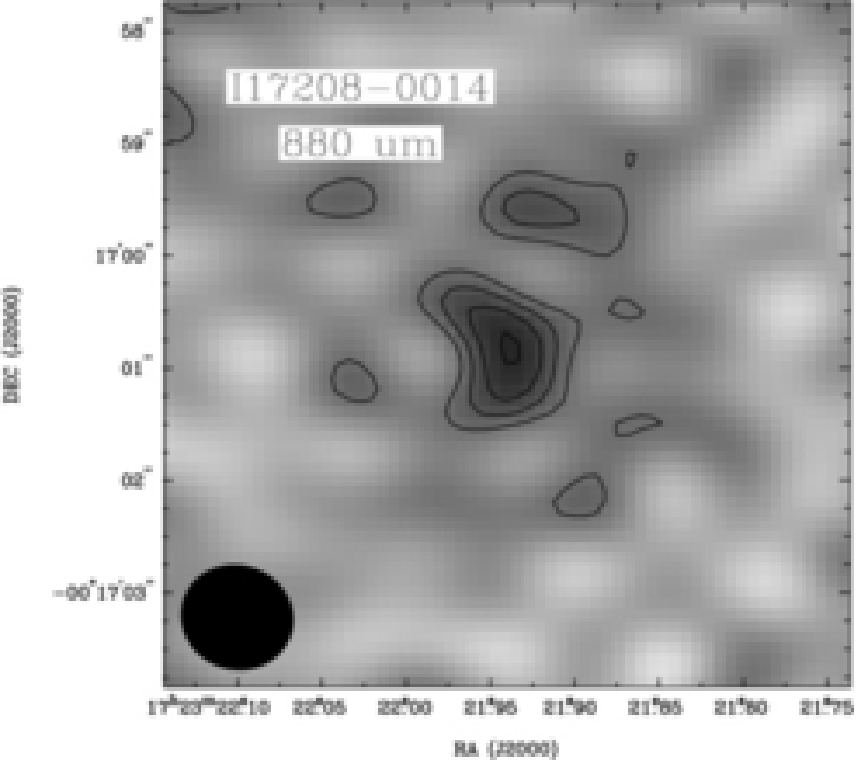}
\caption[I17208co32.mom0.eps]{IRAS 17208-0014 CO J=3-2 and
880 $\mu$m continuum maps. See \S~\ref{sec-obs} for additional
details of the data processing. 
(a) CO J=3-2 moment 0 map. Lowest
contour is $2 \sigma = 11.4 $ Jy beam$^{-1}$ km s$^{-1}$ and contours
increase by factors of 1.5.
Negative contours are shown as dashed lines. The synthesized beam is
shown in the lower left corner of this and subsequent panels.
(b) CO J=3-2 moment 1 map. Contours are 40 km s$^{-1}$ $\times
(-5,-4,-3,-2,-1,0,1,2,3,4,5,6,7)$ relative to $cz$ with negative
contours shown  
as dashed lines. Because of the high noise level, for only this galaxy 
the moment 1 and moment 2 maps were made 
using 40 km s$^{-1}$ channel  
maps with a 4$\sigma$ flux cutoff. 
Note that the negative side of the rotation curve 
peaks at -160 km s$^{-1}$ and then drops to -80 km s$^{-1}$
at the extreme north-west end of the emission.
(c) CO J=3-2 moment 2 map. Contours are 40 km s$^{-1}$ $\times (1,2,3,4,5)$.
This figure plots velocity dispersion, $\sigma_v$, which for a
gaussian line relates to the full-width half-maximum velocity via
$V_{FWHM} = 2.355 \sigma_v$. 
(d) Uncleaned 880 $\mu$m map. Lowest contour
is $2 \sigma = 8.2$ mJy and contours increase in steps of $1 \sigma$.
Negative contours are not shown.
\label{fig-I17208co32}}
\end{figure}

\begin{figure}
\includegraphics[angle=-90,scale=.3]{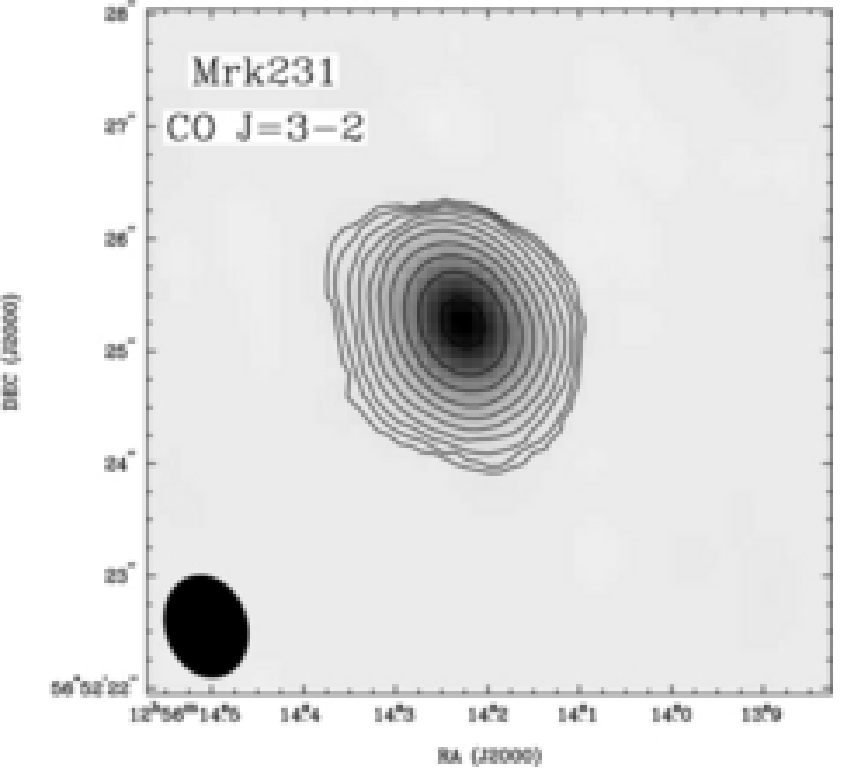}
\includegraphics[angle=-90,scale=.3]{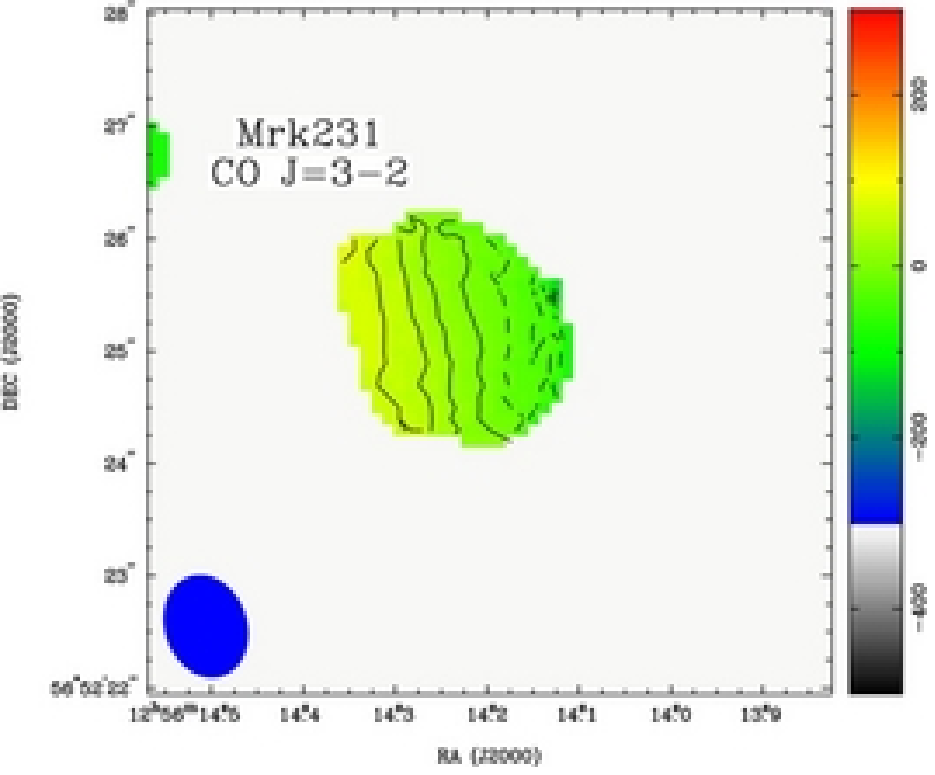}
\includegraphics[angle=-90,scale=.3]{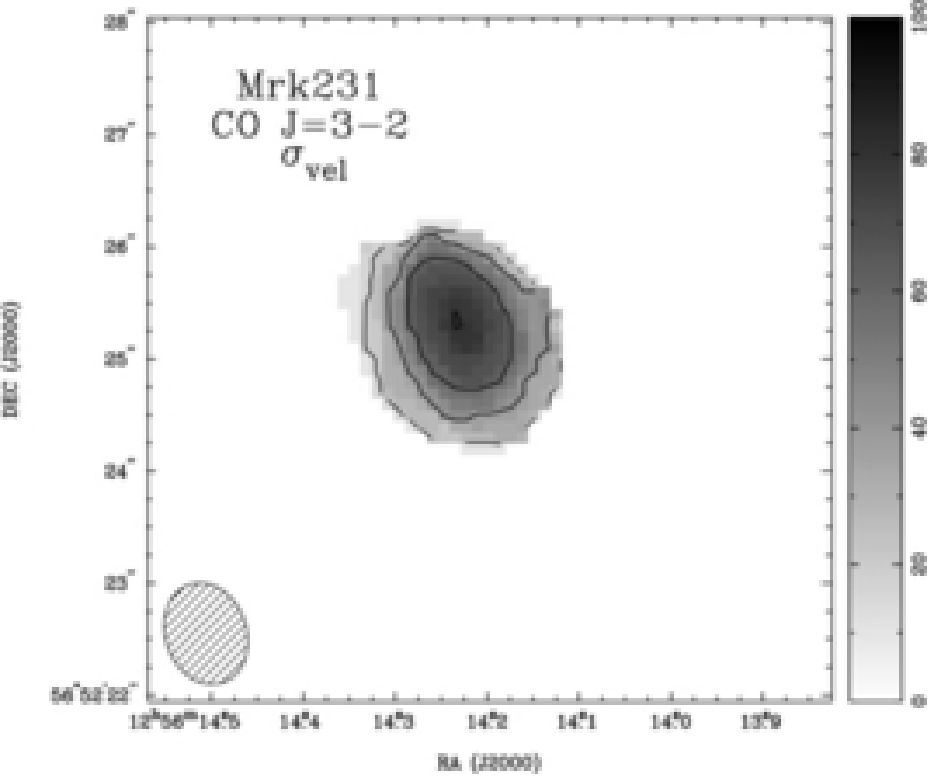}
\includegraphics[angle=-90,scale=.3]{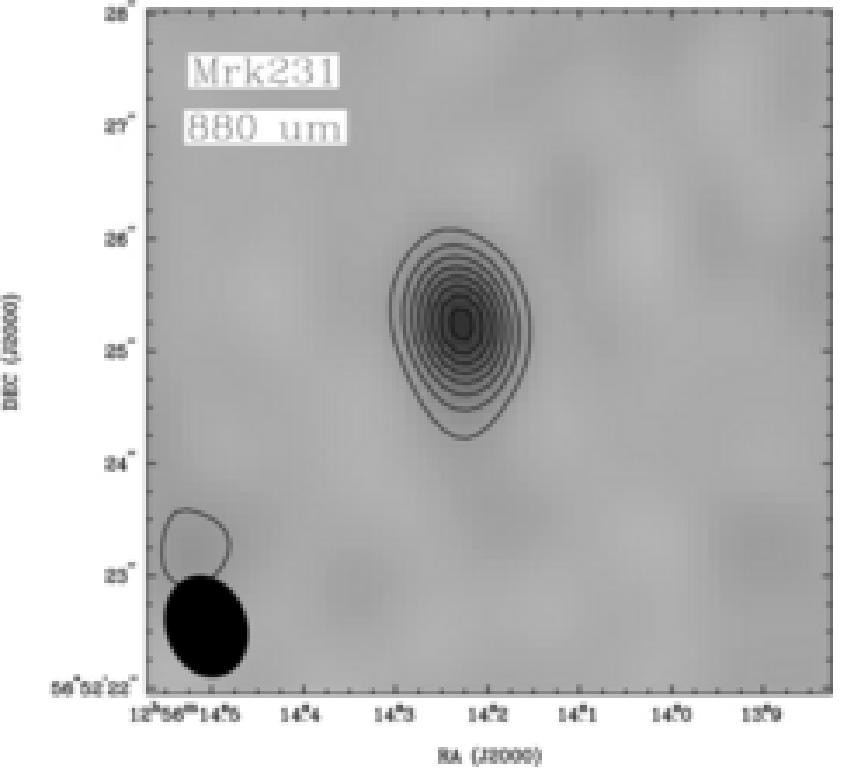}
\includegraphics[angle=-90,scale=.3]{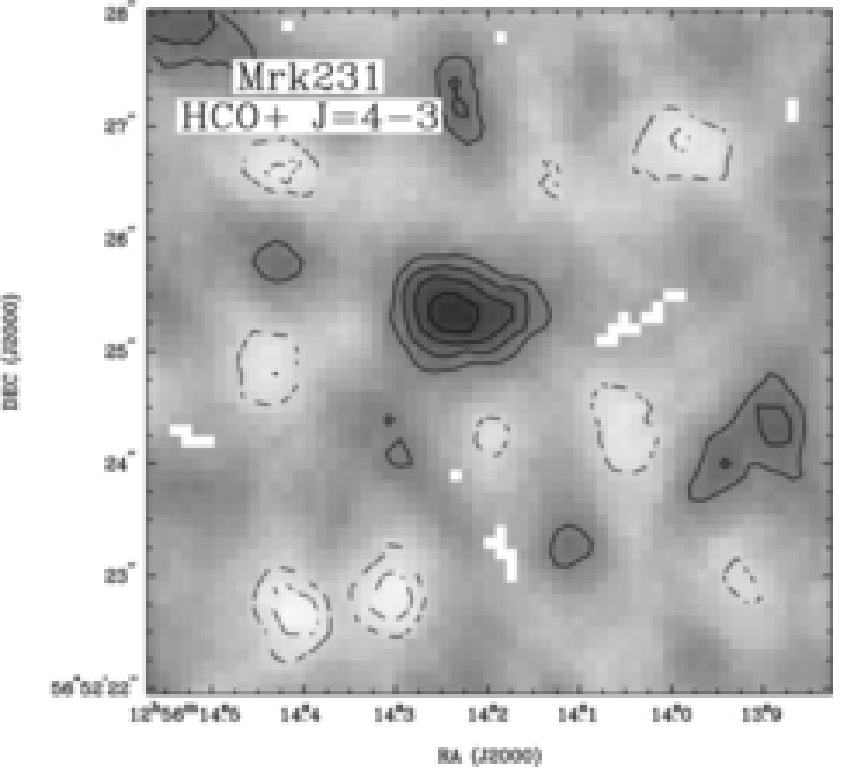}
\caption[Mrk231co32.mom0.eps]{Mrk 231 CO J=3-2 and
880 $\mu$m continuum maps. Notation as in Figure~\ref{fig-I17208co32}.
(a) CO J=3-2 moment 0 map. Lowest
contour is $2 \sigma = 5.2 $ Jy beam$^{-1}$ km s$^{-1}$
and contours increase by factors of 1.5.
(b) CO J=3-2 moment 1 map. Contours are 20 km s$^{-1}$ $\times
(-6,-5,-4,-3,-2,-1,0,1,2,3,4)$ relative to $cz$. 
(c) CO J=3-2 moment 2 map. Contours are 20 km s$^{-1}$ $\times (1,2,3,4)$.
(d) 880 $\mu$m map. Lowest contour
is $2 \sigma = 8.2$ mJy and contours increase in steps of $2 \sigma$.
Because of the strength of the central source, the cleaned continuum
image is shown for this galaxy only.
(e) Uncleaned HCO$^+$ J=4-3 moment 0 map. Lowest
contour is $\pm 2 \sigma = 7.92 $ Jy beam$^{-1}$ km s$^{-1}$ and contours
increase in steps of $1 \sigma$. This image has been
corrected for continuum emission by subtracting the 880 $\mu$m
continuum in the uv plane before imaging. 
\label{fig-Mrk231co32}}
\end{figure}

\begin{figure}
\includegraphics[angle=-90,scale=.35]{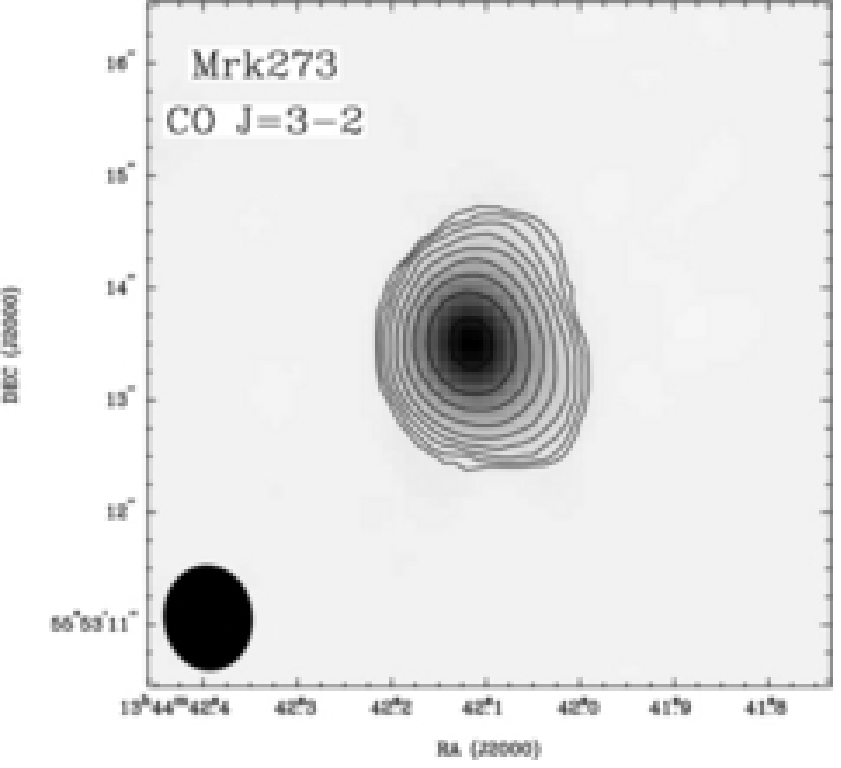}
\includegraphics[angle=-90,scale=.35]{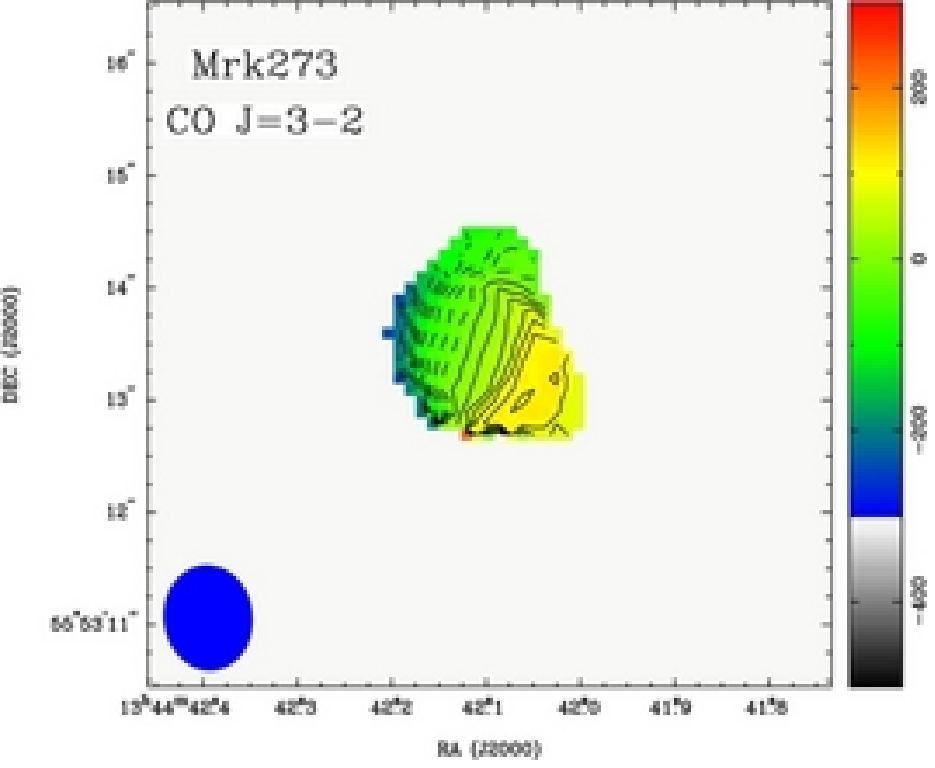}
\includegraphics[angle=-90,scale=.35]{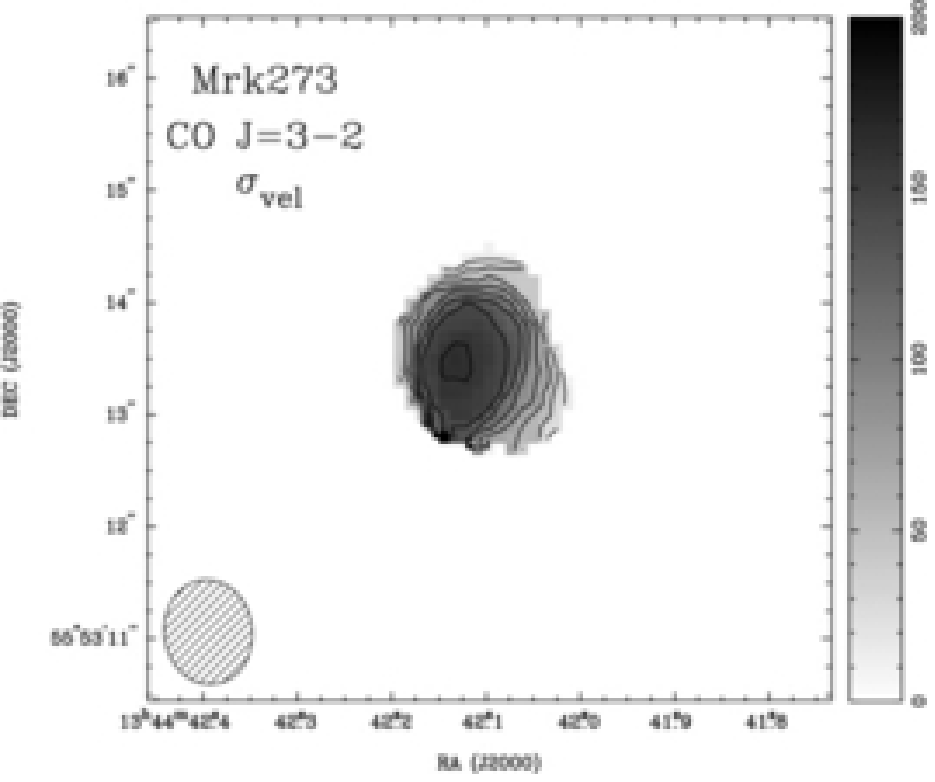}
\includegraphics[angle=-90,scale=.35]{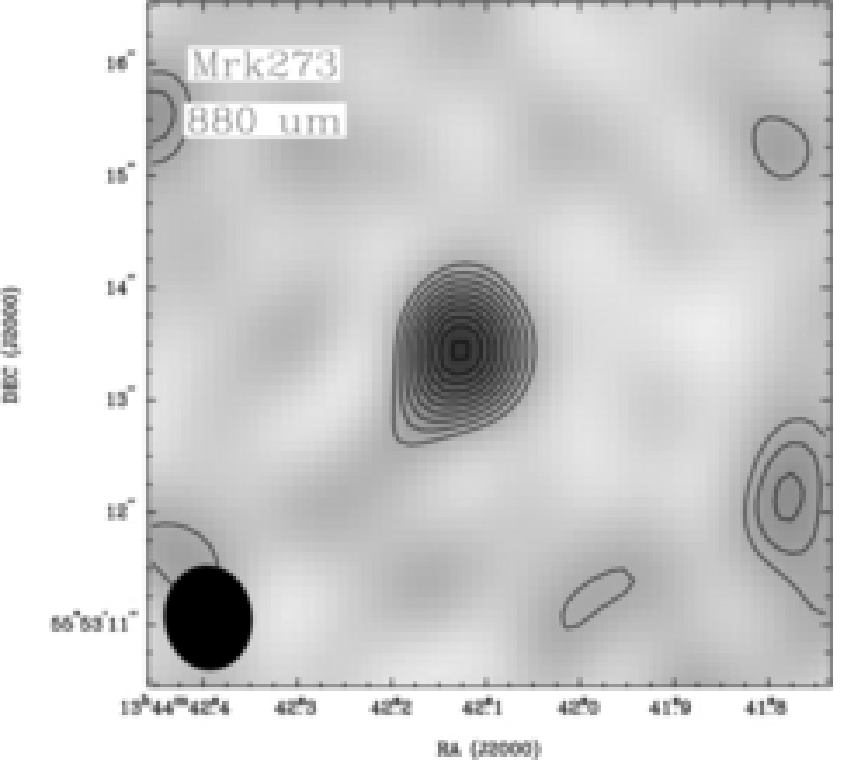}
\caption[Mrk273co32.mom0.eps]{Mrk 273 CO J=3-2 and
880 $\mu$m continuum maps. Notation as in Figure~\ref{fig-I17208co32}.
(a) CO J=3-2 moment 0 map. Lowest
contour is $2 \sigma = 5.2 $ Jy beam$^{-1}$ km s$^{-1}$
and contours
increase by factors of 1.5.
(b) CO J=3-2 moment 1 map. Contours are 20 km s$^{-1}$ $\times
(-11,-10,-9,-8,-7,-6,-5,-4,-3,-2,-1,0,1,2,3,4,5,6)$ relative to $cz$. 
(c) CO J=3-2 moment 2 map. Contours are 20 km s$^{-1}$ $\times (1,2,3,4,5,6,7,8)$.
(d) Uncleaned 880 $\mu$m map. Lowest contour
is $2 \sigma = 10.0$ mJy and contours increase in steps of $1 \sigma$. Residual
sidelobes can be seen at the edges of this figure.
\label{fig-Mrk273co32}}
\end{figure}

\begin{figure}
\includegraphics[angle=-90,scale=.35]{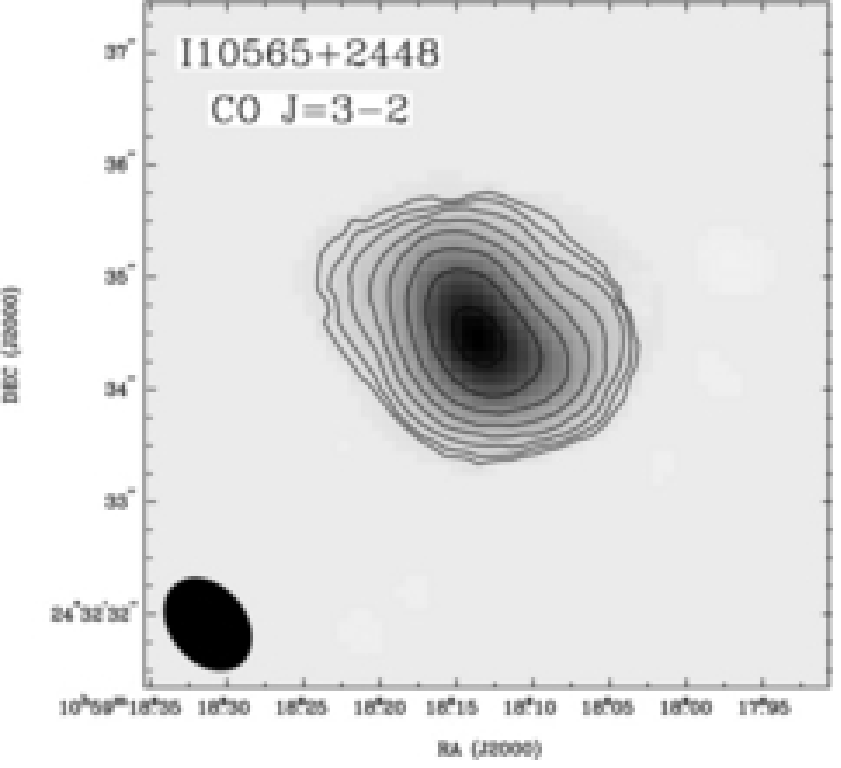}
\includegraphics[angle=-90,scale=.35]{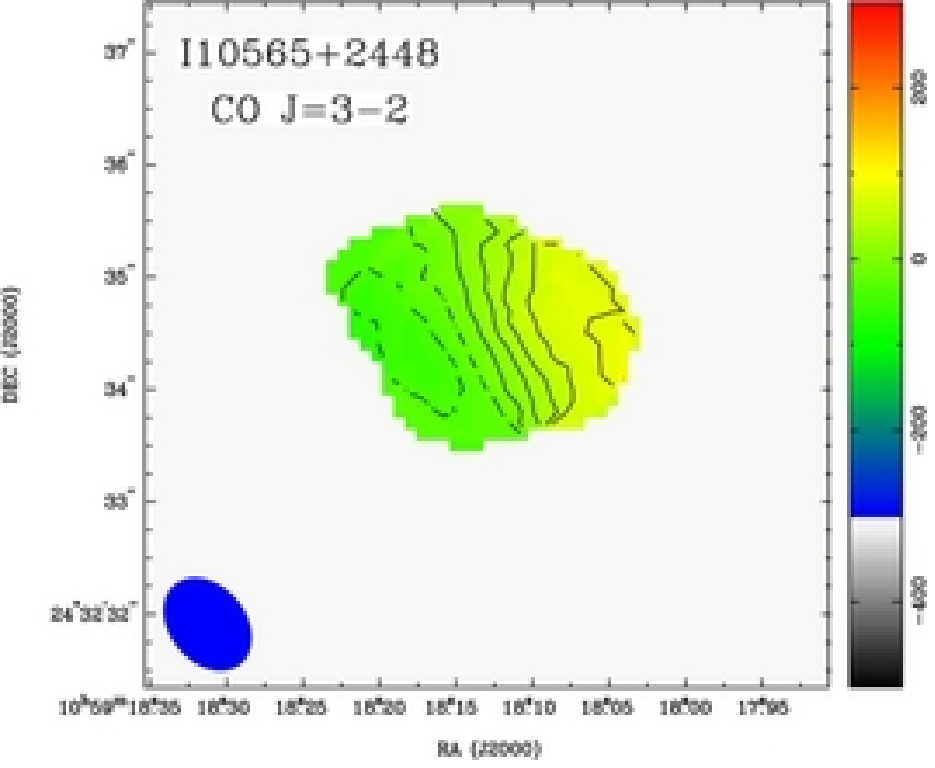}
\includegraphics[angle=-90,scale=.35]{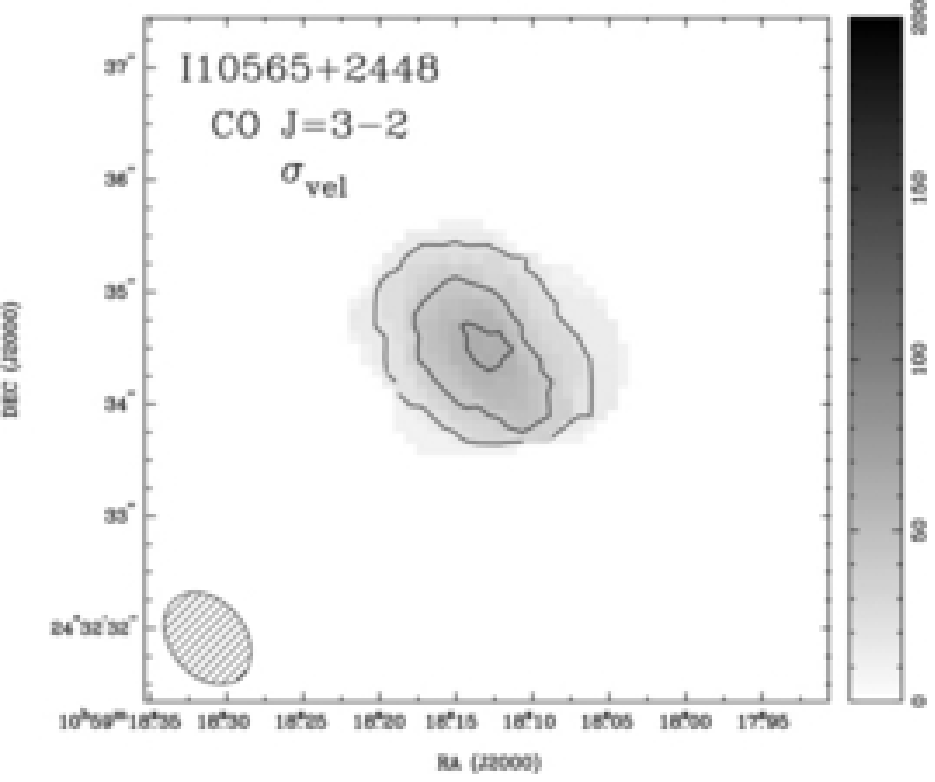}
\includegraphics[angle=-90,scale=.35]{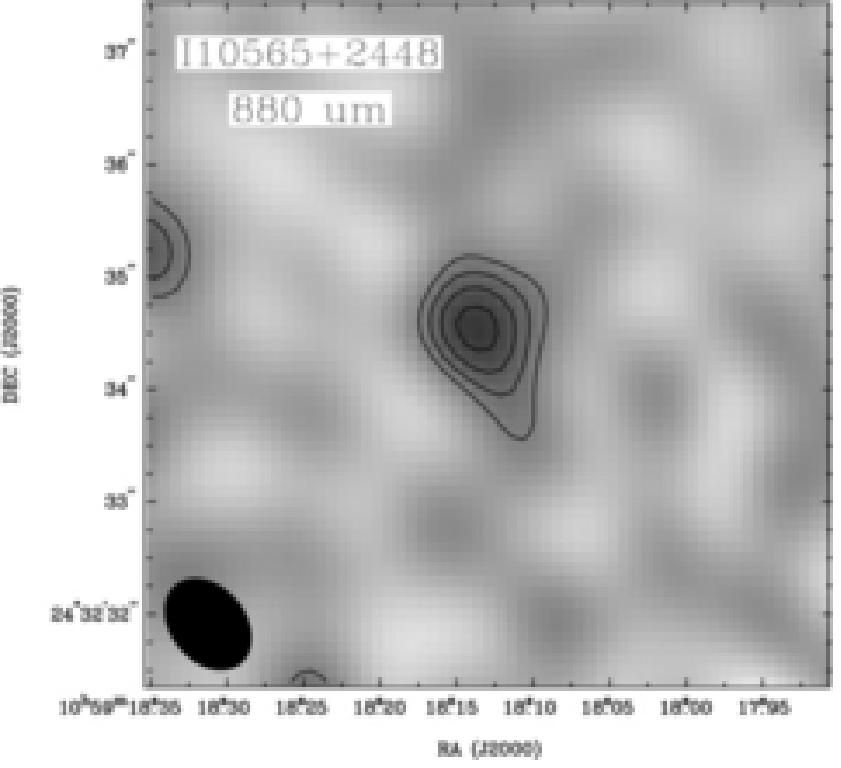}
\caption[I10565co32.mom0.eps]{IRAS 10565 CO J=3-2 and
880 $\mu$m continuum maps. Notation as in Figure~\ref{fig-I17208co32}.
(a) CO J=3-2 moment 0 map. Lowest
contour is $2 \sigma = 4.3 $ Jy beam$^{-1}$ km s$^{-1}$ and contours
increase by factors of 1.5.
(b) CO J=3-2 moment 1 map. Contours are 20 km s$^{-1}$ $\times
(-2,-1,0,1,2,3,4)$ relative to $cz$. 
(c) CO J=3-2 moment 2 map. Contours are 20 km s$^{-1}$ $\times (1,2,3)$.
(d) Uncleaned 880 $\mu$m map. Lowest contour
is $2 \sigma = 5.4$ mJy and contours increase in steps of $1 \sigma$. 
A residual
sidelobe can be seen at the eastern edge of this figure.
\label{fig-I10565co32}}
\end{figure}

\begin{figure}
\includegraphics[angle=-90,scale=.3]{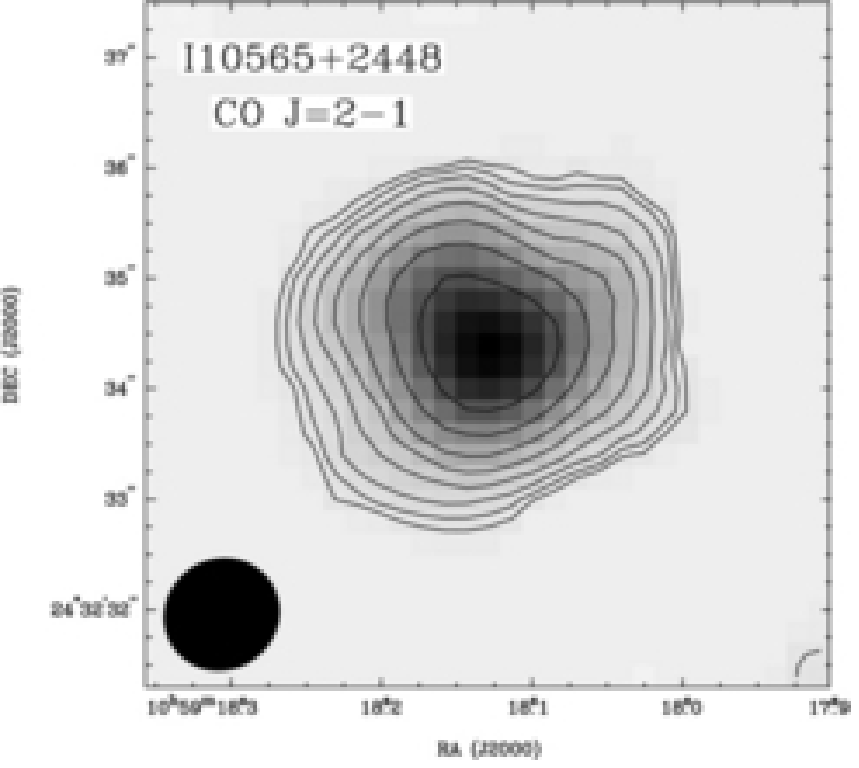}
\includegraphics[angle=-90,scale=.3]{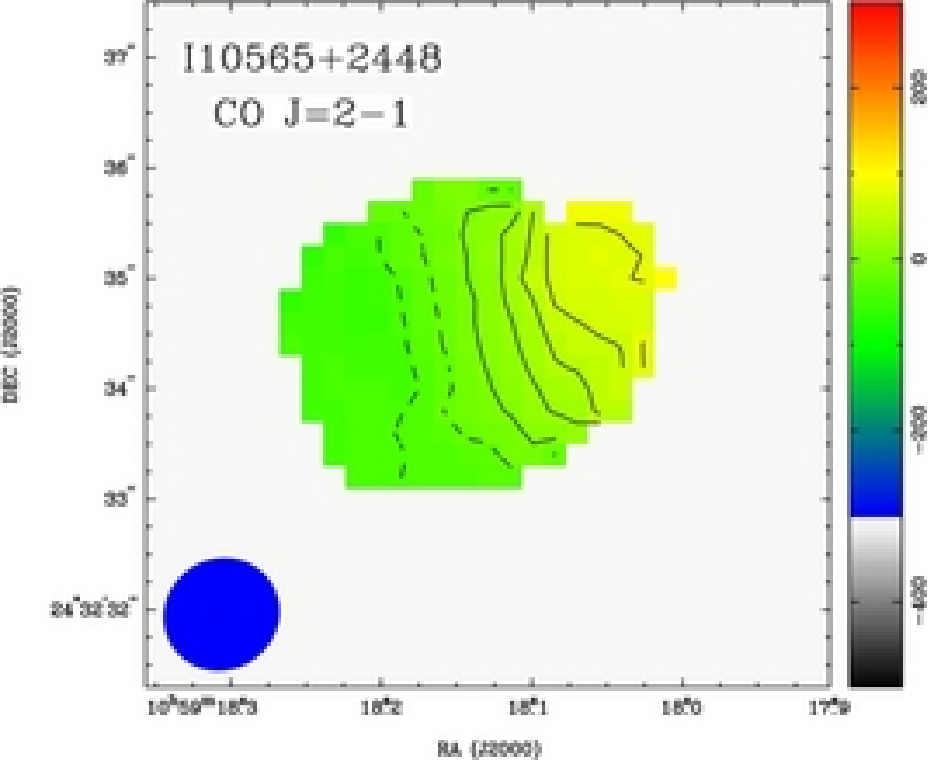}
\includegraphics[angle=-90,scale=.3]{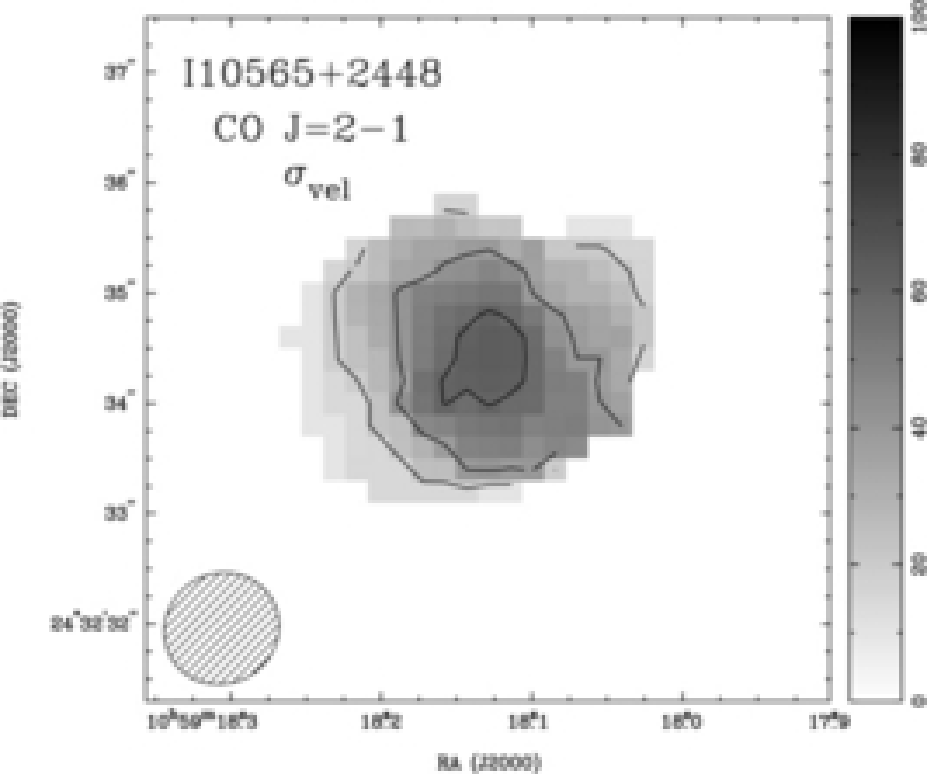}
\includegraphics[angle=-90,scale=.3]{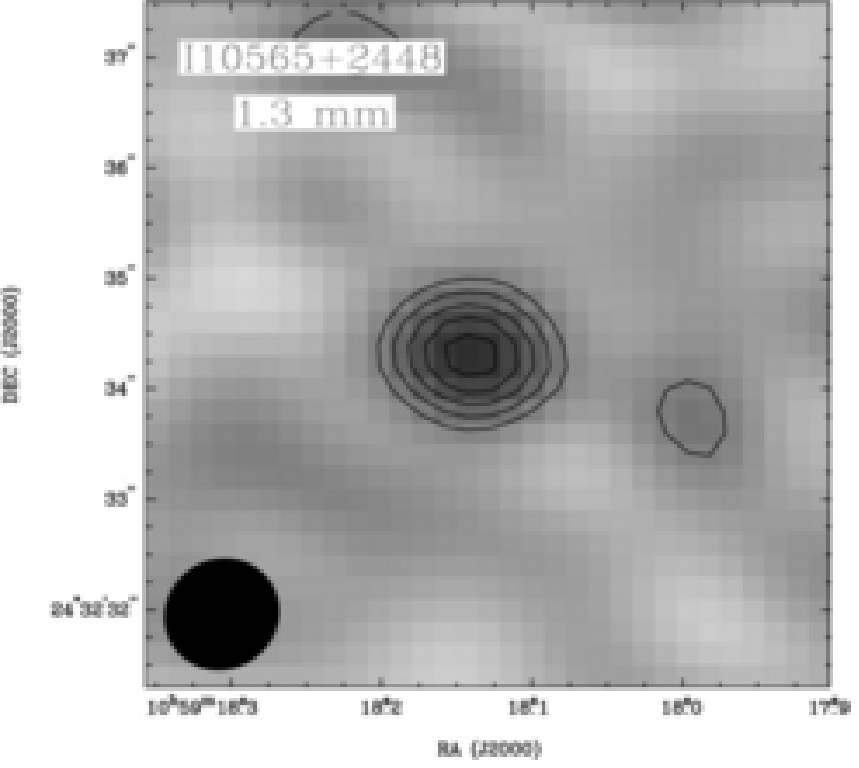}
\caption[I10565co21.mom0.eps]{IRAS 10565 CO J=2-1 and
1.3 mm continuum maps. Notation as in Figure~\ref{fig-I17208co32}.
(a) CO J=2-1 moment 0 map. Lowest
contour is $\pm 2 \sigma = 2.7 $ Jy beam$^{-1}$ km s$^{-1}$ and contours
increase by factors of 1.5. 
(b) CO J=2-1 moment 1 map. Contours are 20 km s$^{-1}$ $\times
(-2,-1,0,1,2,3,4)$ relative to $cz$.
(c) CO J=2-1 moment 2 map. Contours are 20 km s$^{-1}$ $\times (1,2,3)$.
(d) Uncleaned 1.3 mm map. Lowest contour
is $2 \sigma = 2.0$ mJy and contours increase in steps of $1 \sigma$.
\label{fig-I10565co21}}
\end{figure}

\begin{figure}
\includegraphics[angle=-90,scale=.35]{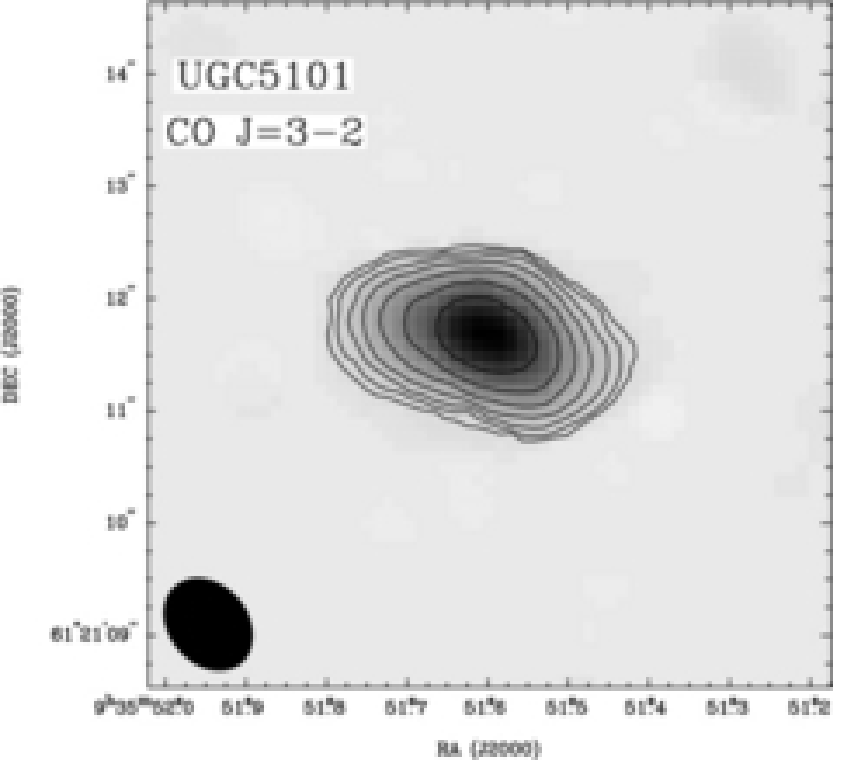}
\includegraphics[angle=-90,scale=.35]{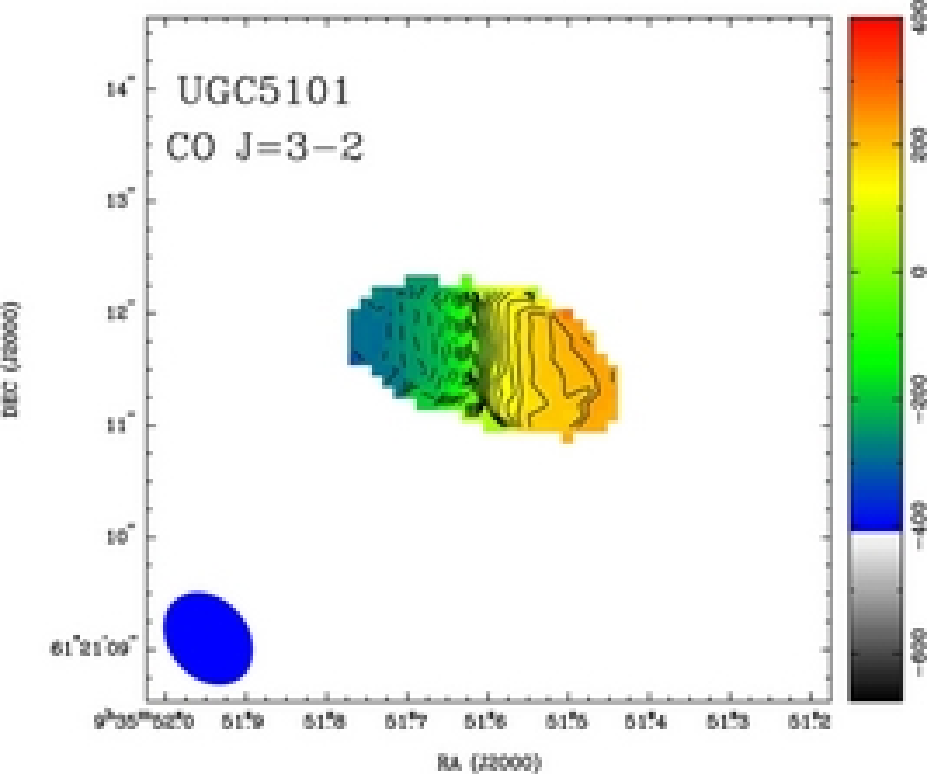}
\includegraphics[angle=-90,scale=.35]{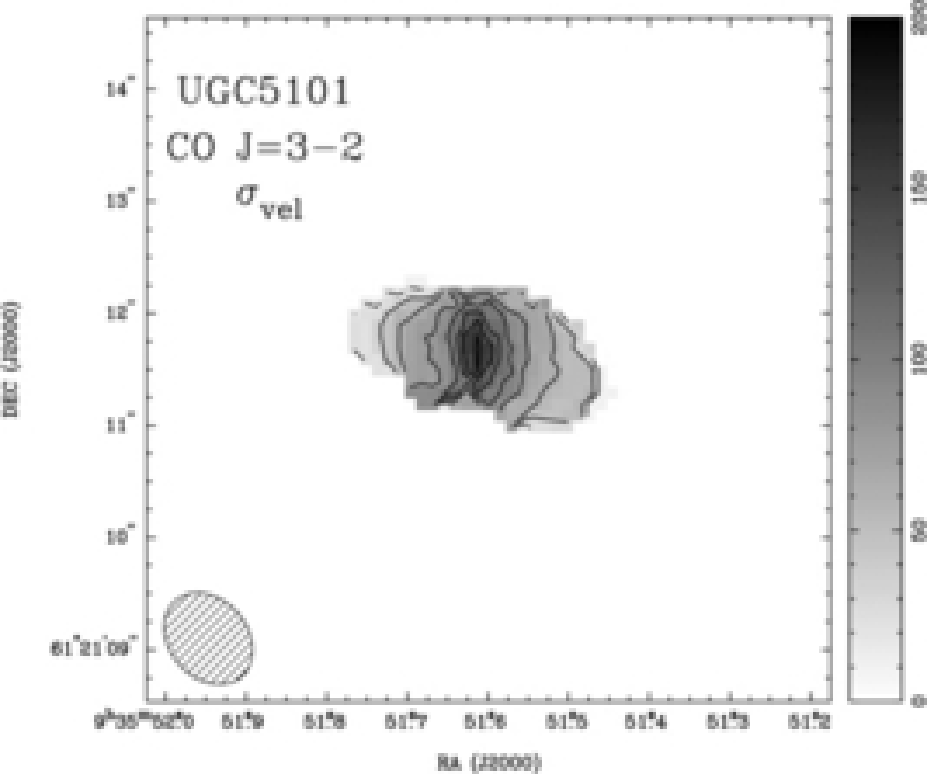}
\includegraphics[angle=-90,scale=.35]{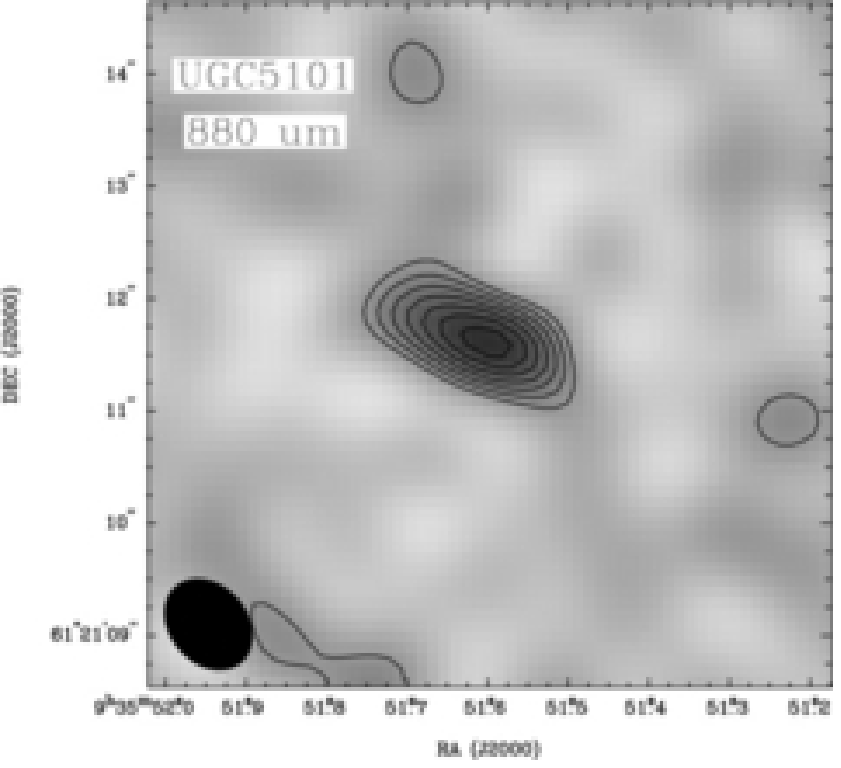}
\caption[UGC5101co32.mom0.eps]{UGC 5101 CO J=3-2 and
880 $\mu$m continuum maps. Notation as in Figure~\ref{fig-I17208co32}.
(a) CO J=3-2 moment 0 map. Lowest
contour is $2 \sigma = 7.6 $ Jy beam$^{-1}$ km s$^{-1}$
 and contours
increase by factors of 1.5.
(b) CO J=3-2 moment 1 map. Contours are 20 km s$^{-1}$ $\times
(-15,-14,-13,...-2,-1,0,1,2,...,10,11)$ relative to $cz$. 
(c) CO J=3-2 moment 2 map. Contours are 20 km s$^{-1}$ $\times (1,2,3,4,5,6,7,8)$.
(d) Uncleaned 880 $\mu$m map. Lowest contour
is $2 \sigma = 5.4$ mJy and contours increase in steps of $1 \sigma$. 
Sidelobes can be seen at the edges of this figure.
\label{fig-UGC5101co32}}
\end{figure}

\begin{figure}
\includegraphics[angle=-90,scale=.3]{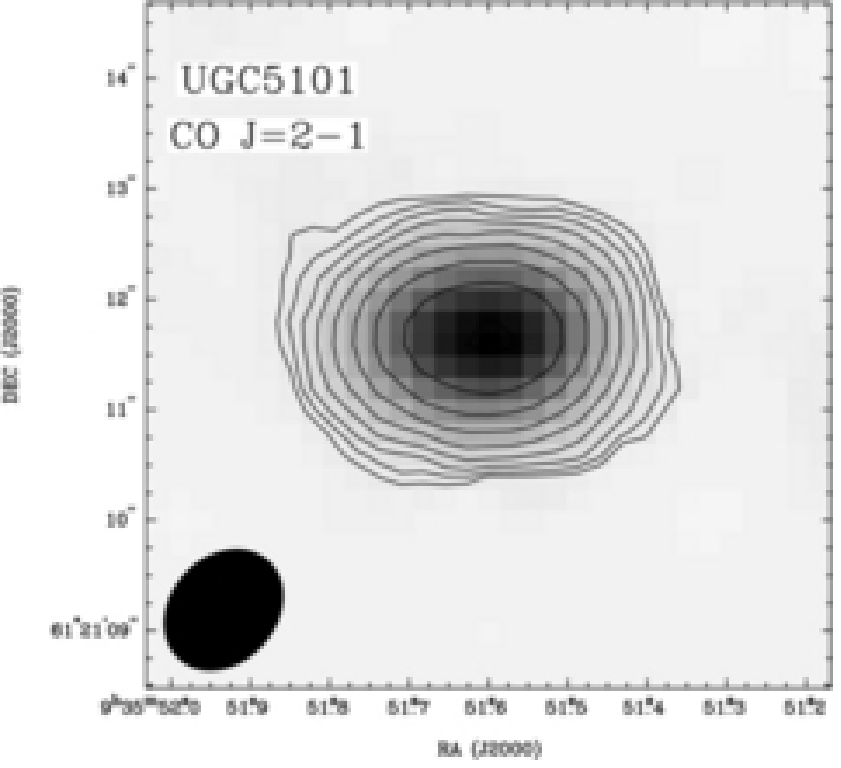}
\includegraphics[angle=-90,scale=.3]{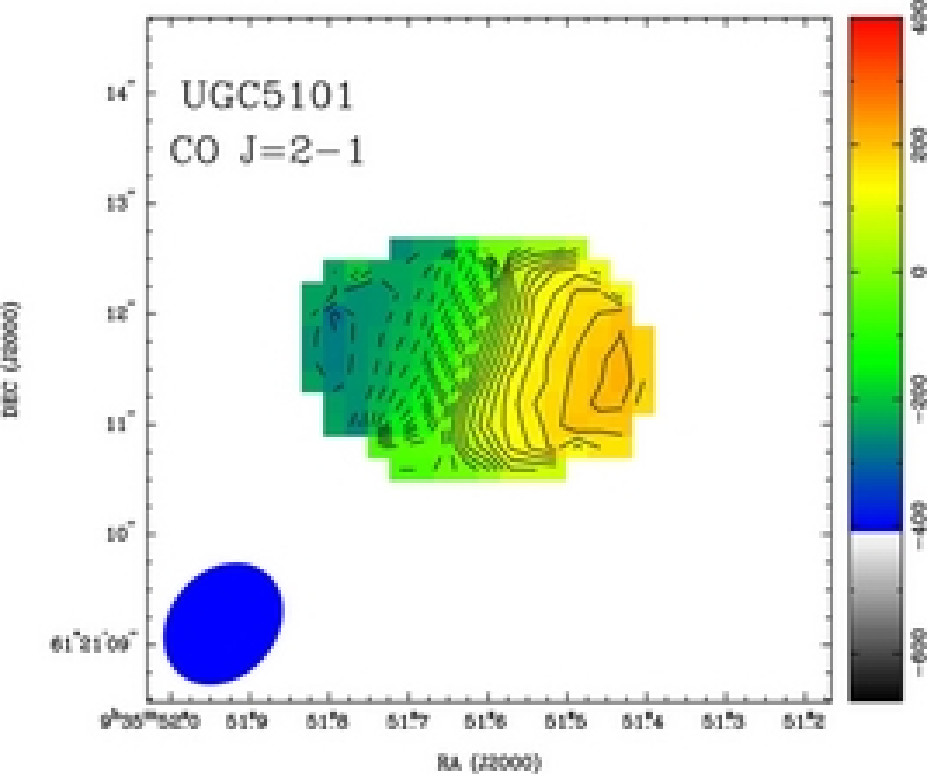}
\includegraphics[angle=-90,scale=.3]{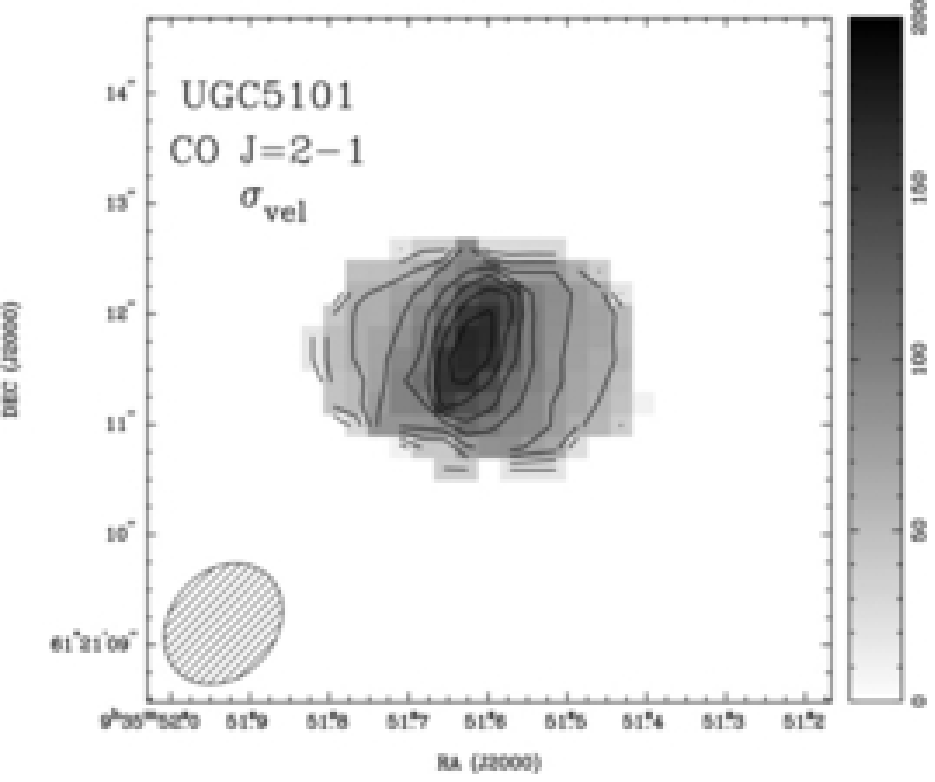}
\includegraphics[angle=-90,scale=.3]{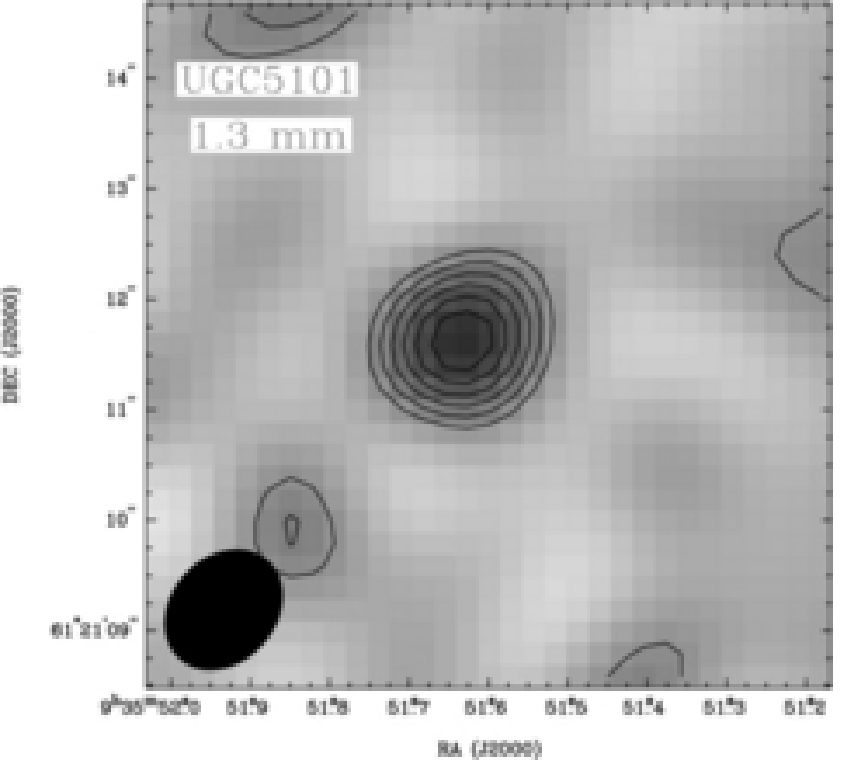}
\caption[UGC5101co21.mom0.eps]{UGC 5101 CO J=2-1 and
1.3 mm continuum maps. Notation as in Figure~\ref{fig-I17208co32}.
(a) CO J=2-1 moment 0 map. Lowest
contour is $\pm 2 \sigma = 4.6 $ Jy beam$^{-1}$ km s$^{-1}$ and contours
increase by factors of 1.5. 
(b) CO J=2-1 moment 1 map. Contours are 20 km s$^{-1}$ $\times
(-13,-12,-11,...-2,-1,0,1,2,...,8,9,10)$ relative to $cz$. 
(c) CO J=2-1 moment 2 map. Contours are 20 km s$^{-1}$ $\times (1,2,3,4,5,6,7,8)$.
(d) Uncleaned 1.3 mm map. Lowest contour
is $2 \sigma = 3.0$ mJy and contours increase in steps of $1 \sigma$.
\label{fig-UGC5101co21}}
\end{figure}

\begin{figure}
\includegraphics[angle=-90,scale=.35]{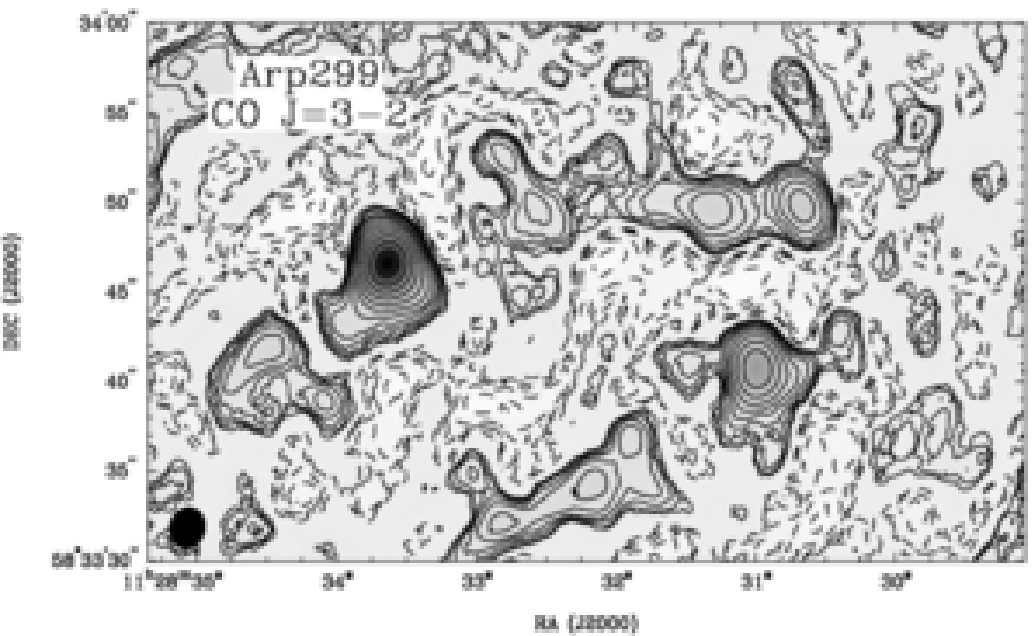}
\includegraphics[angle=-90,scale=.35]{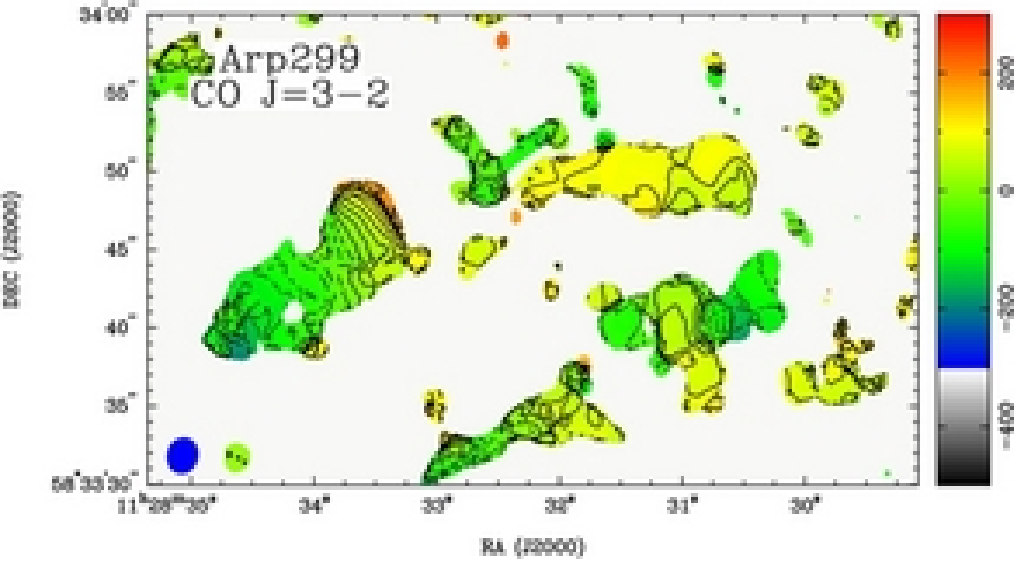}
\includegraphics[angle=-90,scale=.35]{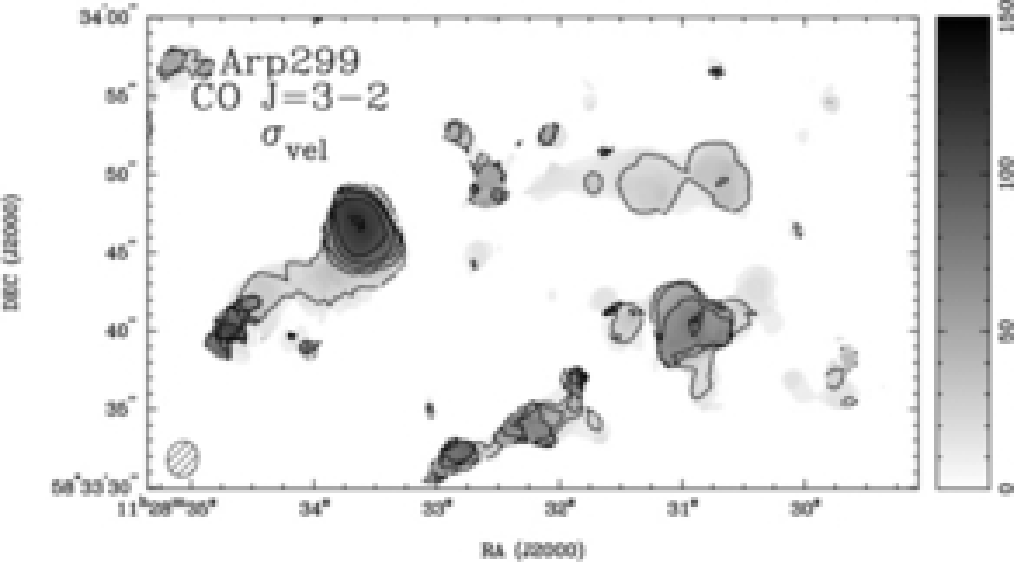}
\includegraphics[angle=-90,scale=.35]{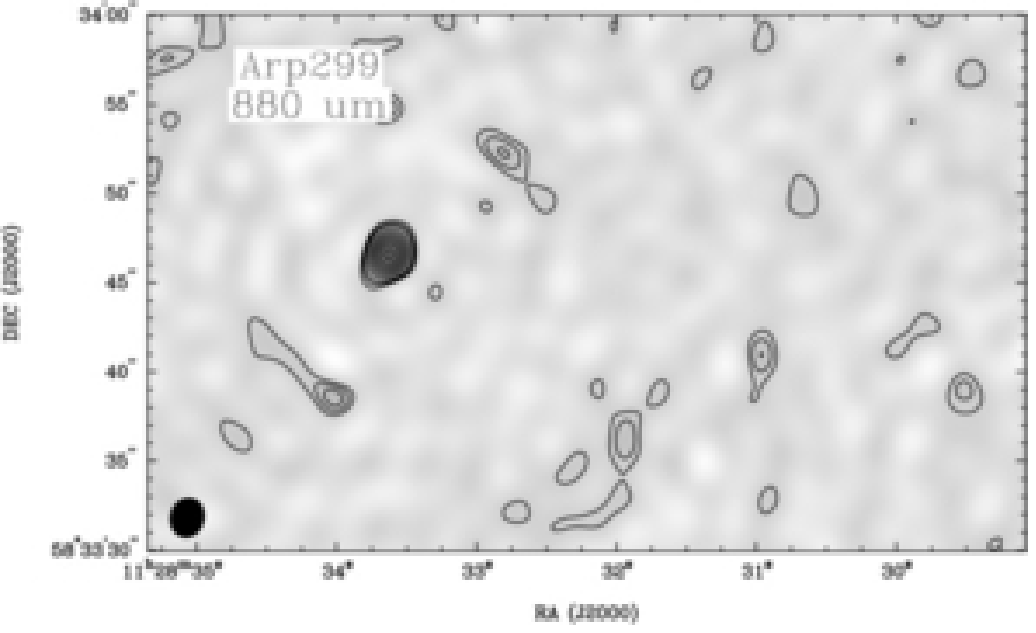}
\includegraphics[angle=-90,scale=.35]{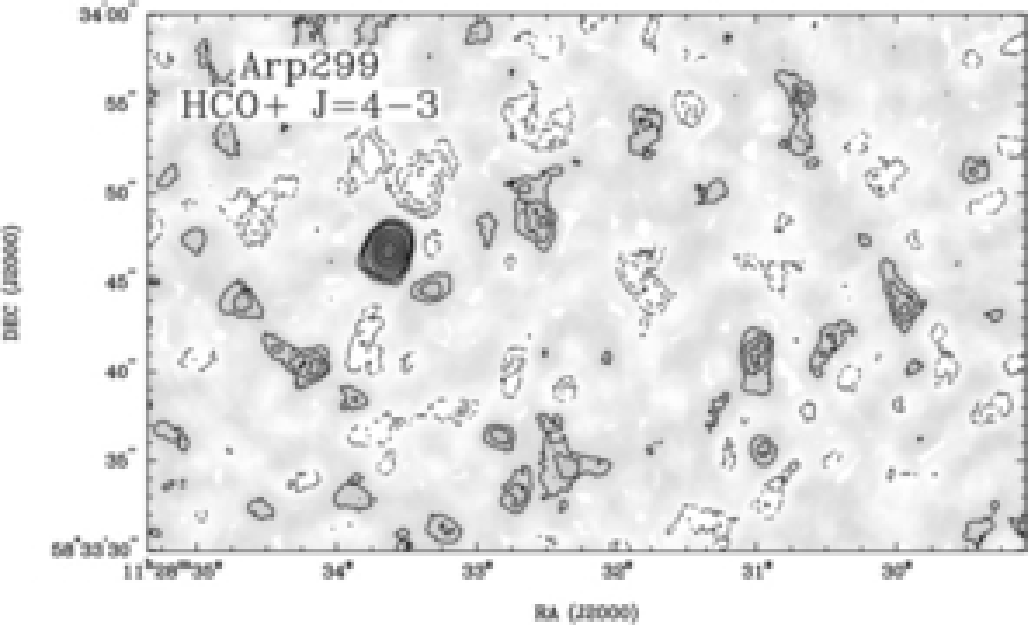}
\caption[Arp299co32.mom0.eps]{Arp 299 CO J=3-2 and
880 $\mu$m continuum maps. Notation as in Figures~\ref{fig-I17208co32}
and \ref{fig-Mrk231co32}.
(a) CO J=3-2 moment 0 map. Lowest
contour is $\pm 2 \sigma = 8.7 $ Jy beam$^{-1}$ km s$^{-1}$ and contours
increase by factors of 1.5.
(b) CO J=3-2 moment 1 map. Contours are 20 km s$^{-1}$ $\times
(-8,-7,-6,-5,-4,-3,-2,-1,0,1,2,3,4,5,6,7,8,9)$ 
relative to $cz$. 
(c) CO J=3-2 moment 2 map. Contours are 20 km s$^{-1}$ $\times (1,2,3,4,5,6)$.
(d) Uncleaned 880 $\mu$m map. Lowest contour
is $2 \sigma = 9.8$ mJy and contours increase in steps of $1 \sigma$
to 4$\sigma$ and then increase in steps of $2 \sigma$.
(e) Uncleaned HCO$^+$ J=4-3 moment 0 map. Lowest
contour is $\pm 2 \sigma = 7.42 $ Jy beam$^{-1}$ km s$^{-1}$ and contours
increase by in steps of $1 \sigma$ to $4\sigma$ and then by steps of
2$\sigma$. 
\label{fig-Arp299co32}}
\end{figure}

\begin{figure}
\includegraphics[angle=-90,scale=.35]{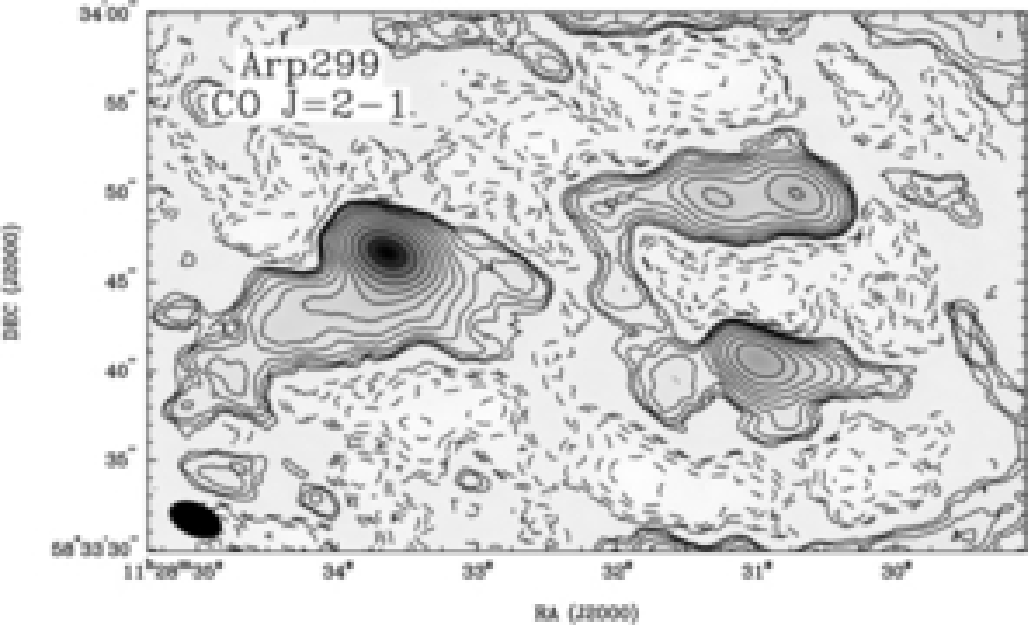}
\includegraphics[angle=-90,scale=.35]{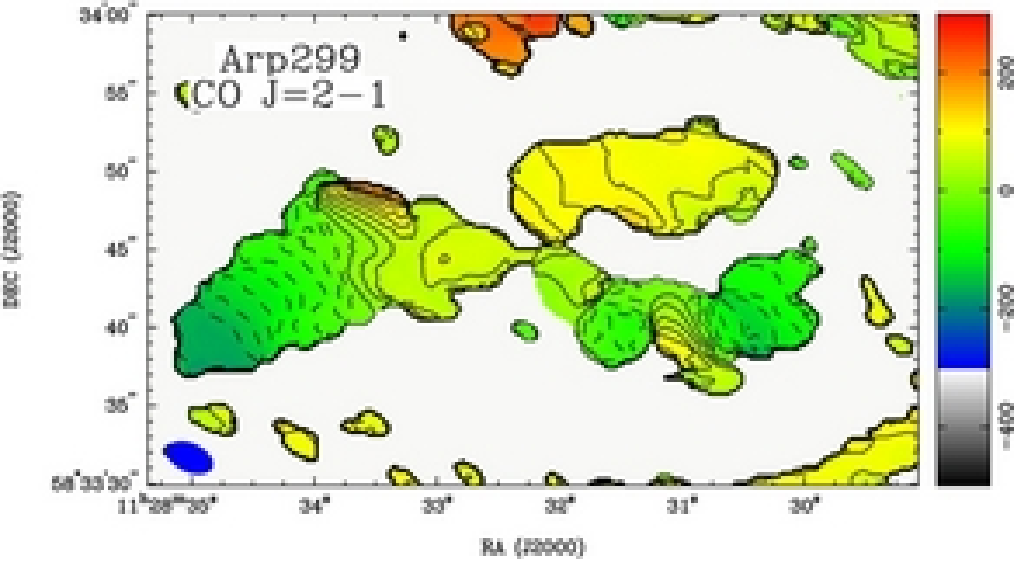}
\includegraphics[angle=-90,scale=.35]{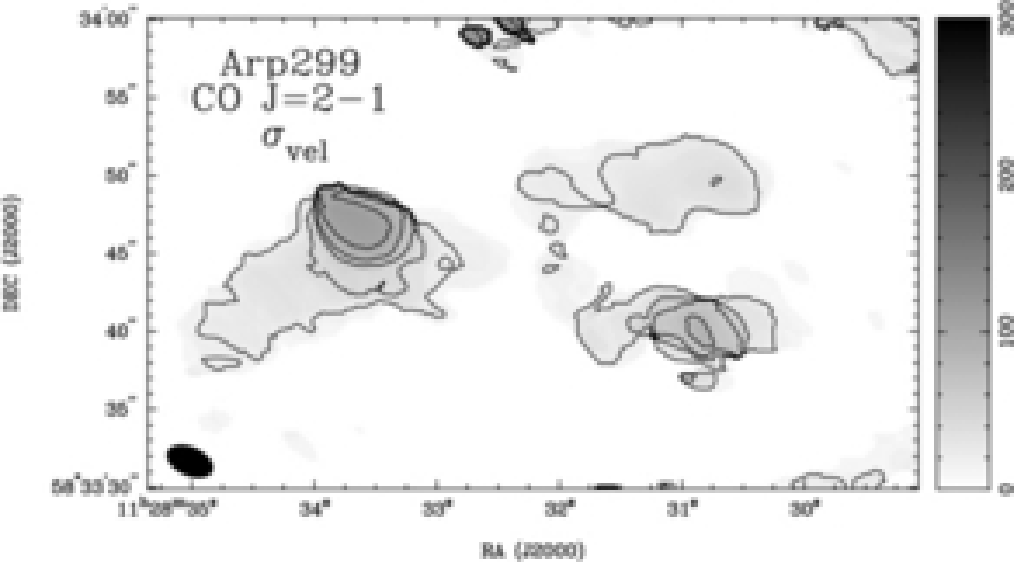}
\includegraphics[angle=-90,scale=.35]{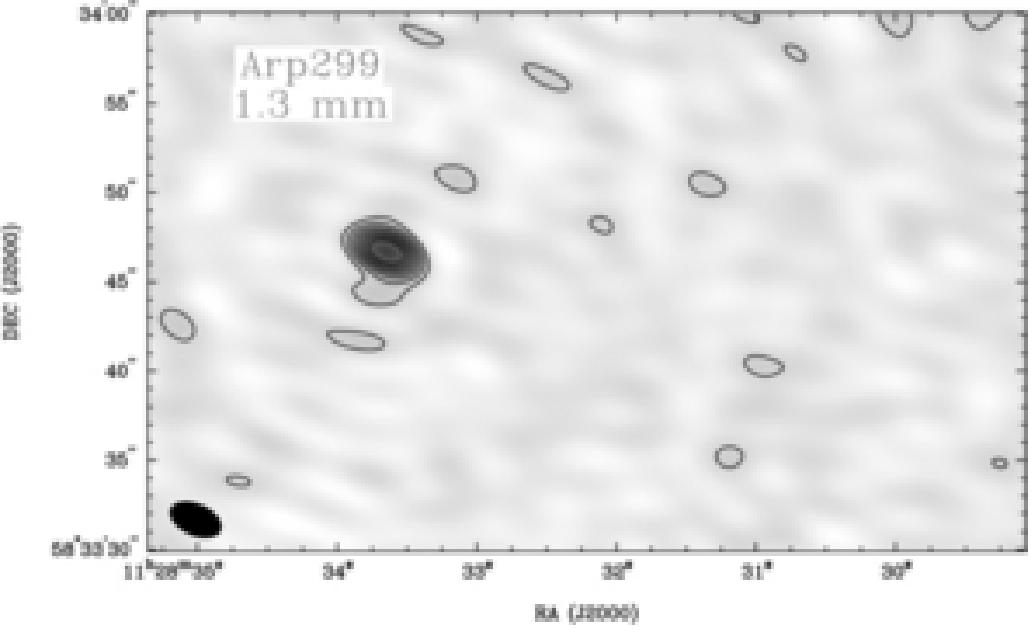}
\includegraphics[angle=-90,scale=.35]{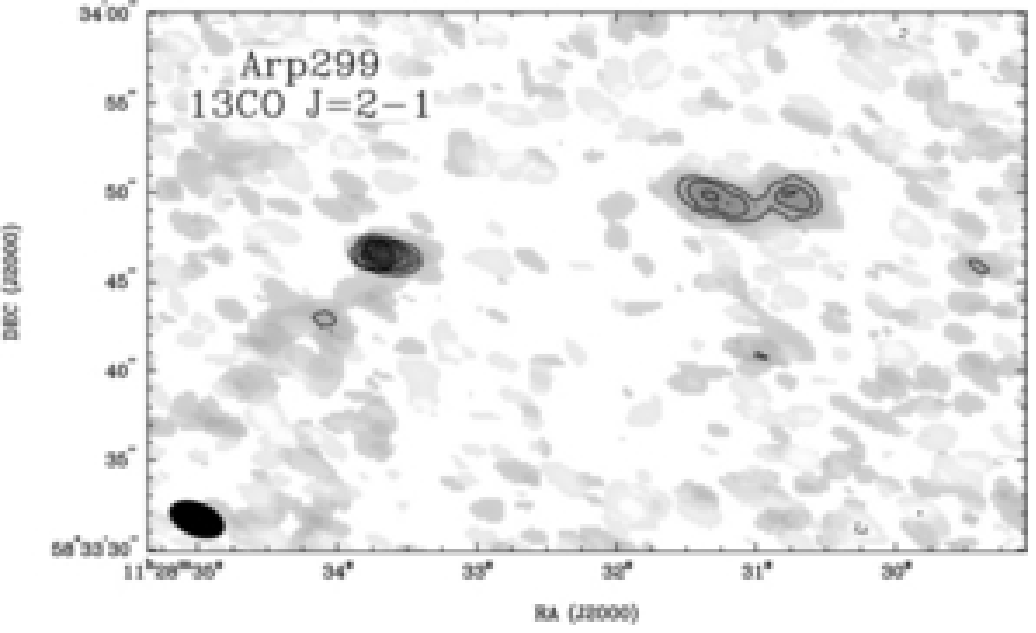}
\caption[Arp299co21.mom0.eps]{Arp 299 CO J=2-1 and
1.3 mm continuum maps. Notation as in Figure~\ref{fig-I17208co32}.
(a) CO J=2-1 moment 0 map. Lowest
contour is $\pm 2 \sigma = 4.6 $ Jy beam$^{-1}$ km s$^{-1}$ and contours
increase by factors of 1.5. 
(b) CO J=2-1 moment 1 map. Contours are 20 km s$^{-1}$ $\times
(-9,-8,-7,-6,-5,-4,-3,-2,-1,0,1,2,3,4,5,6,7,8,9)$ 
relative to $cz$. 
(c) CO J=2-1 moment 2 map. Contours are 20 km s$^{-1}$ $\times (1,2,3,4,5)$.
(d) Uncleaned 1.3 mm map. Lowest contour
is $2 \sigma = 5.4$ mJy and contours increase in steps of $2 \sigma$.
(e) $^{13}$CO J=2-1 moment 0 map. Lowest
contour is $\pm 2 \sigma = 3.9 $ Jy beam$^{-1}$ km s$^{-1}$ and contours
increase in steps of $1 \sigma$. This image has been
corrected for continuum emission by subtracting the 1.3 mm 
continuum in the uv plane before imaging. 
\label{fig-Arp299co21}}
\end{figure}

\begin{figure}
\includegraphics[angle=-90,scale=.35]{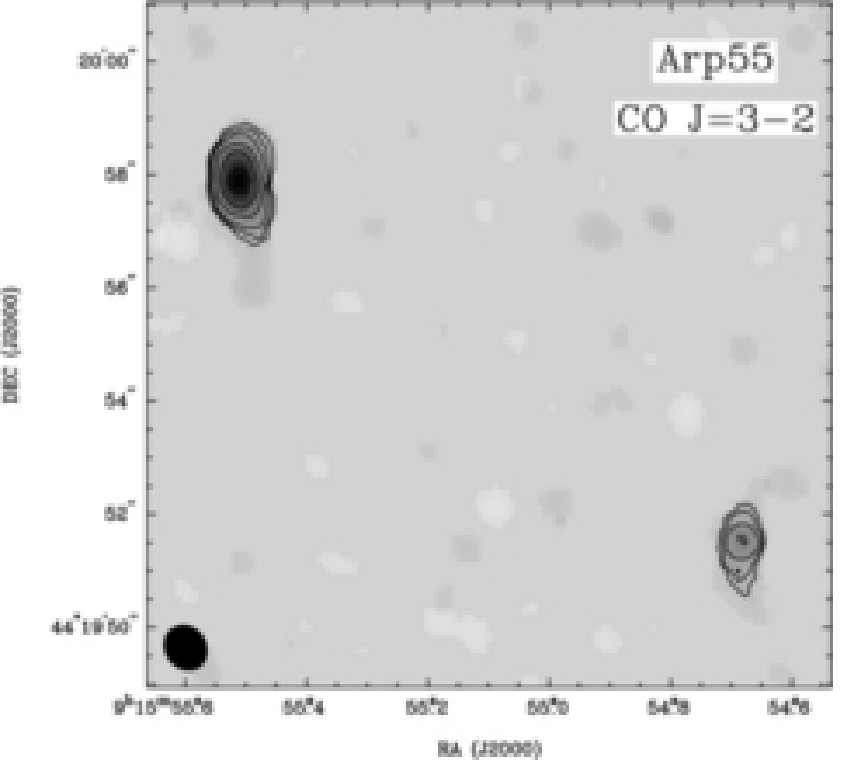}
\includegraphics[angle=-90,scale=.35]{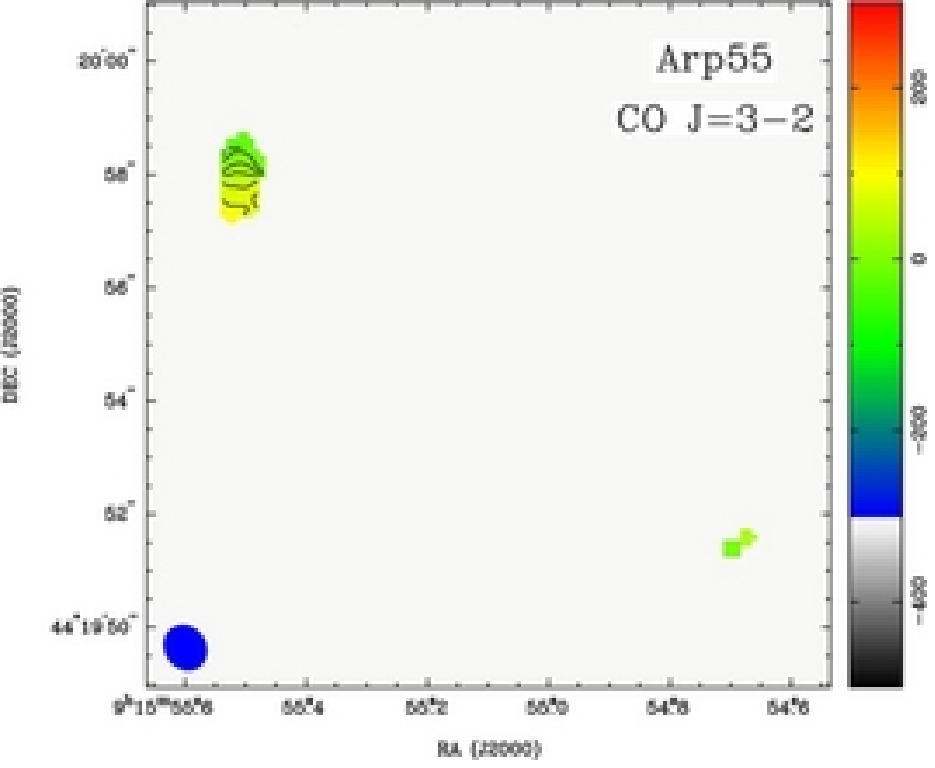}
\includegraphics[angle=-90,scale=.35]{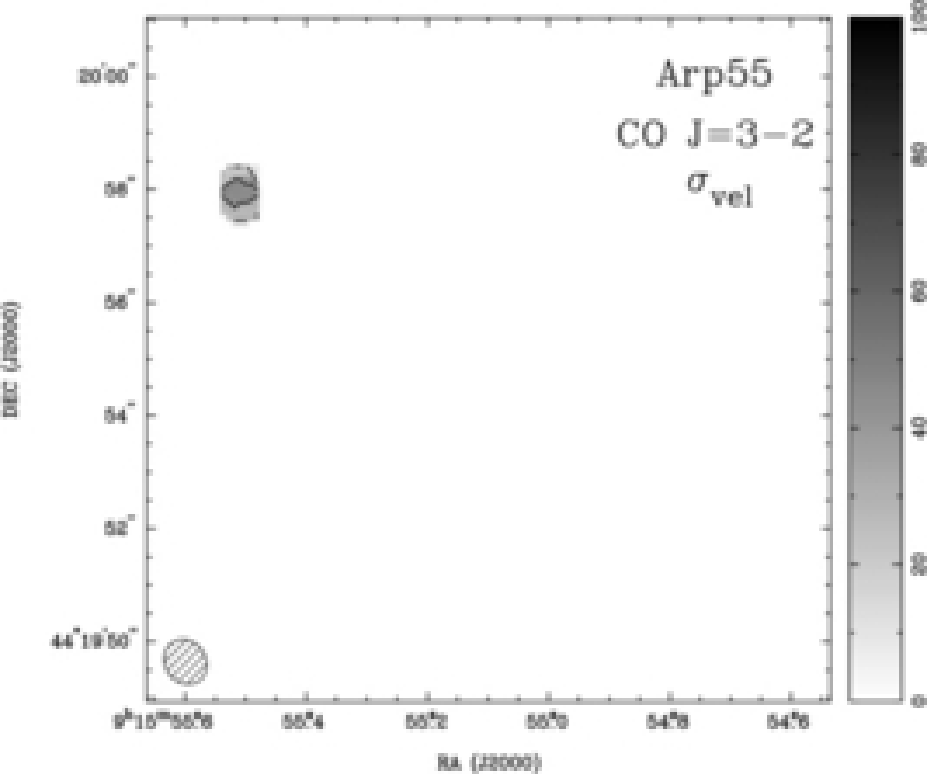}
\includegraphics[angle=-90,scale=.35]{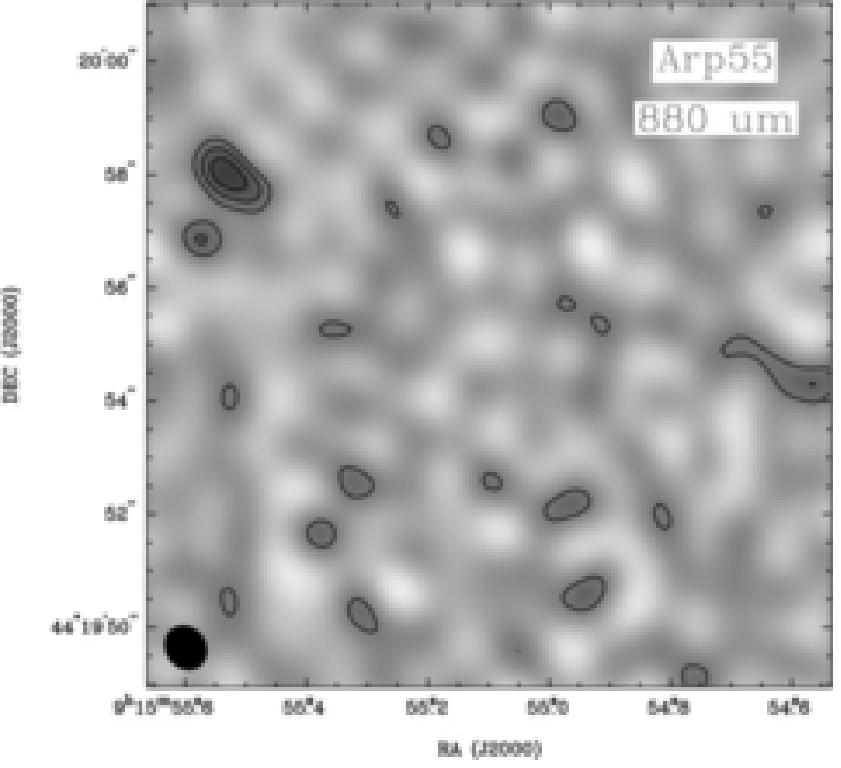}
\caption[Arp55co32.mom0.eps]{Arp 55 CO J=3-2 and
880 $\mu$m continuum maps. Notation as in Figure~\ref{fig-I17208co32}.
(a) CO J=3-2 moment 0 map. Lowest
contour is $2 \sigma = 6.0 $ Jy beam$^{-1}$ km s$^{-1}$ and contours
increase by factors of 1.5.
(b) CO J=3-2 moment 1 map. Contours are 20 km s$^{-1}$ $\times
(-1,0,1,2,3,4)$ relative to $cz$. 
(c) CO J=3-2 moment 2 map. Contours are 20 km s$^{-1}$ $\times (1,2)$.
(d) Uncleaned 880 $\mu$m map. Lowest contour
is $2 \sigma = 8.2$ mJy and contours increase in steps of $1 \sigma$.
\label{fig-Arp55co32}}
\end{figure}

\begin{figure}
\includegraphics[angle=-90,scale=.35]{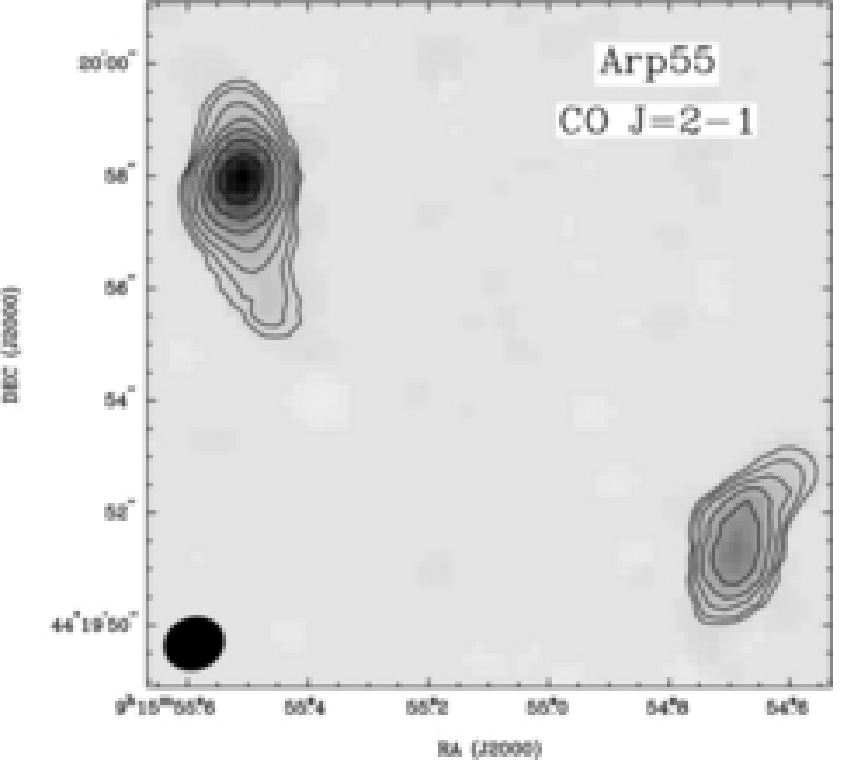}
\includegraphics[angle=-90,scale=.35]{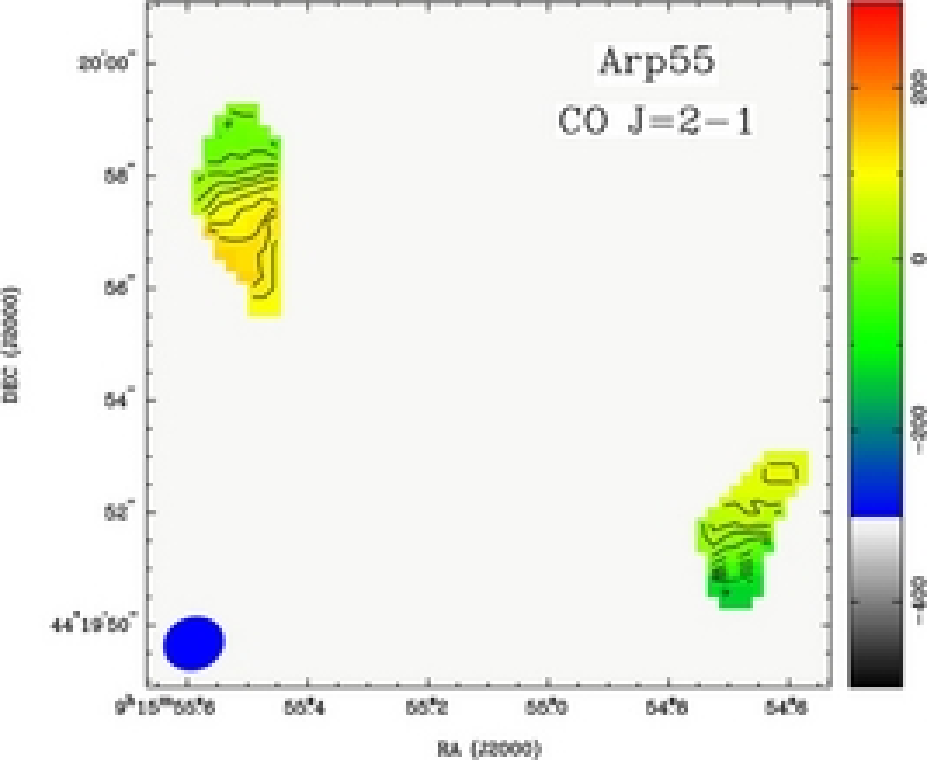}
\includegraphics[angle=-90,scale=.35]{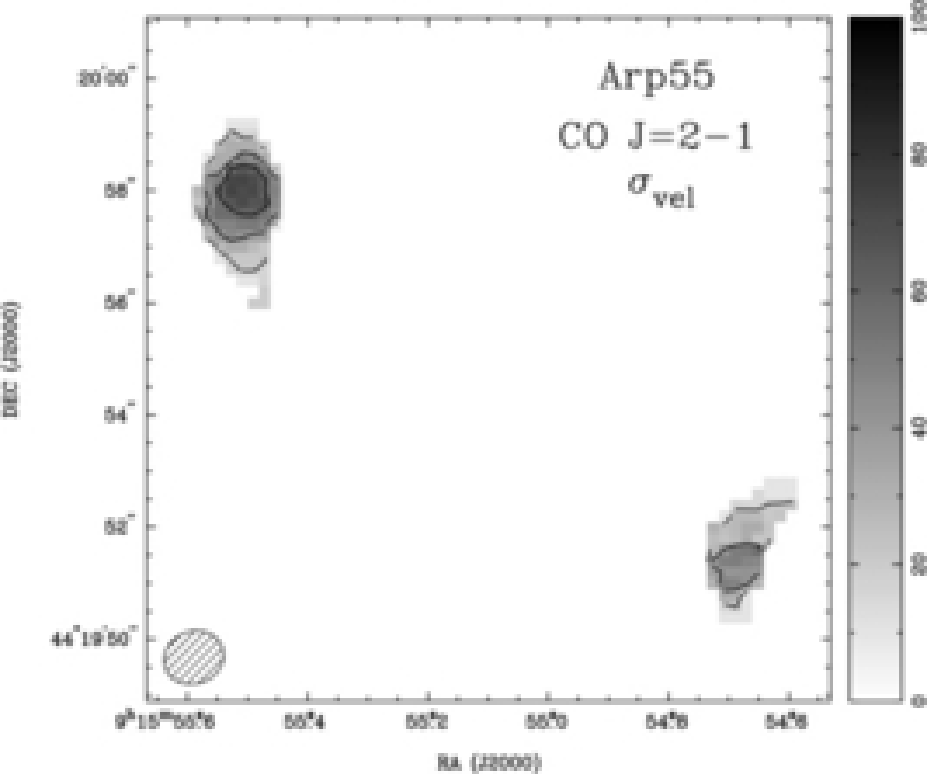}
\caption[Arp55co21.mom0.eps]{Arp 55 CO J=2-1 
maps. Notation as in Figure~\ref{fig-I17208co32}.
(a) CO J=2-1 moment 0 map. Lowest
contour is $2 \sigma = 3.4 $ Jy beam$^{-1}$ km s$^{-1}$ and contours
increase by factors of 1.5.
(b) CO J=2-1 moment 1 map. Contours are 20 km s$^{-1}$ $\times
(-7,-6,-5,-4,-3,-2,-1,0,1,2,3,4,5,6)$ 
relative to $cz$. 
Note that the positive end of the rotation curve in
the north-east component peaks at 120 km/s and then drops to 100 km/s
at the extreme southern end of the emission.
(c) CO J=2-1 moment 2 map. Contours are 20 km s$^{-1}$ $\times (1,2,3)$.
\label{fig-Arp55co21}}
\end{figure}

\begin{figure}
\includegraphics[angle=-90,scale=.3]{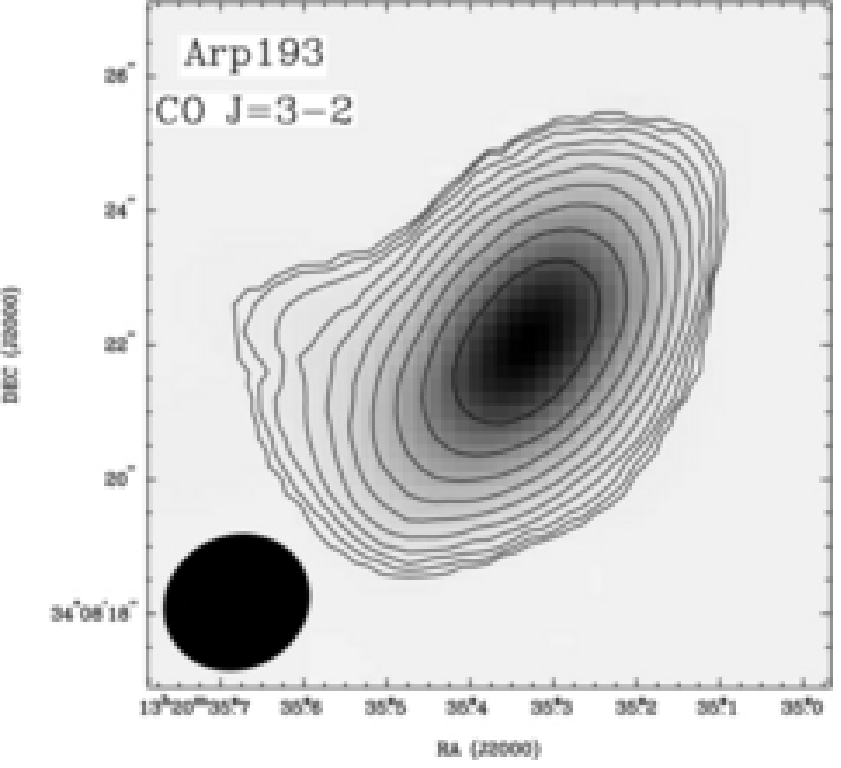}
\includegraphics[angle=-90,scale=.3]{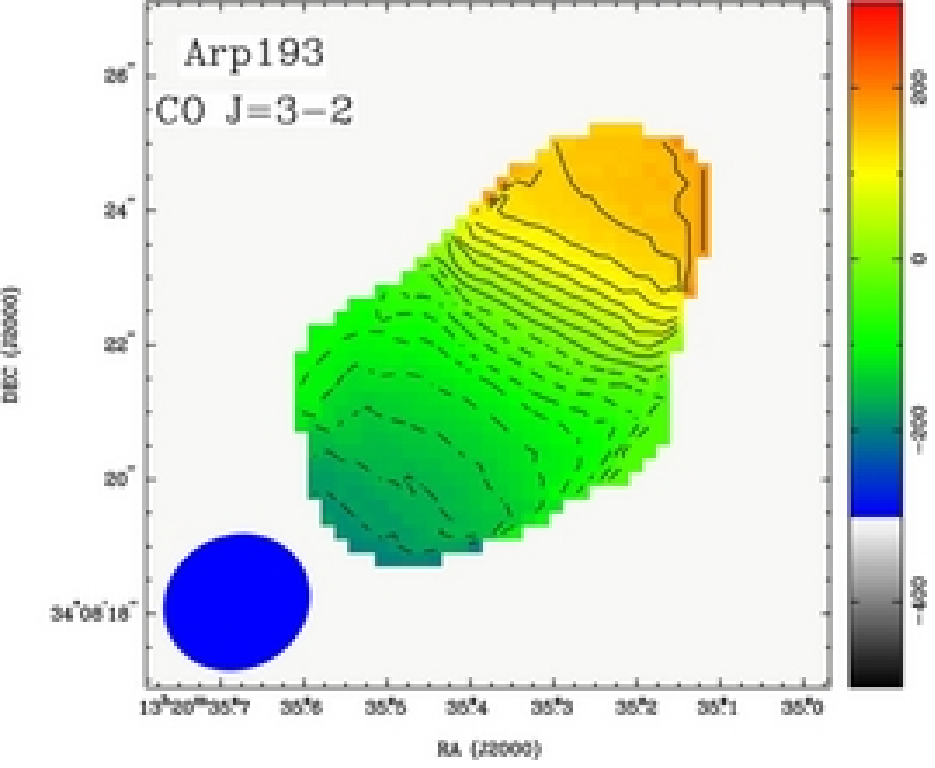}
\includegraphics[angle=-90,scale=.3]{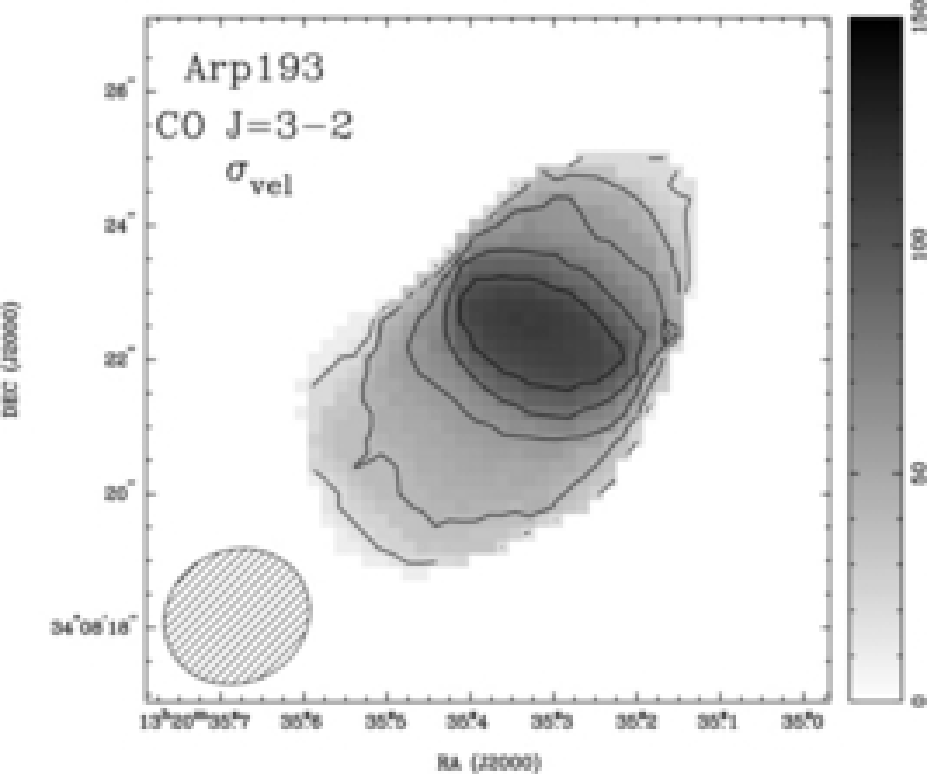}
\includegraphics[angle=-90,scale=.3]{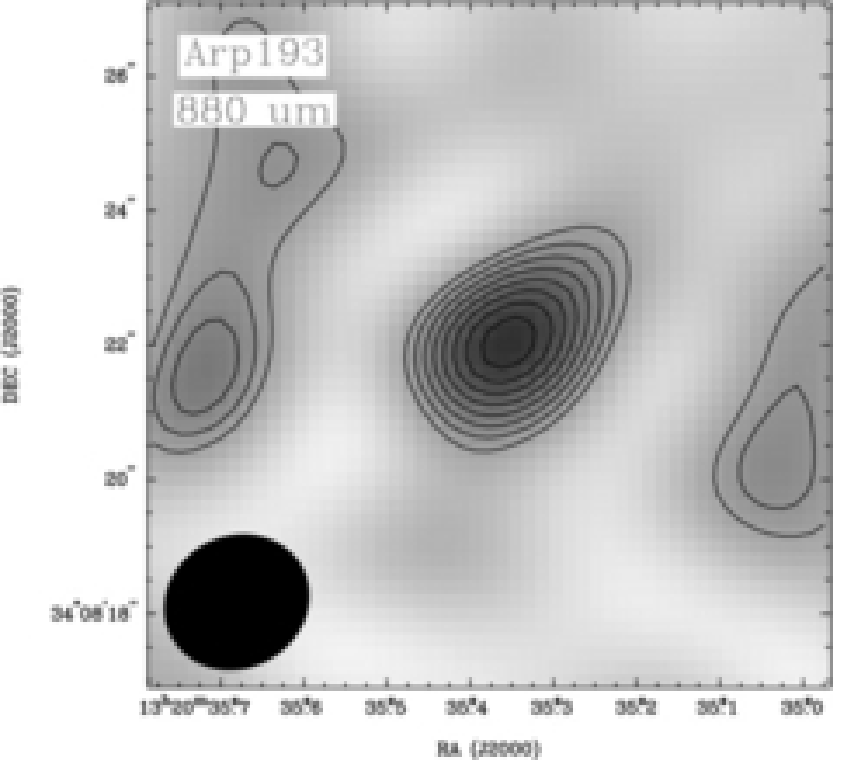}
\includegraphics[angle=-90,scale=.3]{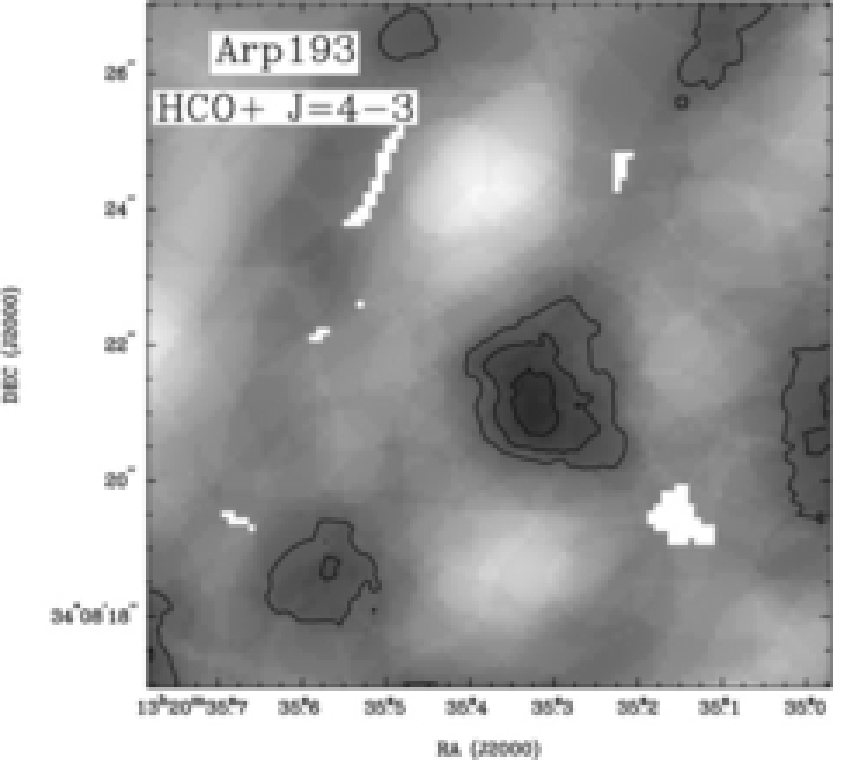}
\caption[Arp193co32.mom0.eps]{Arp 193 CO J=3-2 and
880 $\mu$m continuum maps. Notation as in Figures~\ref{fig-I17208co32}
and~\ref{fig-Mrk231co32}.
(a) CO J=3-2 moment 0 map. Lowest
contour is $2 \sigma = 8.5 $ Jy beam$^{-1}$ km s$^{-1}$ and contours
increase by factors of 1.5.
(b) CO J=3-2 moment 1 map. Contours are 20 km s$^{-1}$ $\times
(-9,-8,-7,-6,-5,-4,-3,-2,-1,0,1,2,3,4,5,6,7,8,9)$ 
relative to $cz$. 
(c) CO J=3-2 moment 2 map. Contours are 20 km s$^{-1}$ $\times (1,2,3,4,5)$.
(d) Uncleaned 880 $\mu$m map. Lowest contour
is $2 \sigma = 8.0$ mJy and contours increase in steps of $1 \sigma$. 
Sidelobes can be seen at the edges of this figure.
(e) Uncleaned HCO$^+$ J=4-3 moment 0 map. Lowest
contour is $2 \sigma = 8.84 $ Jy beam$^{-1}$ km s$^{-1}$ and contours
increase by in steps of $1 \sigma$. 
\label{fig-Arp193co32}}
\end{figure}

\begin{figure}
\includegraphics[angle=-90,scale=.35]{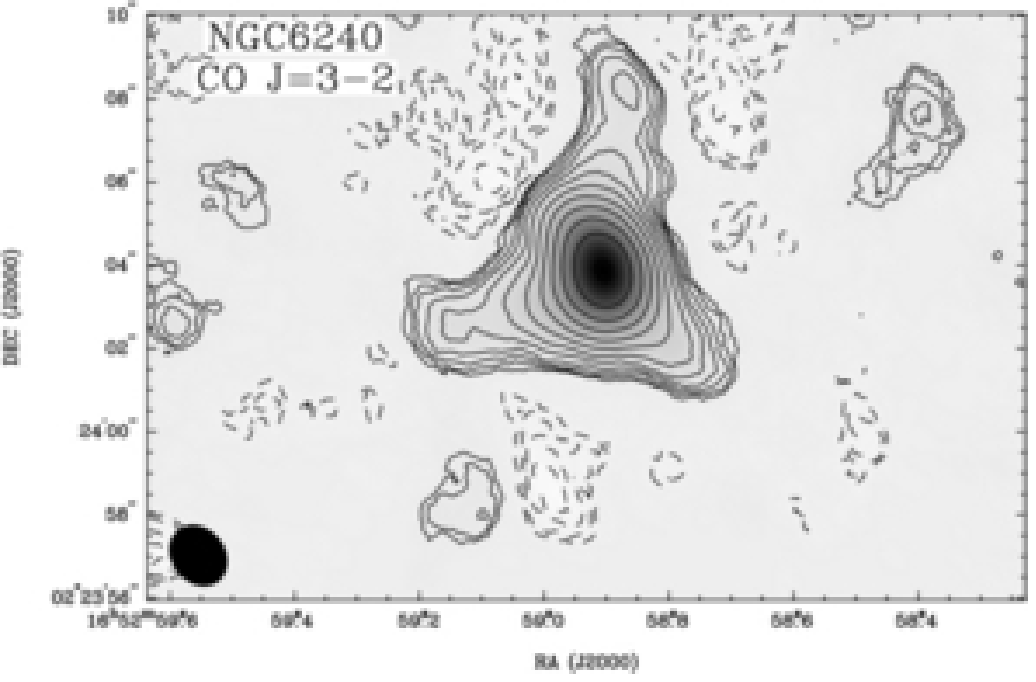}
\includegraphics[angle=-90,scale=.35]{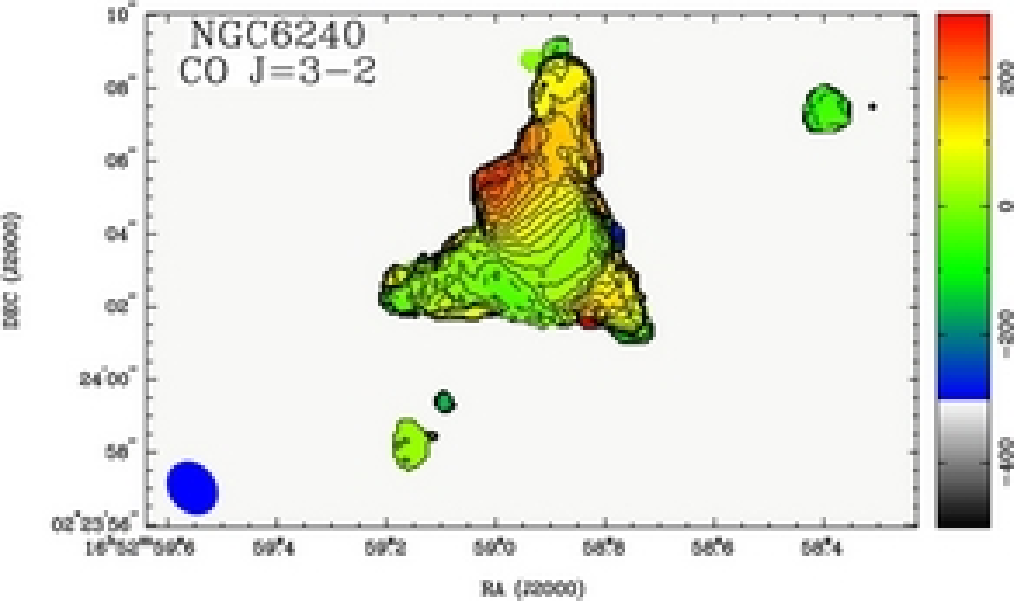}
\includegraphics[angle=-90,scale=.35]{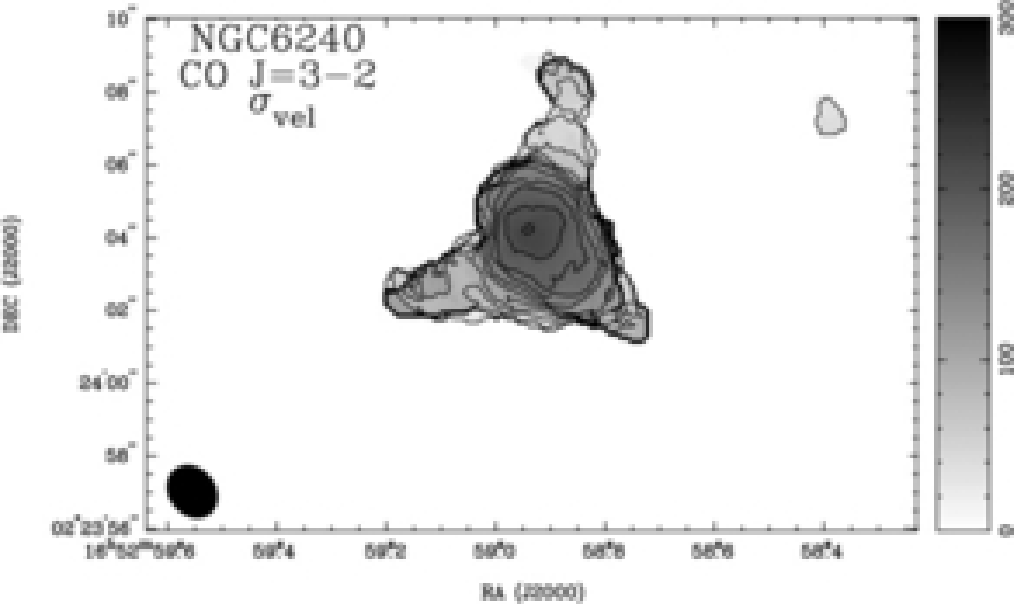}
\includegraphics[angle=-90,scale=.35]{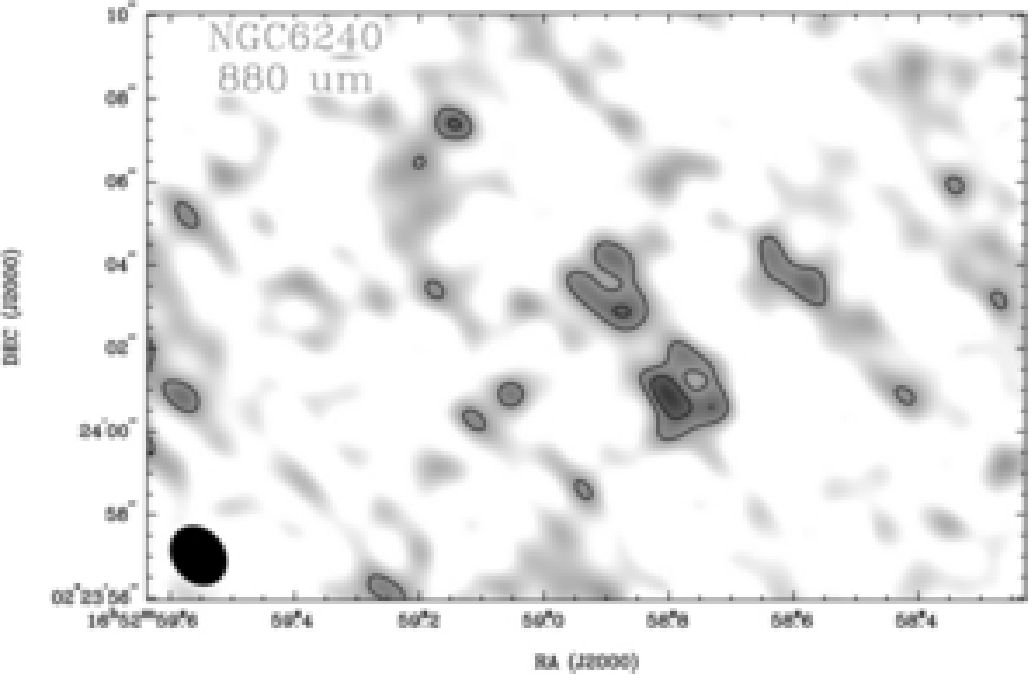}
\includegraphics[angle=-90,scale=.35]{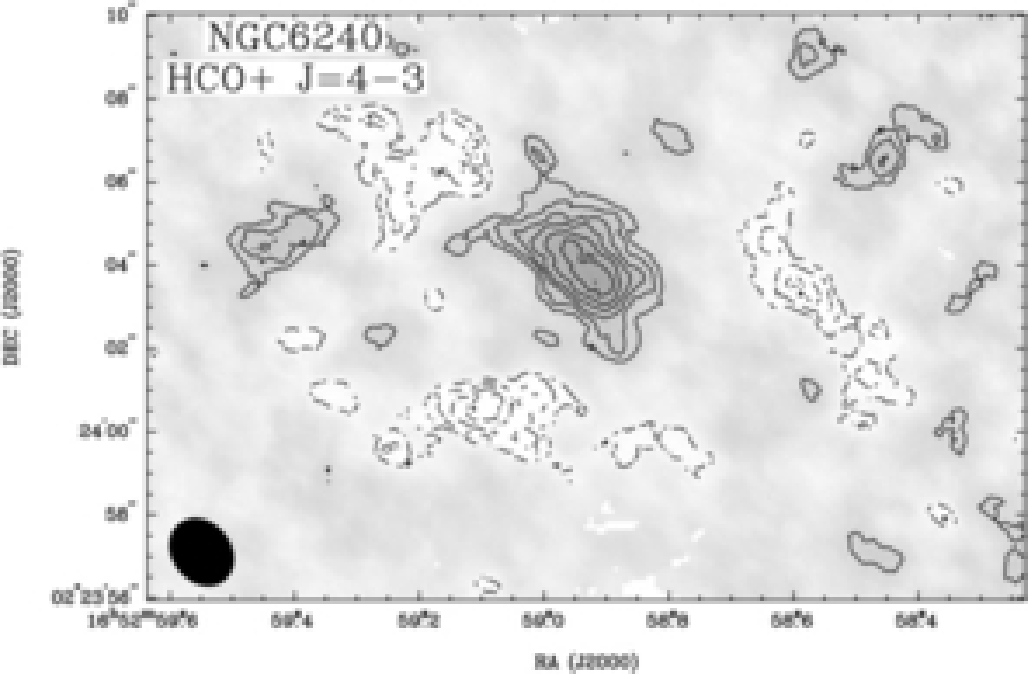}
\caption[NGC6240co32.mom0.eps]{NGC 6240 CO J=3-2 and
880 $\mu$m continuum maps. Notation as in Figures~\ref{fig-I17208co32}
and \ref{fig-Mrk231co32}.
(a) CO J=3-2 moment 0 map. Lowest
contour is $2 \sigma = 12.8 $ Jy beam$^{-1}$ km s$^{-1}$ and contours
increase by factors of 1.5.
(b) CO J=3-2 moment 1 map. Contours are 20 km s$^{-1}$ $\times
(-2,-1,0,1,2,3,4,5,6,7,8,9,10,11,12,13)$ 
relative to $cz$. 
(c) CO J=3-2 moment 2 map. Contours are 20 km s$^{-1}$ $\times
(1,2,3,4,5,6,7,8,9,10,11)$. 
(d) Uncleaned 880 $\mu$m map. Lowest contour
is $2 \sigma = 17$ mJy and contours increase in steps of $1 \sigma$. 
(e) Uncleaned HCO$^+$ J=4-3 moment 0 map. Lowest
contour is $\pm 2 \sigma = 10.22 $ Jy beam$^{-1}$ km s$^{-1}$ and contours
increase in steps of $1 \sigma$. 
\label{fig-NGC6240co32}}
\end{figure}

\begin{figure}
\includegraphics[angle=-90,scale=.3]{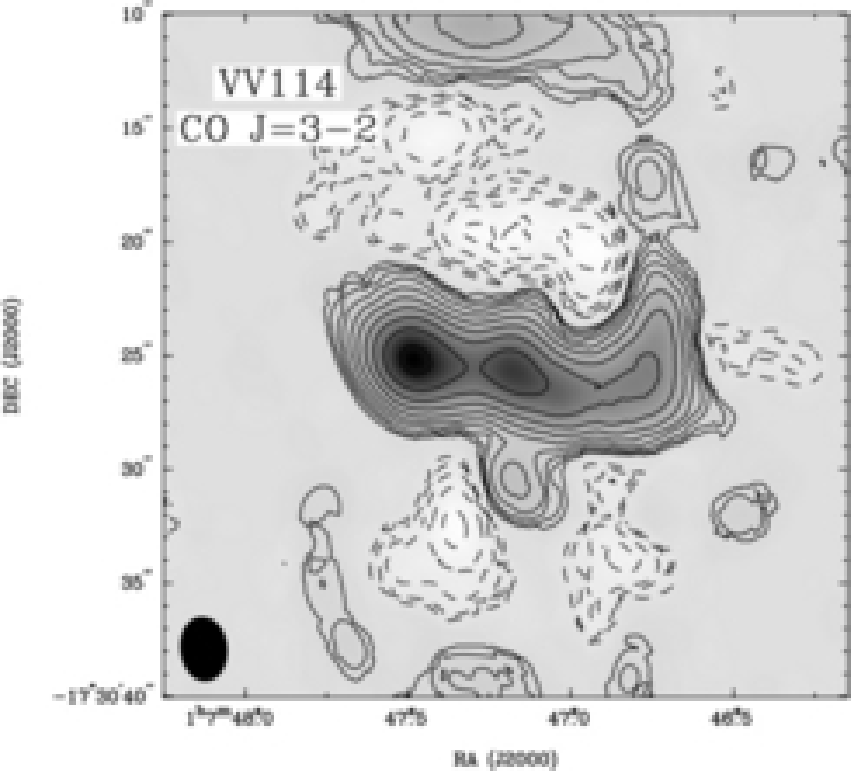}
\includegraphics[angle=-90,scale=.3]{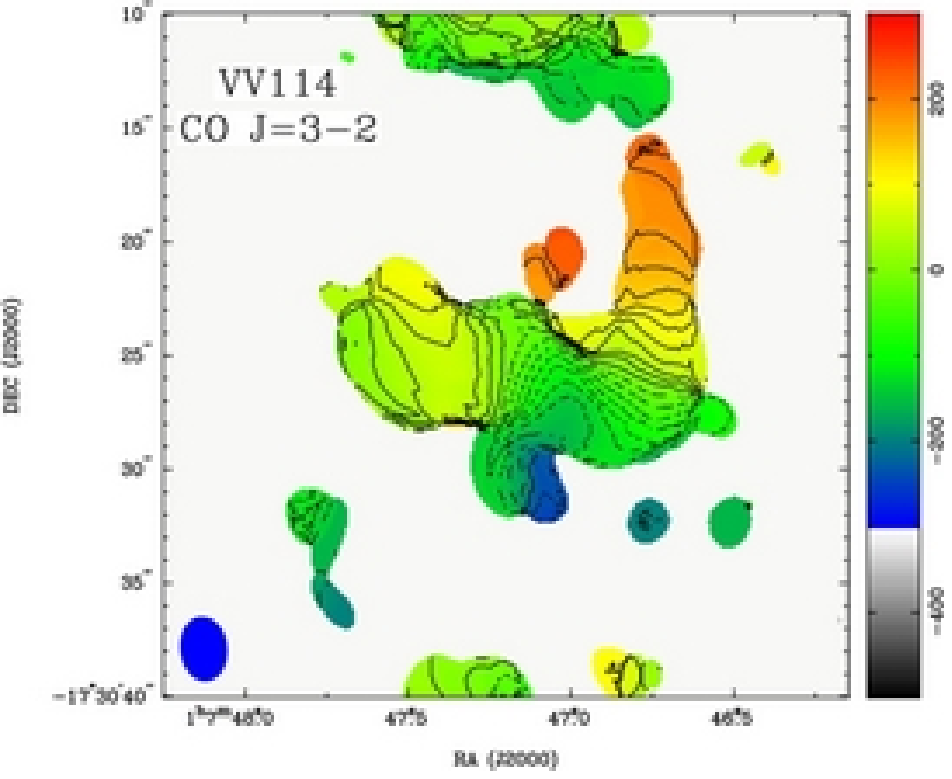}
\includegraphics[angle=-90,scale=.3]{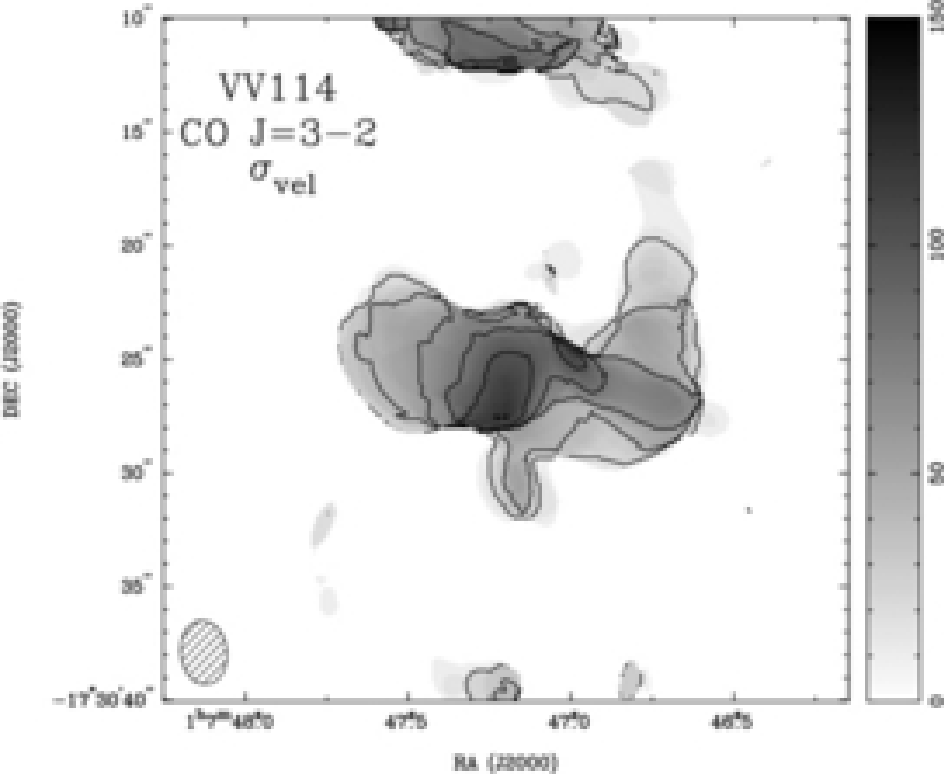}
\includegraphics[angle=-90,scale=.3]{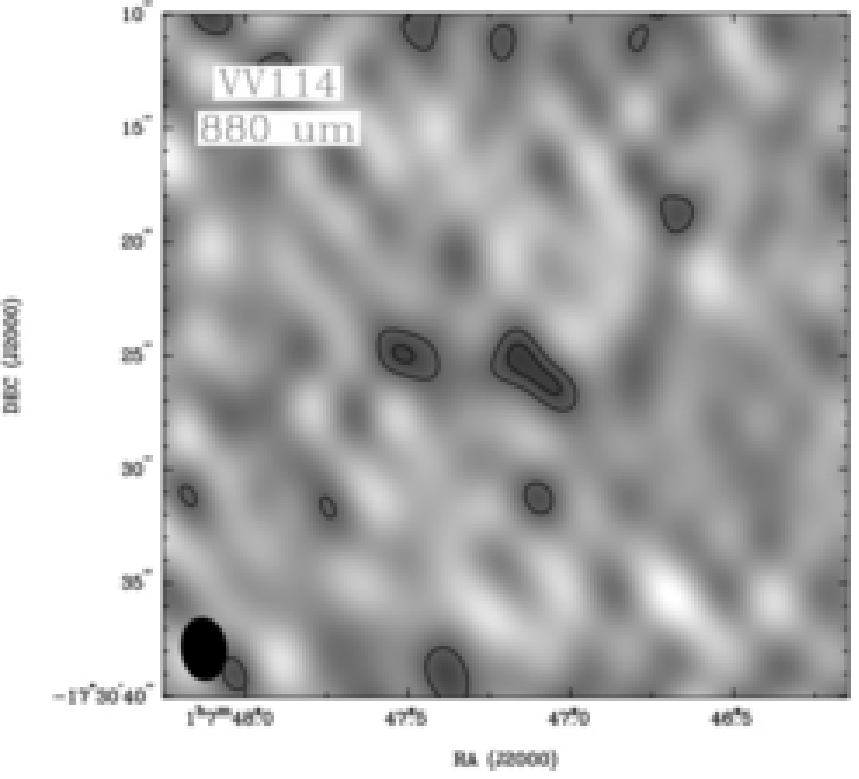}
\includegraphics[angle=-90,scale=.3]{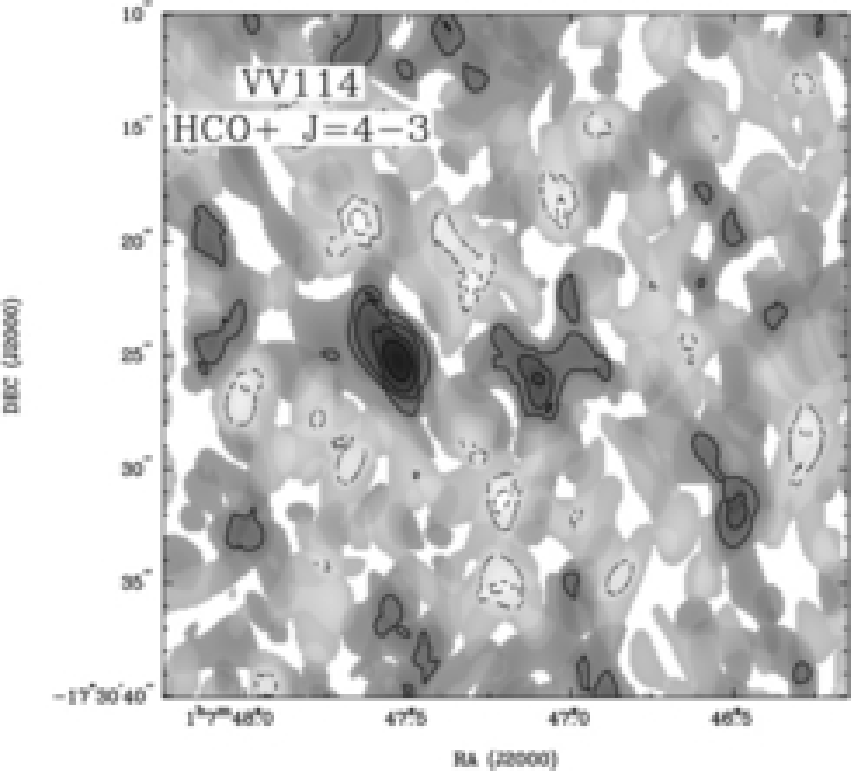}
\caption[VV114co32.mom0.eps]{VV 114 CO J=3-2 and
880 $\mu$m continuum maps. Notation as in Figures~\ref{fig-I17208co32}
and \ref{fig-Mrk231co32}.
(a) CO J=3-2 moment 0 map. Lowest
contour is $\pm 2 \sigma = 7.0 $ Jy beam$^{-1}$ km s$^{-1}$ and contours
increase by factors of 1.5. The regions of positive emission at the
extreme northern and southern edges of the image are artifacts.
(b) CO J=3-2 moment 1 map. Contours are 20 km s$^{-1}$ $\times
(-12,-11,-10,...,-2,-1,0,1,2,...,9,10,11)$ 
relative to $cz$. 
(c) CO J=3-2 moment 2 map. Contours are 20 km s$^{-1}$ $\times (1,2,3,4,5)$.
(d) Uncleaned 880 $\mu$m map. Lowest contour
is $2 \sigma = 7.8$ mJy and contours increase in steps of $1 \sigma$.
(e) HCO$^+$ J=4-3 moment 0 map. Lowest
contour is $\pm 2 \sigma = 3.68 $ Jy beam$^{-1}$ km s$^{-1}$ and contours
increase in steps of $1 \sigma$. The eastern component contains
60\% of the flux given in Table~\ref{tbl-fluxes}. 
\label{fig-VV114co32}}
\end{figure}

\begin{figure}
\includegraphics[angle=-90,scale=.3]{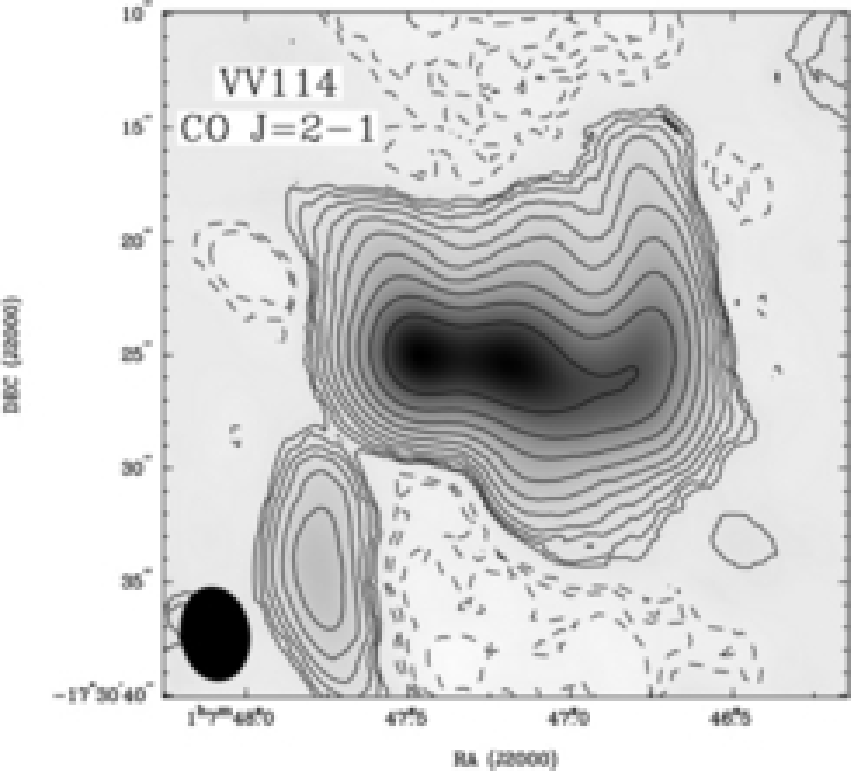}
\includegraphics[angle=-90,scale=.3]{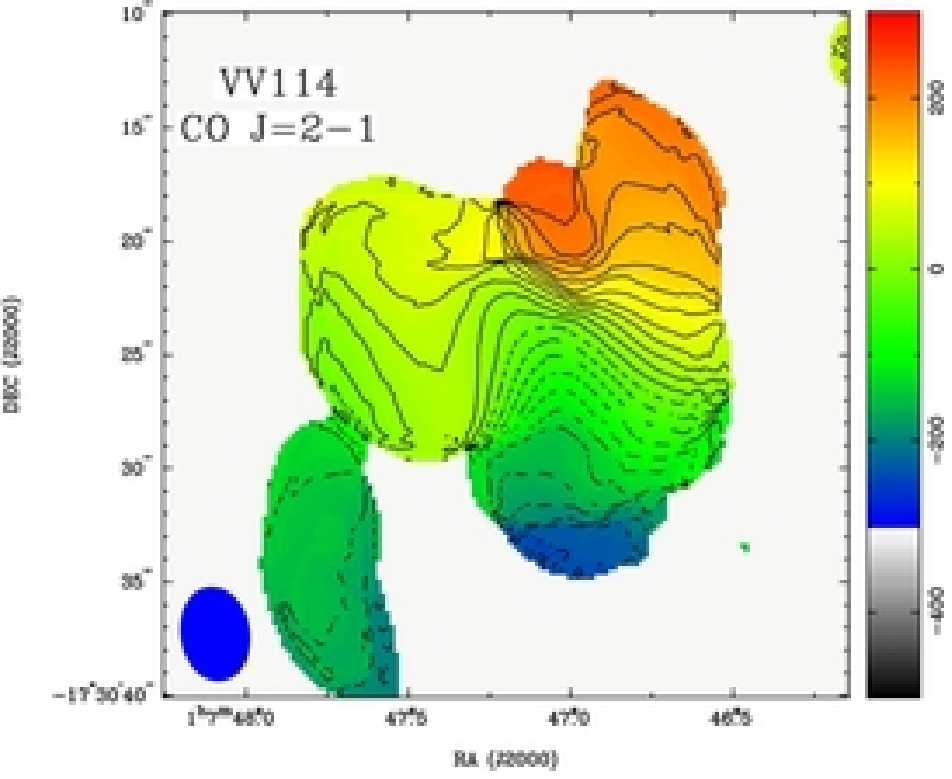}
\includegraphics[angle=-90,scale=.3]{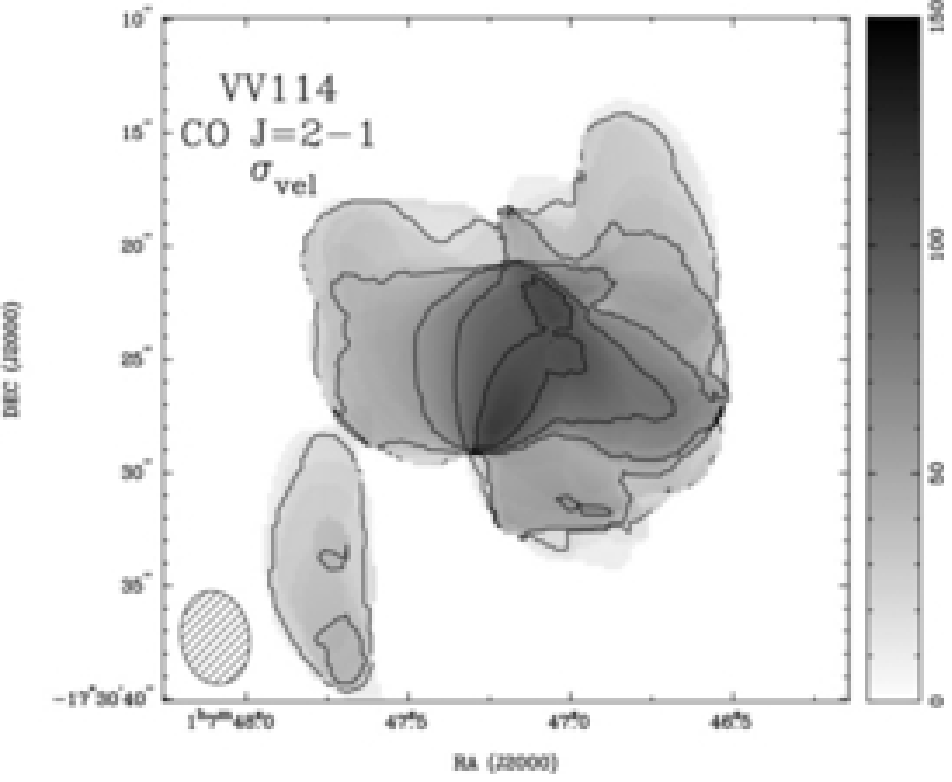}
\includegraphics[angle=-90,scale=.3]{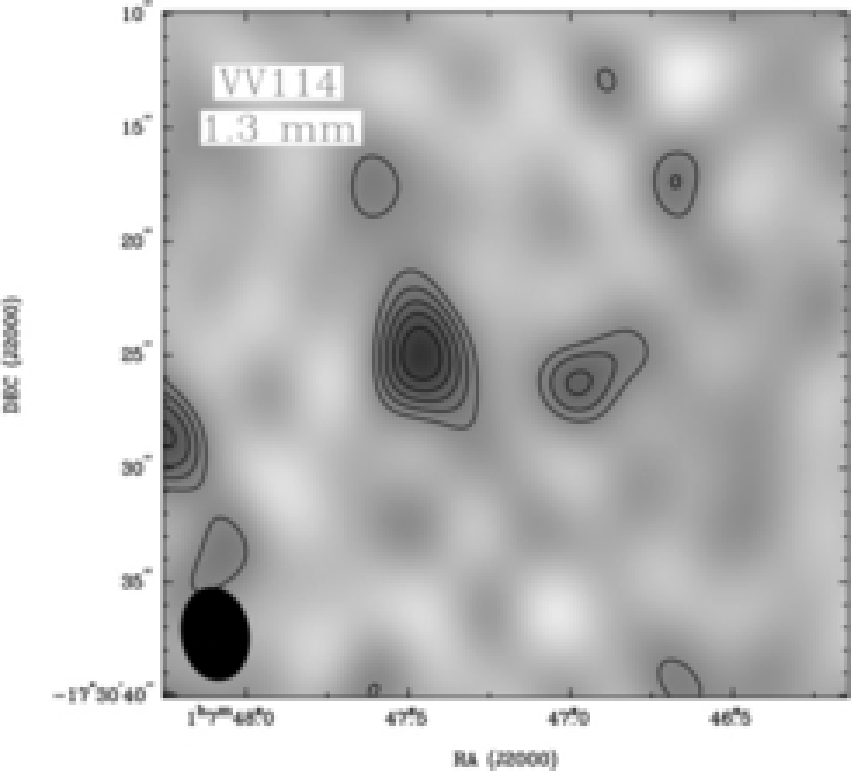}
\includegraphics[angle=-90,scale=.3]{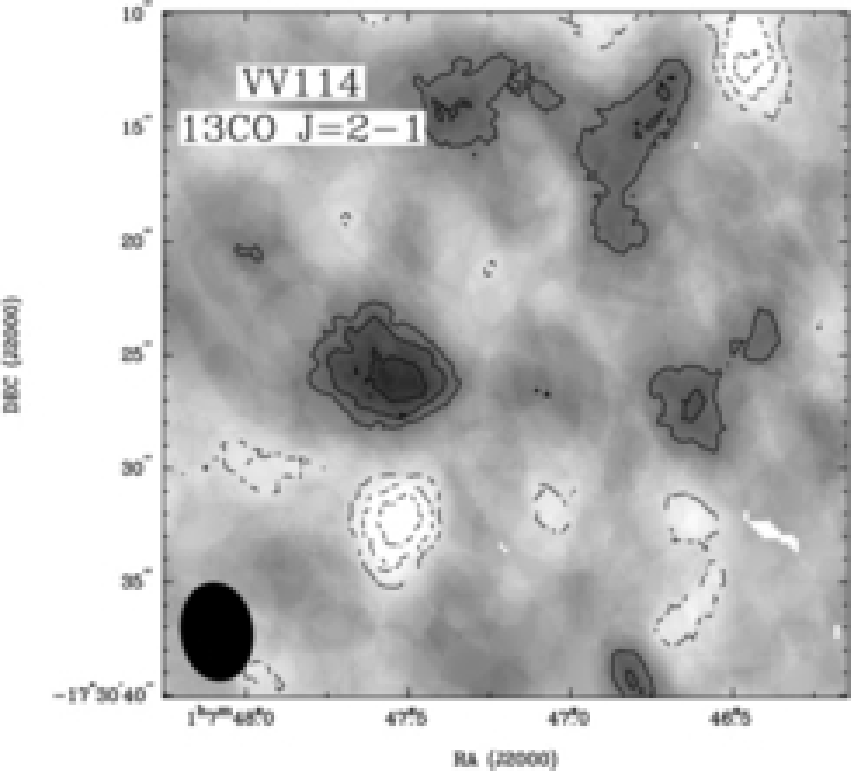}
\caption[VV114co21.mom0.eps]{VV 114 CO J=2-1 and
1.3 mm continuum maps. Notation as in Figure~\ref{fig-I17208co32} 
and~\ref{fig-Arp299co21}.
(a) CO J=2-1 moment 0 map. Lowest
contour is $\pm 2 \sigma = 3.5 $ Jy beam$^{-1}$ km s$^{-1}$ and contours
increase by factors of 1.5. (This image also has strong artifacts 
similar to those seen in Figure~\ref{fig-VV114co32} but they are
outside the field of view shown here.)
(b) CO J=2-1 moment 1 map. Contours are 20 km s$^{-1}$ $\times
(-12,-11,-10,...,-2,-1,0,1,2,...,9,10,11)$ 
relative to $cz$. 
(c) CO J=2-1 moment 2 map. Contours are 20 km s$^{-1}$ $\times (1,2,3,4,5)$.
(d) Uncleaned 1.3 mm map. Lowest contour
is $2 \sigma = 2.6$ mJy and contours increase in steps of $1 \sigma$.
(e) $^{13}$CO J=2-1 moment 0 map. Lowest
contour is $\pm 2 \sigma = 3.2 $ Jy beam$^{-1}$ km s$^{-1}$ and contours
increase in steps of $1 \sigma$. 
\label{fig-VV114co21}}
\end{figure}

\begin{figure}
\includegraphics[angle=-90,scale=.5]{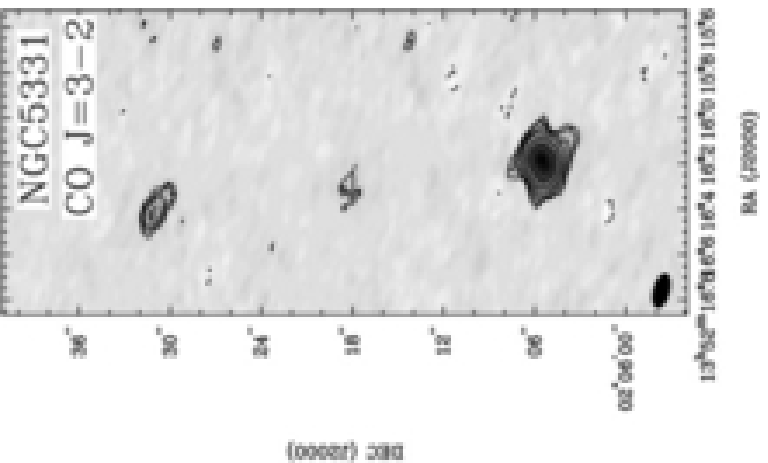}
\includegraphics[angle=-90,scale=.5]{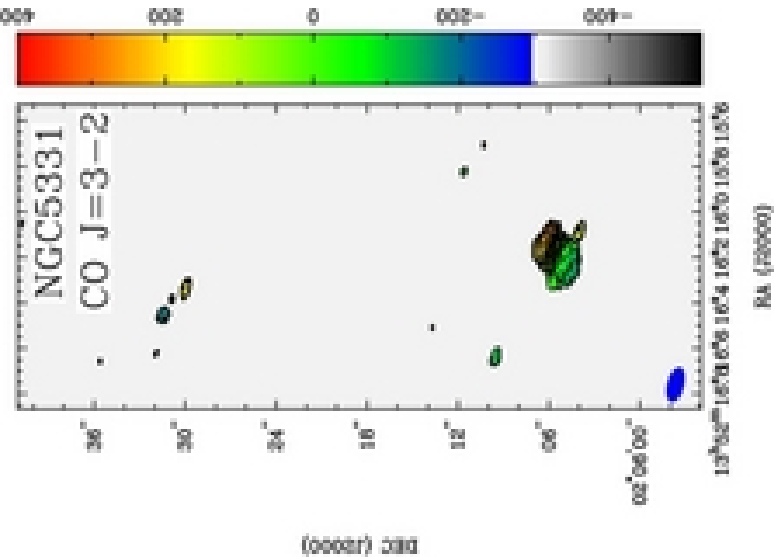}
\includegraphics[angle=-90,scale=.5]{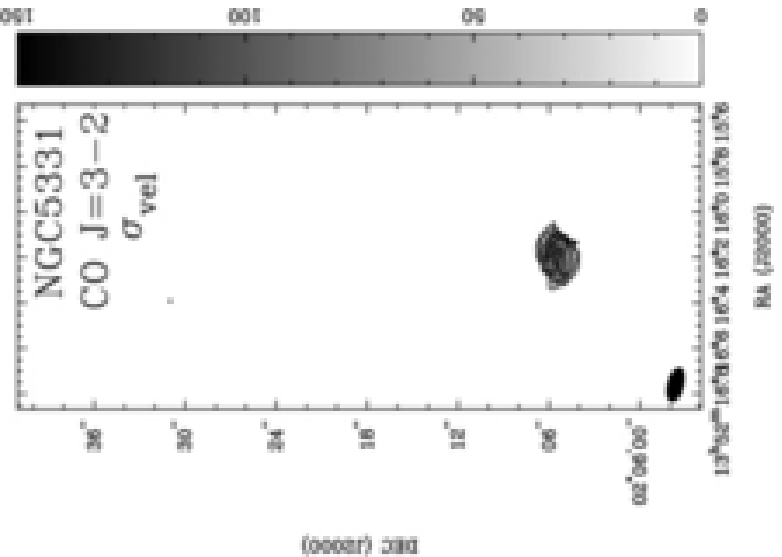}
\includegraphics[angle=-90,scale=.5]{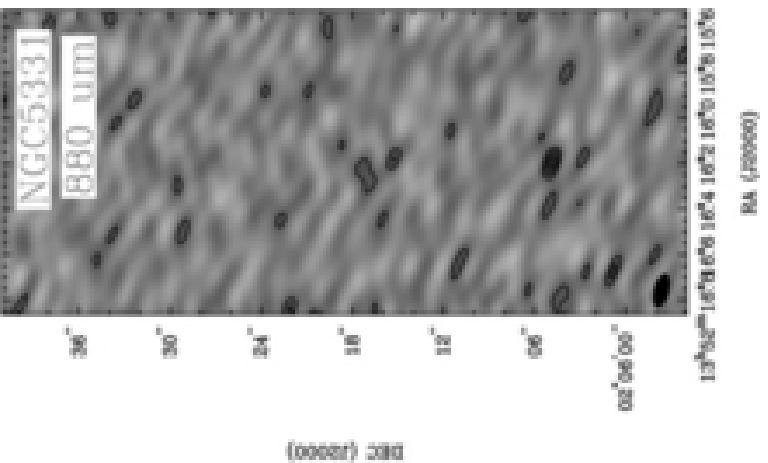}
\caption[NGC5331co32.mom0.eps]{NGC 5331 CO J=3-2 and
880 $\mu$m continuum maps. Notation as in Figure~\ref{fig-I17208co32}.
(a) CO J=3-2 moment 0 map. Lowest
contour is $\pm 2 \sigma = 21.6 $ Jy beam$^{-1}$ km s$^{-1}$ and contours
increase by factors of 1.5. 
(b) CO J=3-2 moment 1 map. Contours are 20 km s$^{-1}$ $\times
(-8,-7,-6,-5,...,-2,-1,0,1,2,...,14,15,16)$ relative to $cz$. 
(c) CO J=3-2 moment 2 map. Contours are 20 km s$^{-1}$ $\times (1,2,3,4,5)$.
(d) Uncleaned 880 $\mu$m map. Lowest contour
is $2 \sigma = 12$ mJy and contours increase in steps of $1 \sigma$.
\label{fig-NGC5331co32}}
\end{figure}

\begin{figure}
\includegraphics[angle=-90,scale=.4]{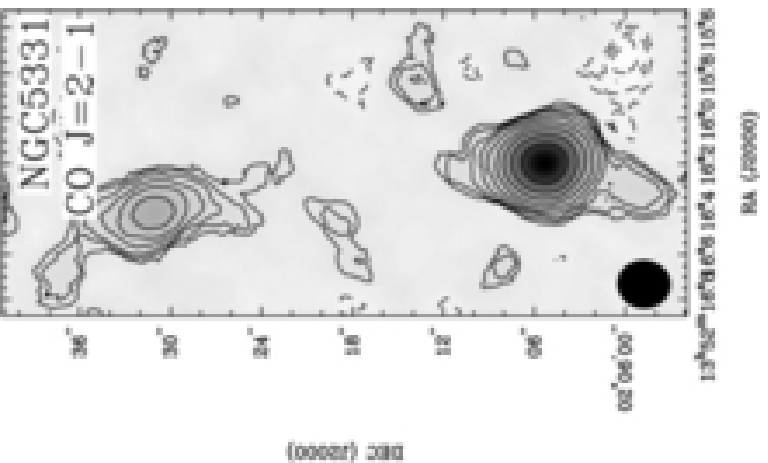}
\includegraphics[angle=-90,scale=.4]{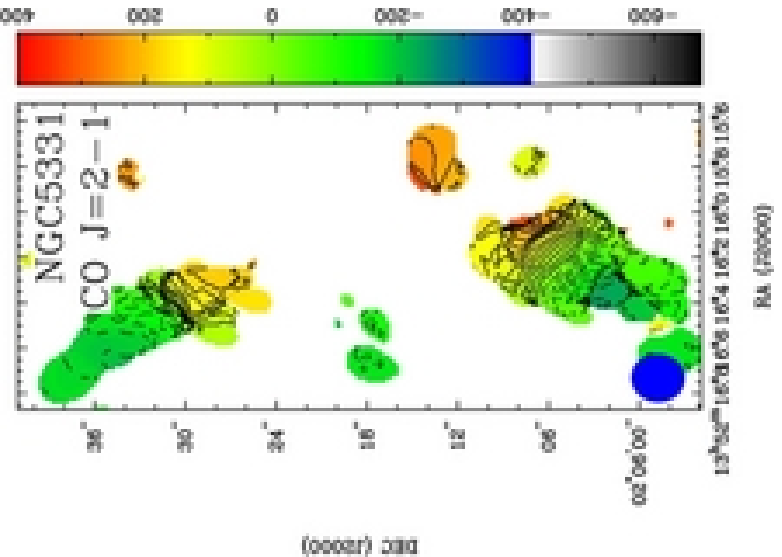}
\includegraphics[angle=-90,scale=.4]{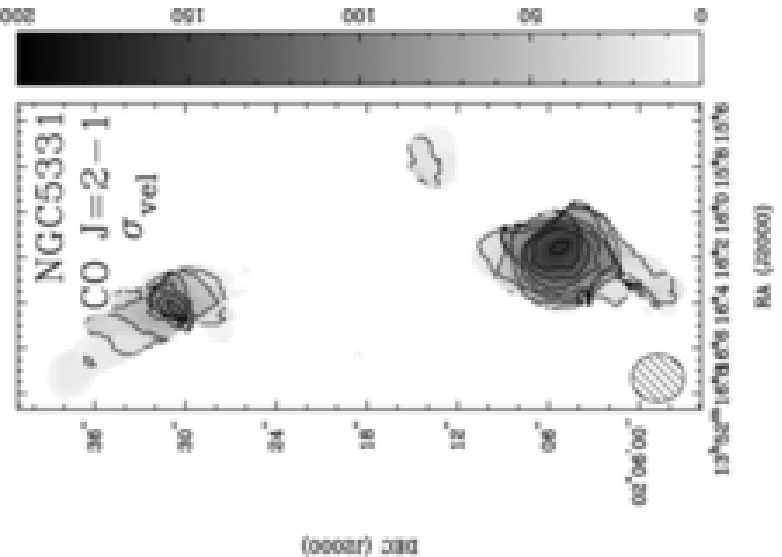}
\includegraphics[angle=-90,scale=.4]{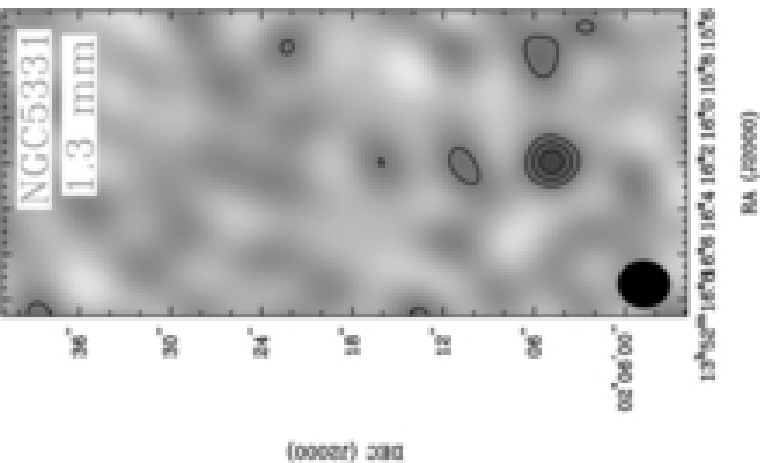}
\includegraphics[angle=-90,scale=.4]{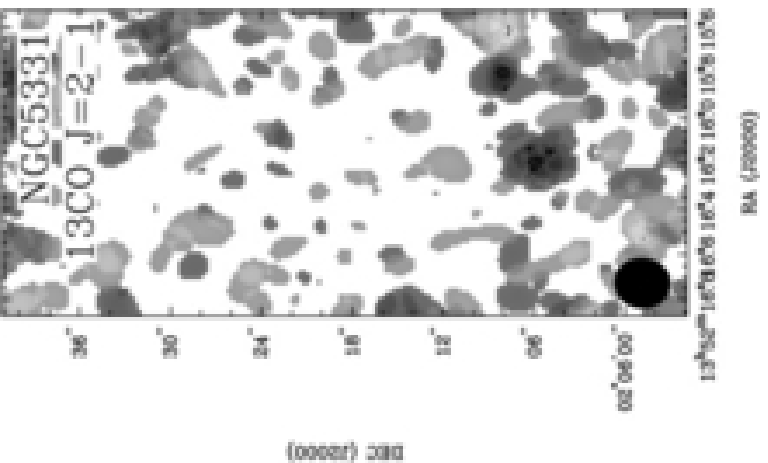}
\caption[NGC5331co21.mom0.eps]{NGC 5331 CO J=2-1 and
1.3 mm continuum maps. Notation as in Figure~\ref{fig-I17208co32}
and~\ref{fig-Arp299co21}.
(a) CO J=2-1 moment 0 map. Lowest
contour is $\pm 2 \sigma = 5.1 $ Jy beam$^{-1}$ km s$^{-1}$ and contours
increase by factors of 1.5. 
(b) CO J=2-1 moment 1 map. Contours are 20 km s$^{-1}$ $\times
(-12,-11,-10,...,-2,-1,0,1,2,...,12,13,14)$ relative to $cz$. 
(c) CO J=2-1 moment 2 map. Contours are 20 km s$^{-1}$ $\times (1,2,3,4,5,6,7,8)$.
(d) Uncleaned 1.3 mm map. Lowest contour
is $2 \sigma = 2.6$ mJy and contours increase in steps of $1 \sigma$.
(e) $^{13}$CO J=2-1 moment 0 map. The 
contour is $2 \sigma = 4.6 $ Jy beam$^{-1}$ km s$^{-1}$.
\label{fig-NGC5331co21}}
\end{figure}

\begin{figure}
\includegraphics[angle=-90,scale=.3]{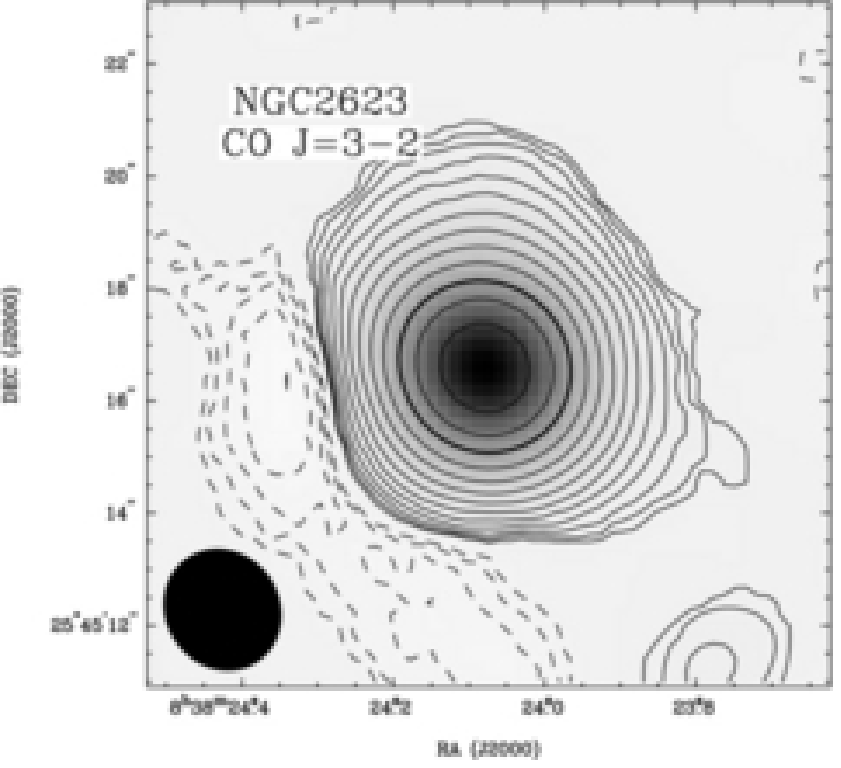}
\includegraphics[angle=-90,scale=.3]{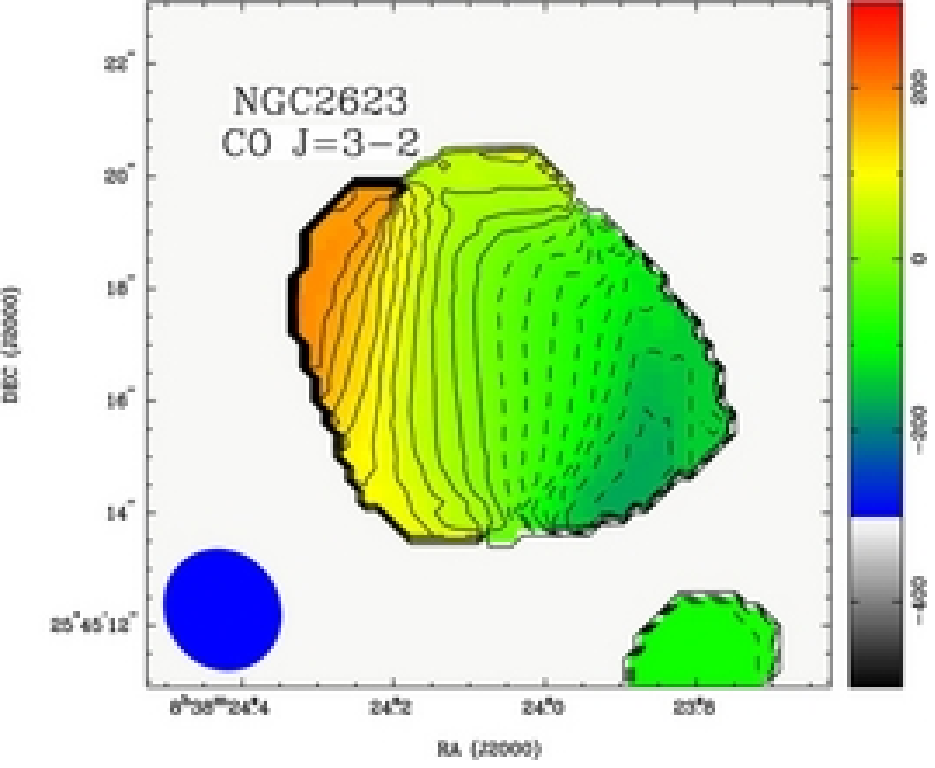}
\includegraphics[angle=-90,scale=.3]{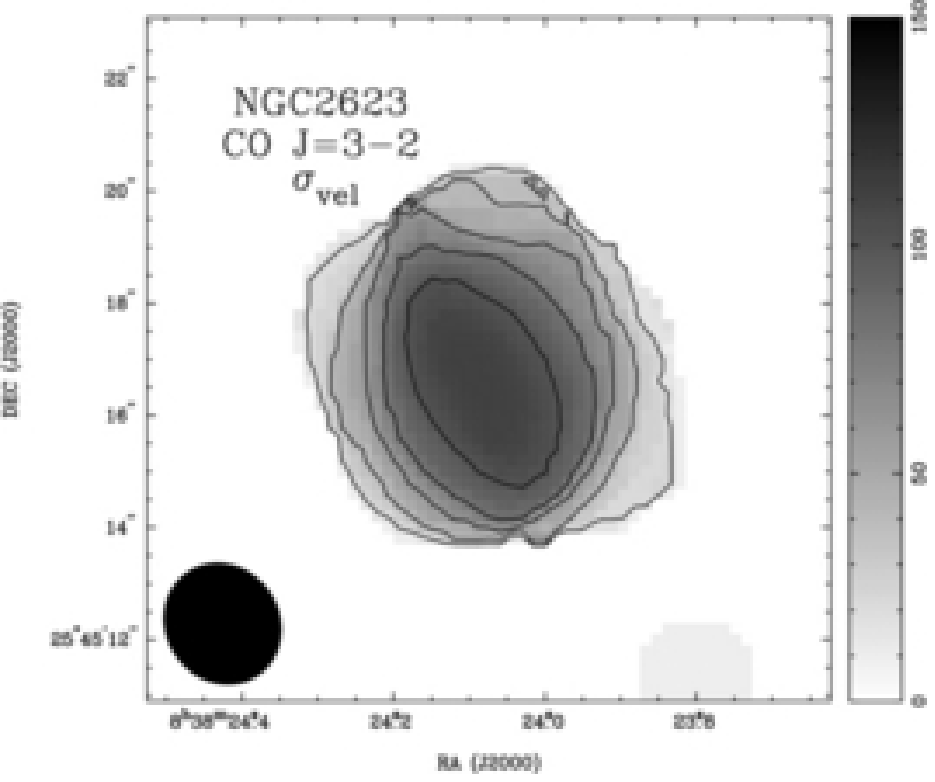}
\includegraphics[angle=-90,scale=.3]{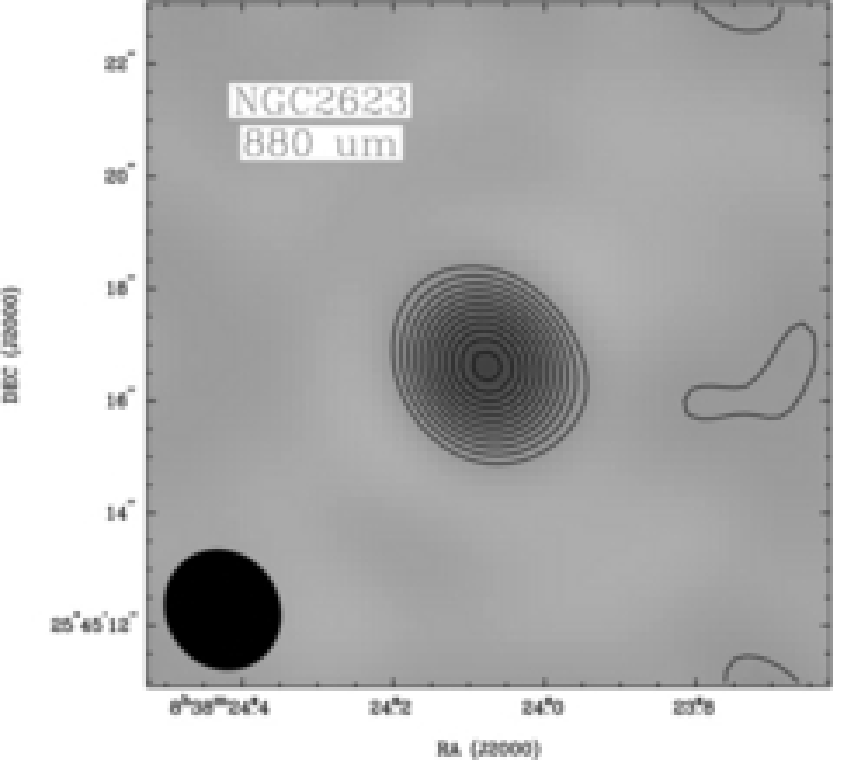}
\includegraphics[angle=-90,scale=.3]{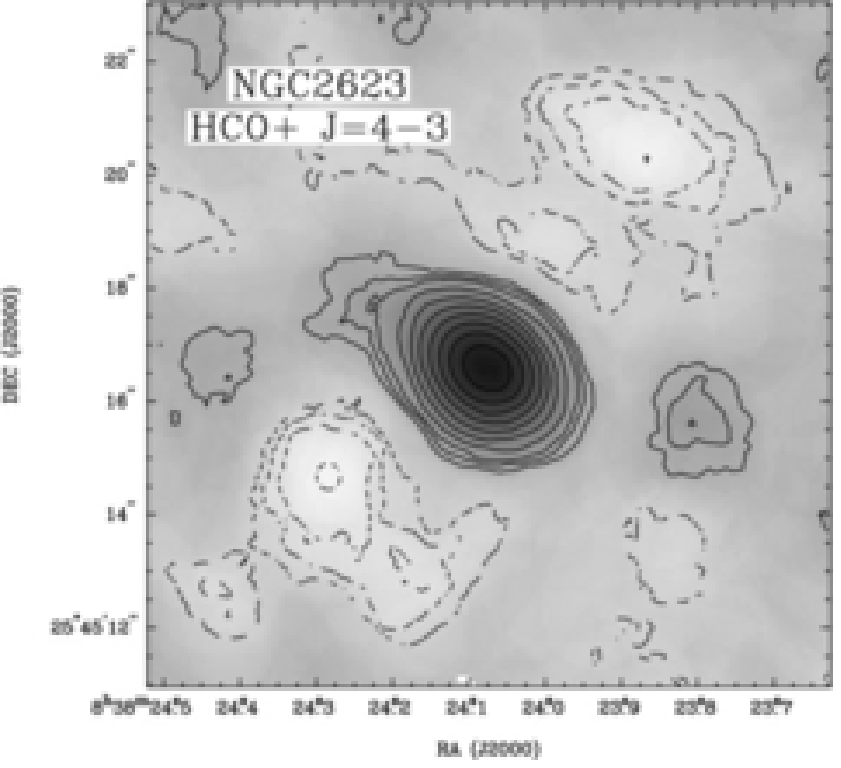}
\caption[NGC2623co32.mom0.eps]{NGC 2623 CO J=3-2 and
880 $\mu$m continuum maps. Notation as in Figures~\ref{fig-I17208co32}
and ~\ref{fig-Mrk231co32}.
(a) CO J=3-2 moment 0 map. Lowest
contour is $\pm 2 \sigma = 2.7 $ Jy beam$^{-1}$ km s$^{-1}$ and contours
increase by factors of 1.5. 
(b) CO J=3-2 moment 1 map. Contours are 20 km s$^{-1}$ $\times
(-8,-7,-6,-5,-4,-3,-2,-1,0,1,2,3,4,5,6,7,8,9)$ relative to $cz$. 
(c) CO J=3-2 moment 2 map. Contours are 20 km s$^{-1}$ $\times (1,2,3,4,5)$.
(d) Uncleaned 880 $\mu$m map. Lowest contour
is $2 \sigma = 4$ mJy and contours increase in steps of $2 \sigma$.
(e) Uncleaned HCO$^+$ J=4-3 moment 0 map. Lowest
contour is $\pm 2 \sigma = 2.60 $ Jy beam$^{-1}$ km s$^{-1}$ and contours
increase in steps of $1 \sigma$ to $4\sigma$ and then by steps of
2$\sigma$. 
\label{fig-NGC2623co32}}
\end{figure}

\begin{figure}
\includegraphics[angle=-90,scale=.3]{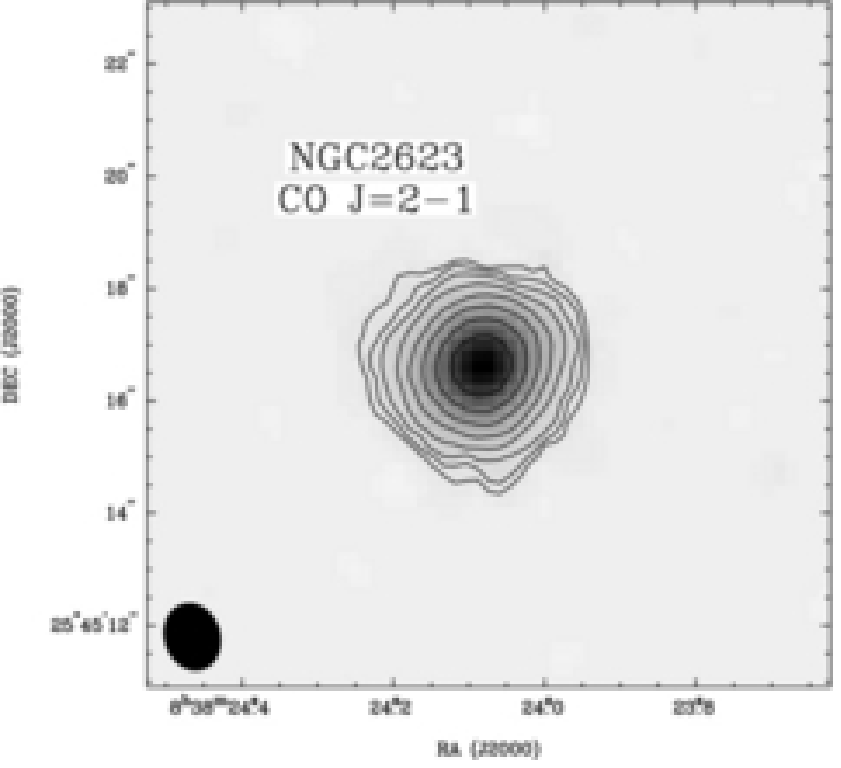}
\includegraphics[angle=-90,scale=.3]{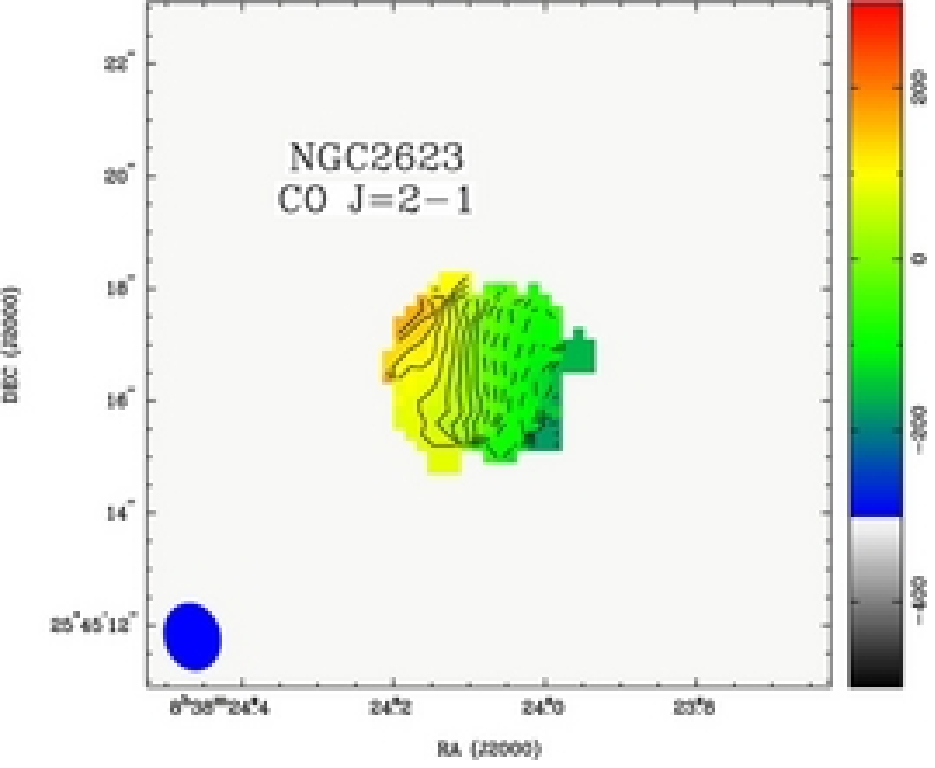}
\includegraphics[angle=-90,scale=.3]{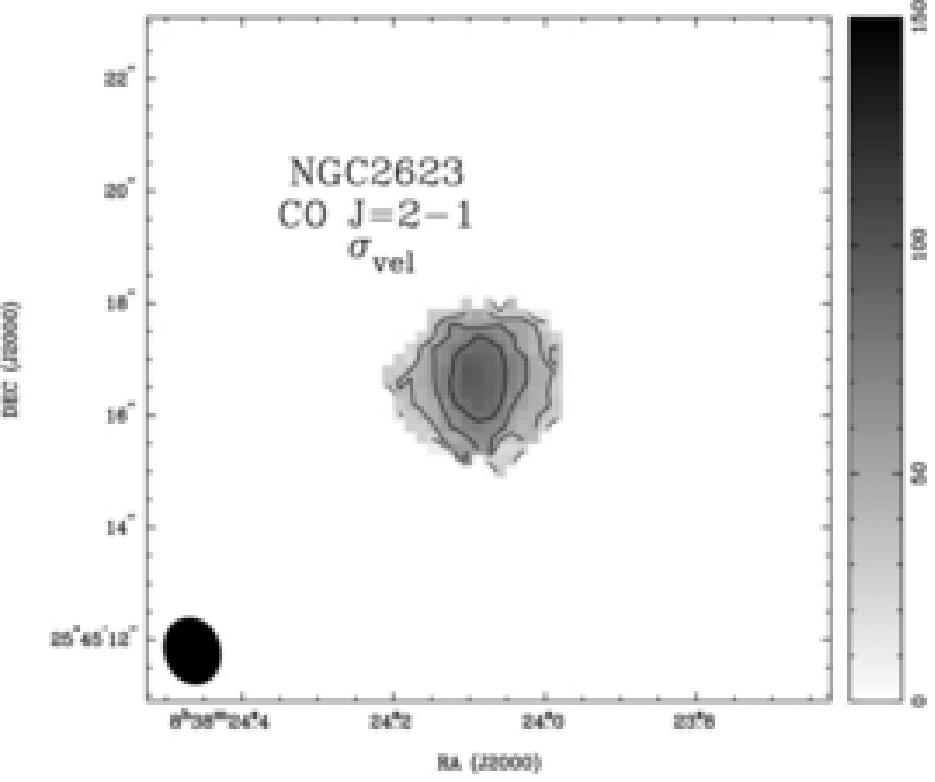}
\includegraphics[angle=-90,scale=.3]{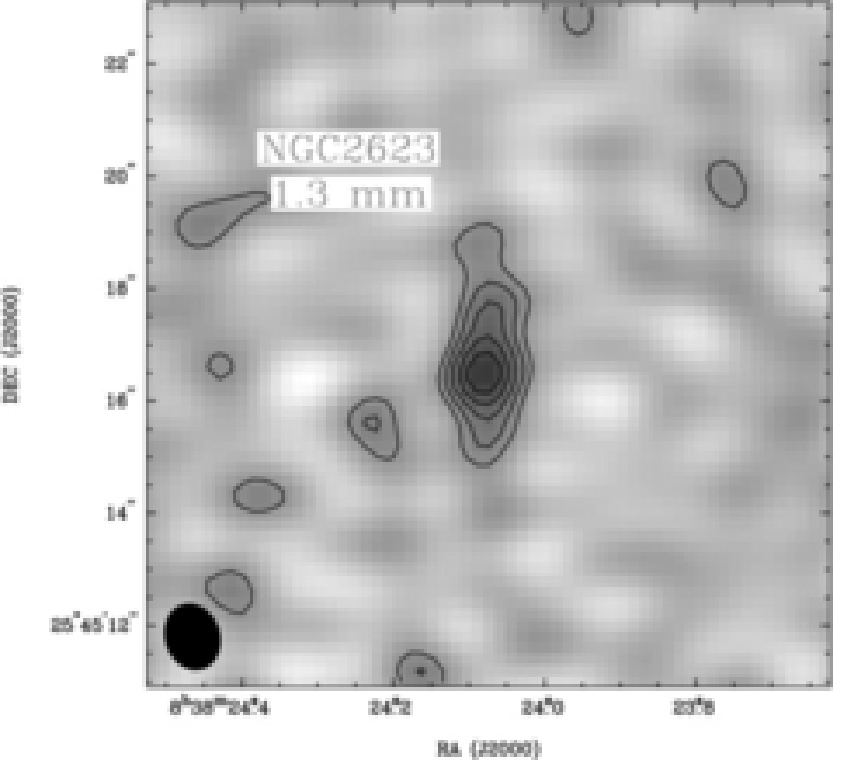}
\caption[NGC2623co21.mom0.eps]{NGC 2623 CO J=2-1 and
1.3 mm continuum maps. Notation as in Figure~\ref{fig-I17208co32}.
(a) CO J=2-1 moment 0 map. Lowest
contour is $2 \sigma = 4.0 $ Jy beam$^{-1}$ km s$^{-1}$ and contours
increase by factors of 1.5. 
(b) CO J=2-1 moment 1 map. Contours are 20 km s$^{-1}$ $\times
(-9,-8,-7,-6,-5,-4,-3,-2,-1,0,1,2,3,4,5,6,7,8)$ relative to $cz$. 
(c) CO J=2-1 moment 2 map. Contours are 20 km s$^{-1}$ $\times (1,2,3,4)$.
(d) Uncleaned 1.3 mm map. Lowest contour
is $2 \sigma = 2.6$ mJy and contours increase in steps of $1 \sigma$.
\label{fig-NGC2623co21}}
\end{figure}

\begin{figure}
\includegraphics[angle=-90,scale=.3]{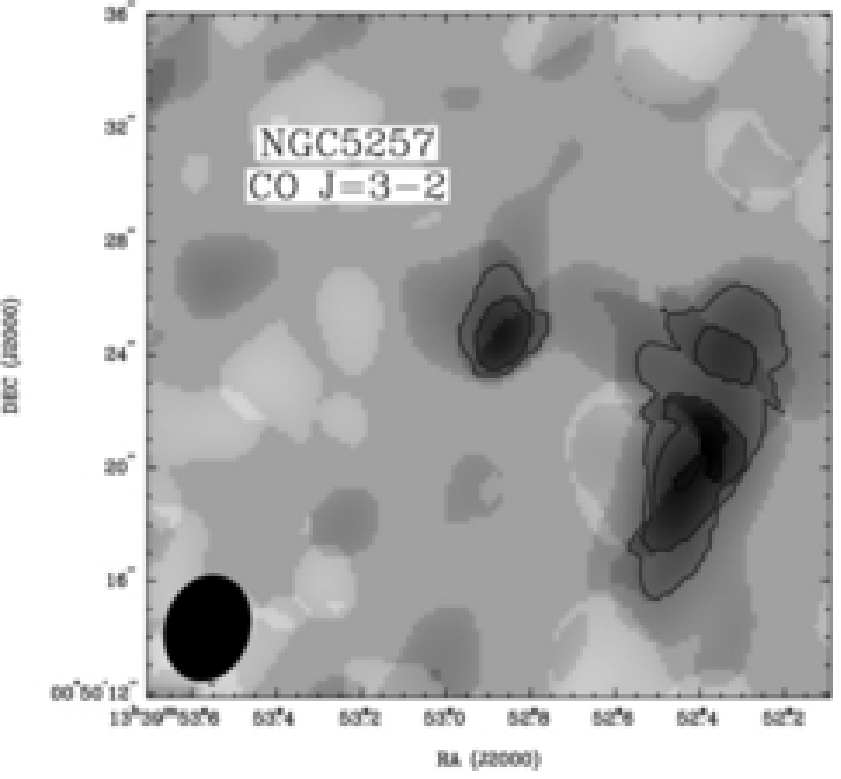}
\includegraphics[angle=-90,scale=.3]{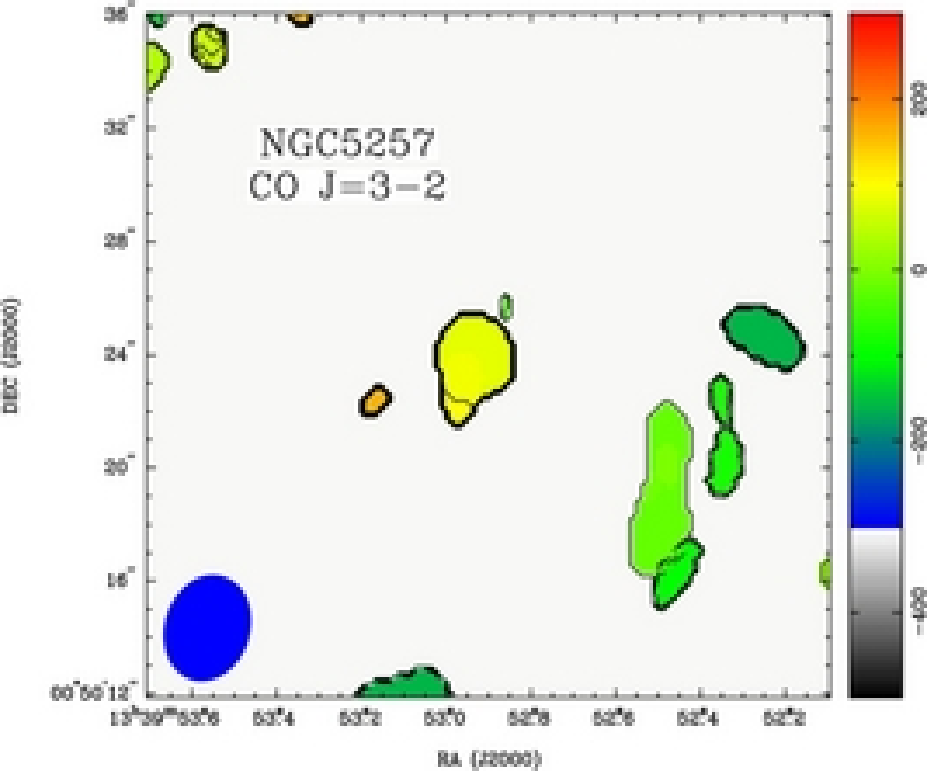}
\includegraphics[angle=-90,scale=.3]{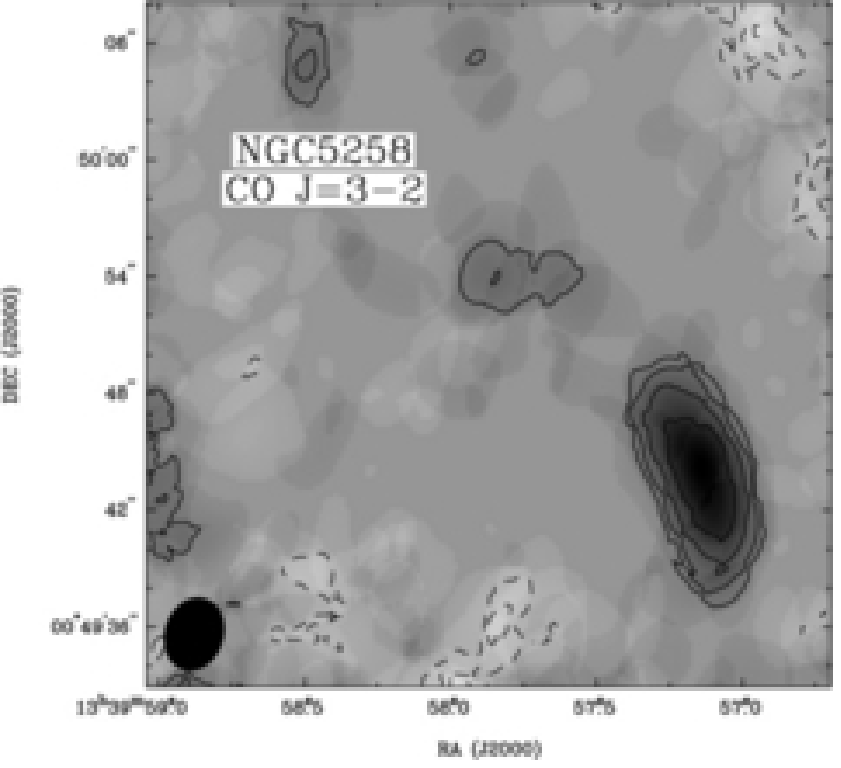}
\includegraphics[angle=-90,scale=.3]{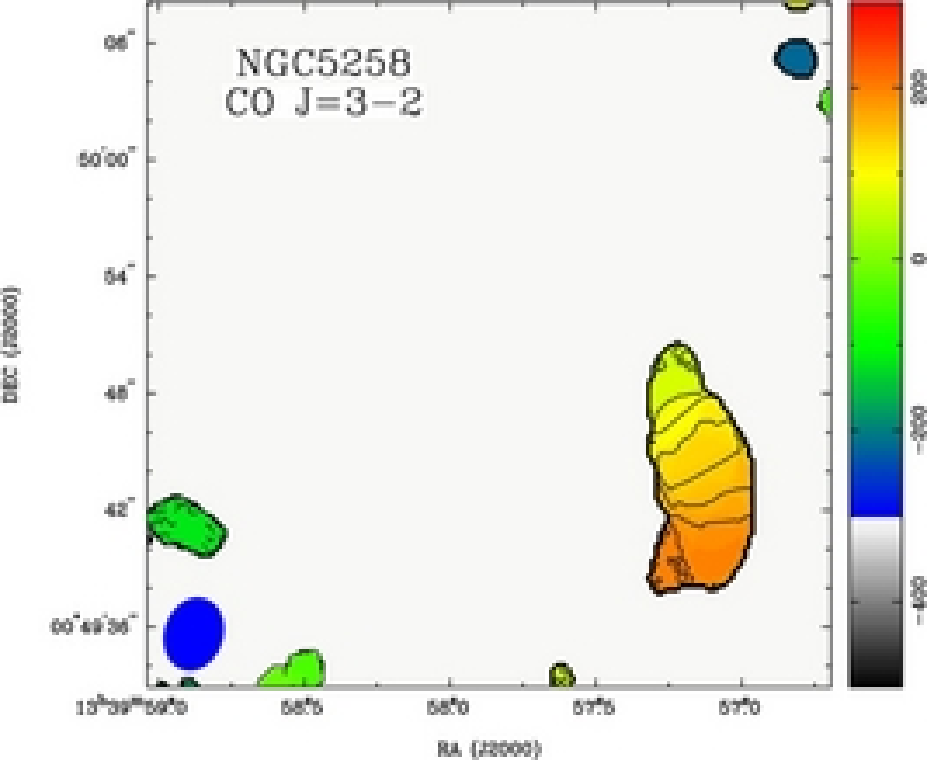}
\includegraphics[angle=-90,scale=.3]{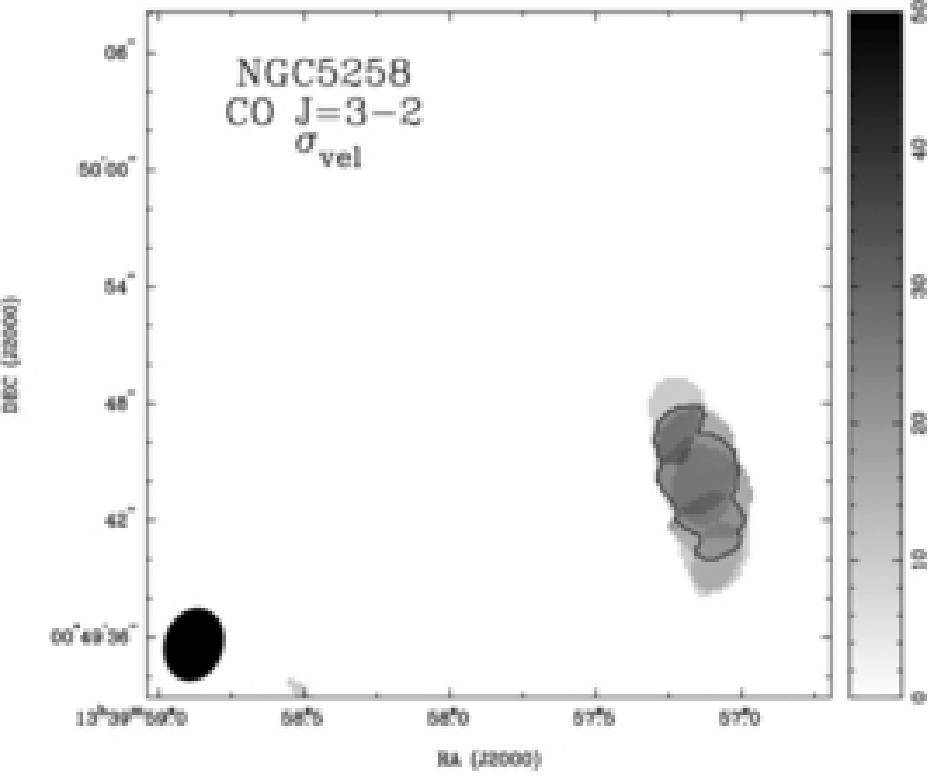}
\includegraphics[angle=-90,scale=.3]{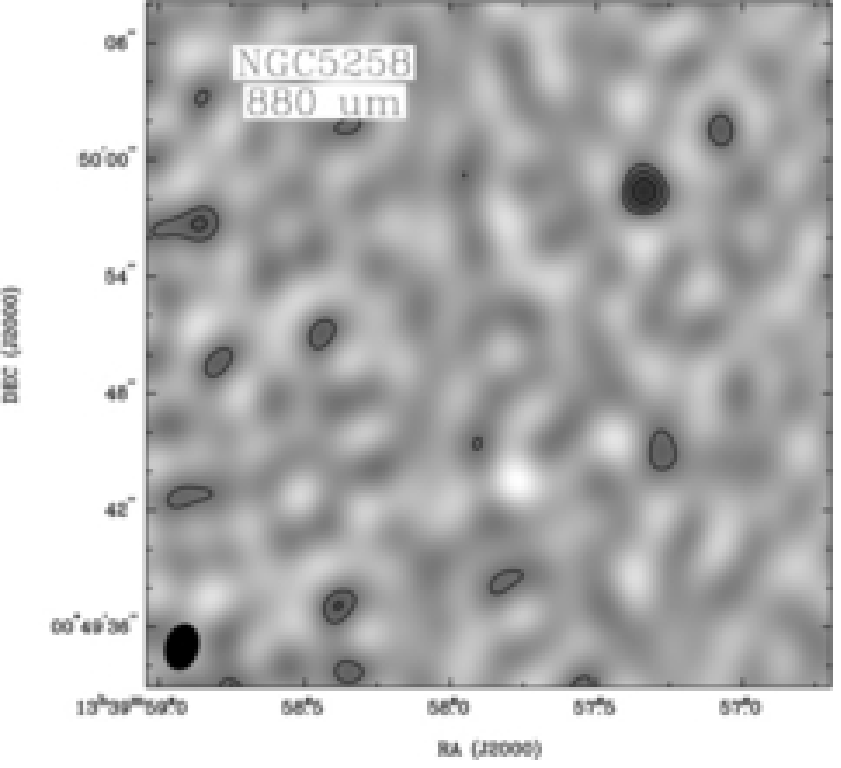}
\caption[NGC5258co32.mom0.eps]{NGC 5257/8 CO J=3-2 and
880 $\mu$m continuum maps. Notation as in Figure~\ref{fig-I17208co32}.
(a) CO J=3-2 moment 0 map for NGC 5257. Lowest
contour is $2 \sigma = 22.1 $ Jy beam$^{-1}$ km s$^{-1}$ and contours
increase by factors of 1.5. 
(b) CO J=3-2 moment 1 map for 
NGC 5257. Contours are 20 km s$^{-1}$ $\times
(0,1,2,34,5)$ relative to $cz$.
(c) CO J=3-2 moment 0 map for NGC 5258. Lowest
contour is $\pm 2 \sigma = 33.7 $ Jy beam$^{-1}$ km s$^{-1}$ and contours
increase by factors of 1.5. 
(d) CO J=3-2 moment 1 map for 
NGC 5258. Contours are 20 km s$^{-1}$ $\times
(3,4,5,6,7,8,9,10)$ relative to $cz$. 
(e) CO J=3-2 moment 2 map for NGC 5258. Contour is 20 km s$^{-1}$.
(f) Uncleaned 880 $\mu$m map. Lowest contour
is $2 \sigma = 30$ mJy and contours increase in steps of $1 \sigma$.
\label{fig-NGC5258co32}}
\end{figure}

\begin{figure}
\includegraphics[angle=-90,scale=.35]{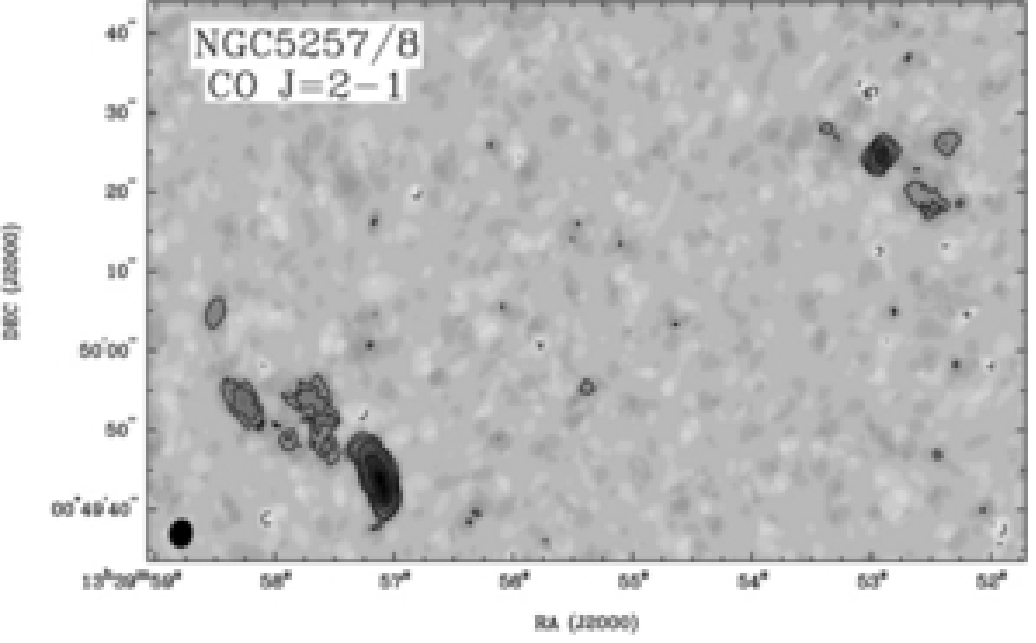}
\includegraphics[angle=-90,scale=.35]{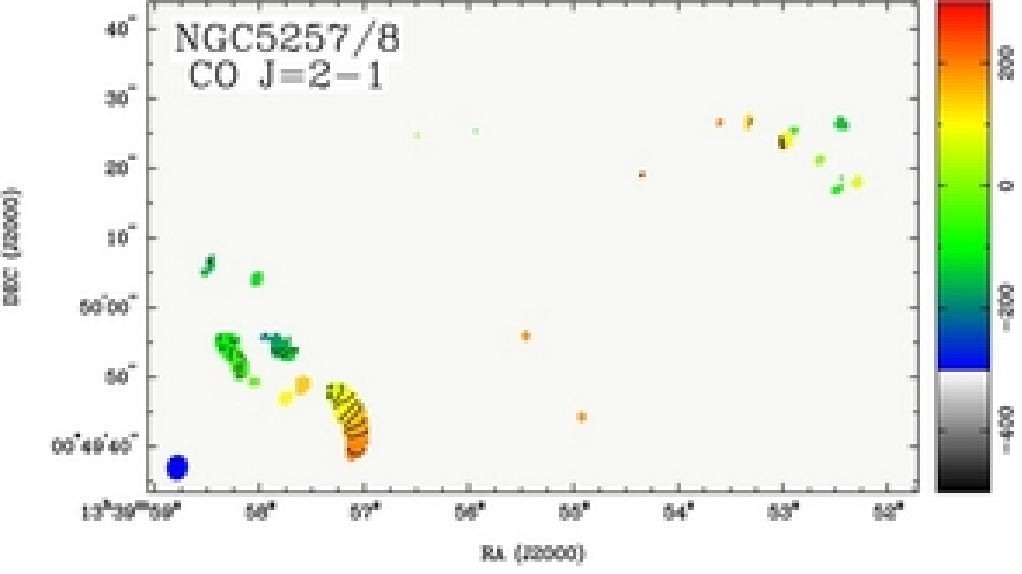}
\includegraphics[angle=-90,scale=.35]{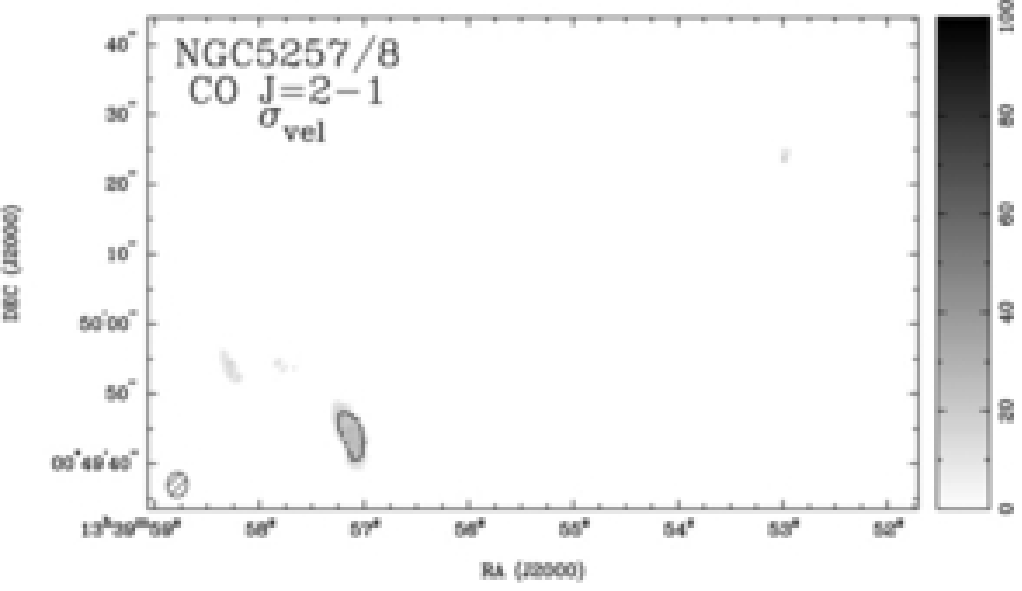}
\caption[NGC5257co21.mom0.eps]{NGC 5257/8 CO J=2-1 
maps. Notation as in Figure~\ref{fig-I17208co32}.
(a) CO J=2-1 moment 0 map. Lowest
contour is $2 \sigma = 10.8 $ Jy beam$^{-1}$ km s$^{-1}$ and contours
increase by factors of 1.5. 
(b) CO J=2-1 moment 1 map. Contours are 20 km s$^{-1}$ $\times
(-8,-7,-6,...,-2,-1,0,1,2,...,8,9,10)$ 
relative to $cz$. 
(c) CO J=2-1 moment 2 map. Contour is 20 km s$^{-1}$.
\label{fig-NGC5257co21}}
\end{figure}

\begin{figure}
\includegraphics[angle=-90,scale=.3]{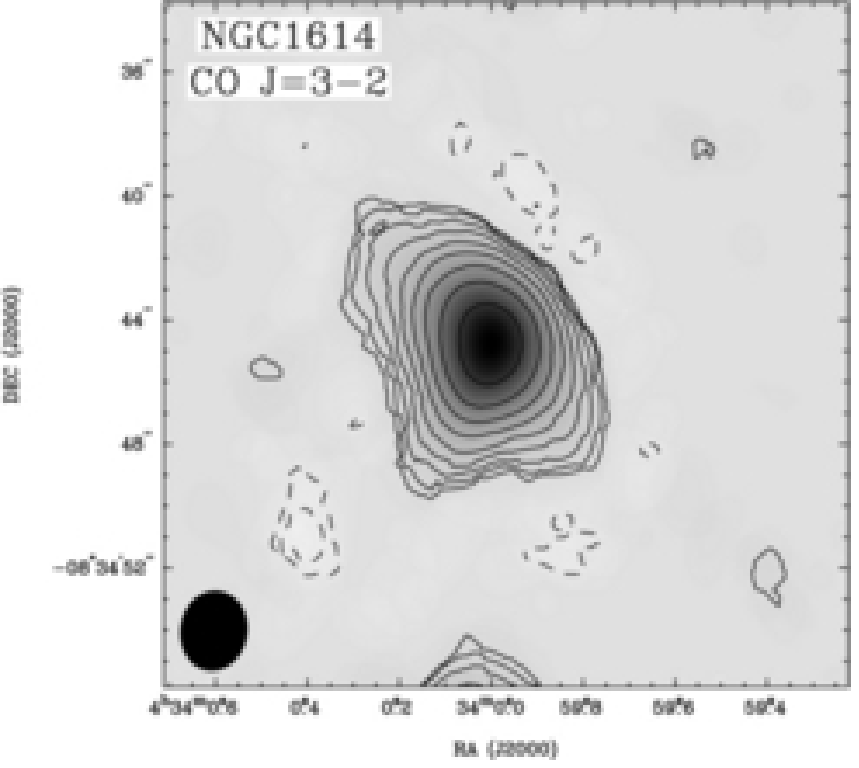}
\includegraphics[angle=-90,scale=.3]{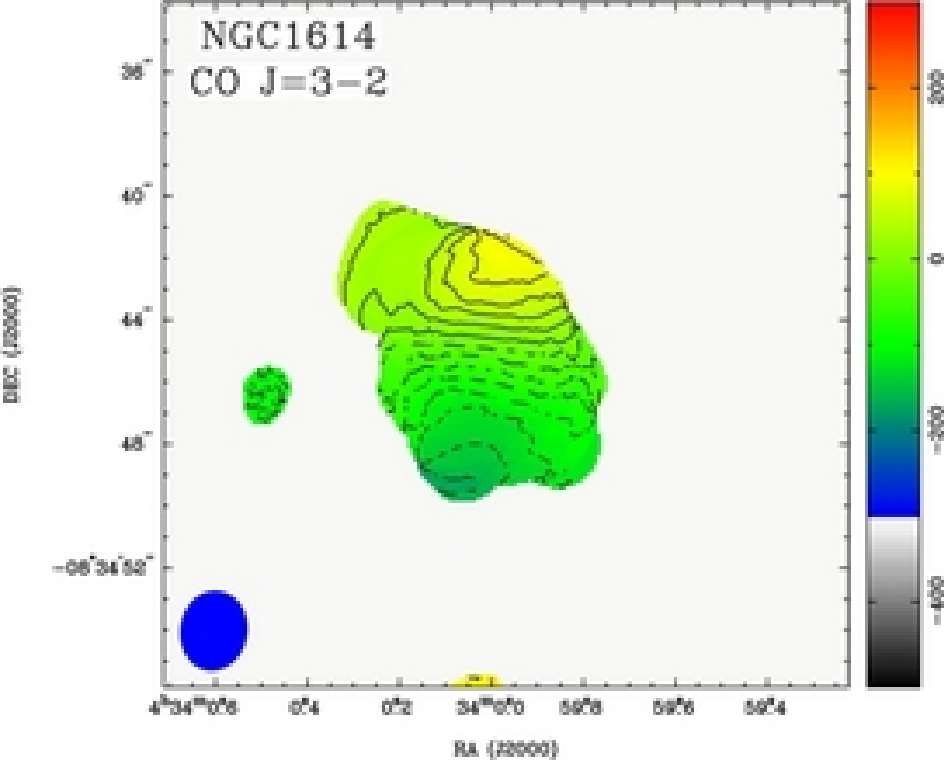}
\includegraphics[angle=-90,scale=.3]{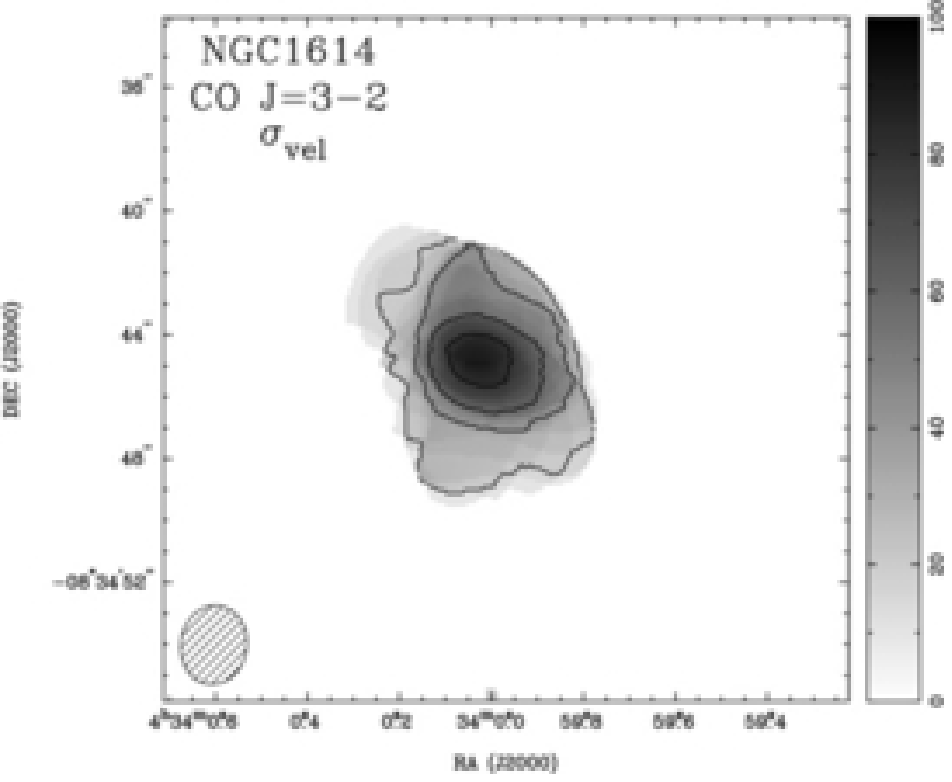}
\includegraphics[angle=-90,scale=.3]{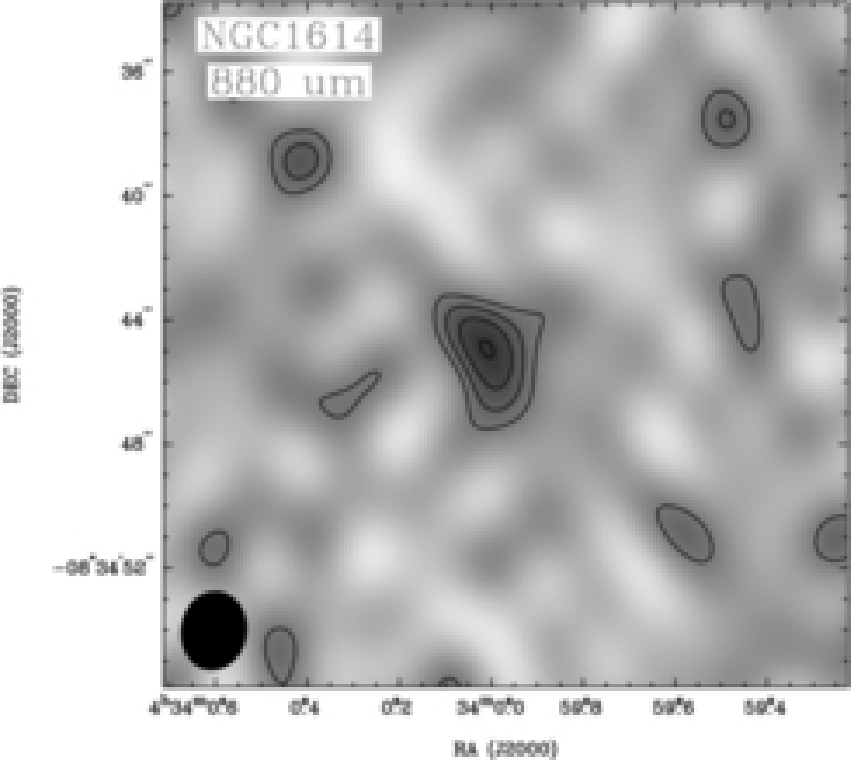}
\includegraphics[angle=-90,scale=.3]{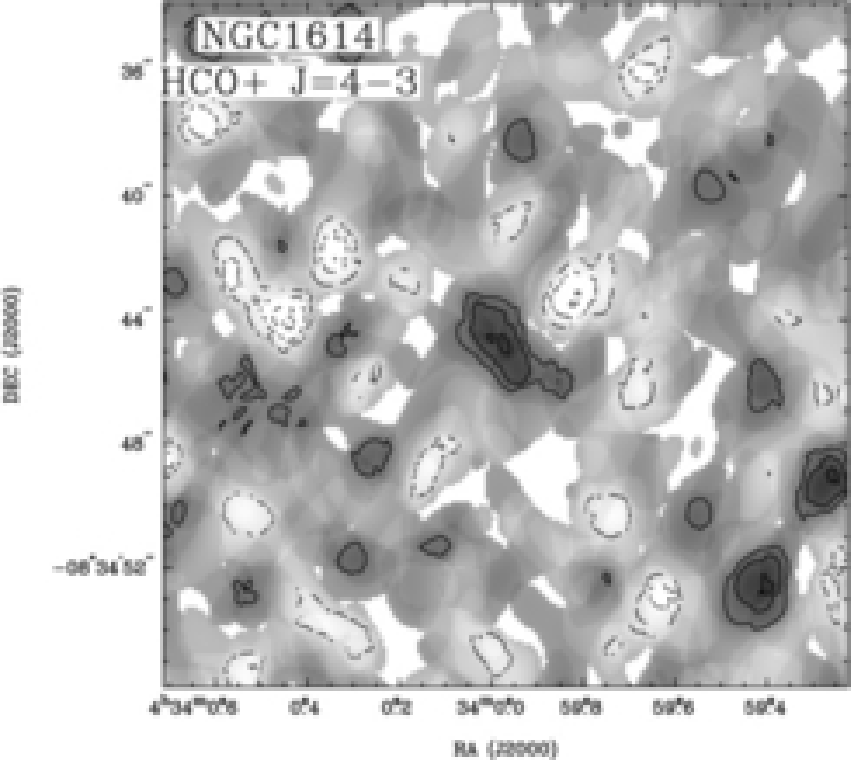}
\caption[NGC1614co32.mom0.eps]{NGC 1614 CO J=3-2 and
880 $\mu$m continuum maps. Notation as in Figures~\ref{fig-I17208co32}
and \ref{fig-Mrk231co32}.
(a) CO J=3-2 moment 0 map. Lowest
contour is $\pm 2 \sigma = 8.2 $ Jy beam$^{-1}$ km s$^{-1}$ and contours
increase by factors of 1.5. 
(b) CO J=3-2 moment 1 map. Contours are 20 km s$^{-1}$ $\times
(-8,-7,-6,-5,-4,-3,-2,-1,0,1,2,3,4,5)$ 
relative to $cz$. 
(c) CO J=3-2 moment 2 map. Contours are 20 km s$^{-1}$ $\times (1,2,3,4)$.
(d) Uncleaned 880 $\mu$m map. Lowest contour
is $2 \sigma = 8.2$ mJy and contours increase in steps of $1 \sigma$.
(e) Uncleaned HCO$^+$ J=4-3 moment 0 map. Lowest
contour is $\pm 2 \sigma = 6.24 $ Jy beam$^{-1}$ km s$^{-1}$ and contours
increase in steps of $1 \sigma$. 
\label{fig-NGC1614co32}}
\end{figure}

\begin{figure}
\includegraphics[angle=-90,scale=.3]{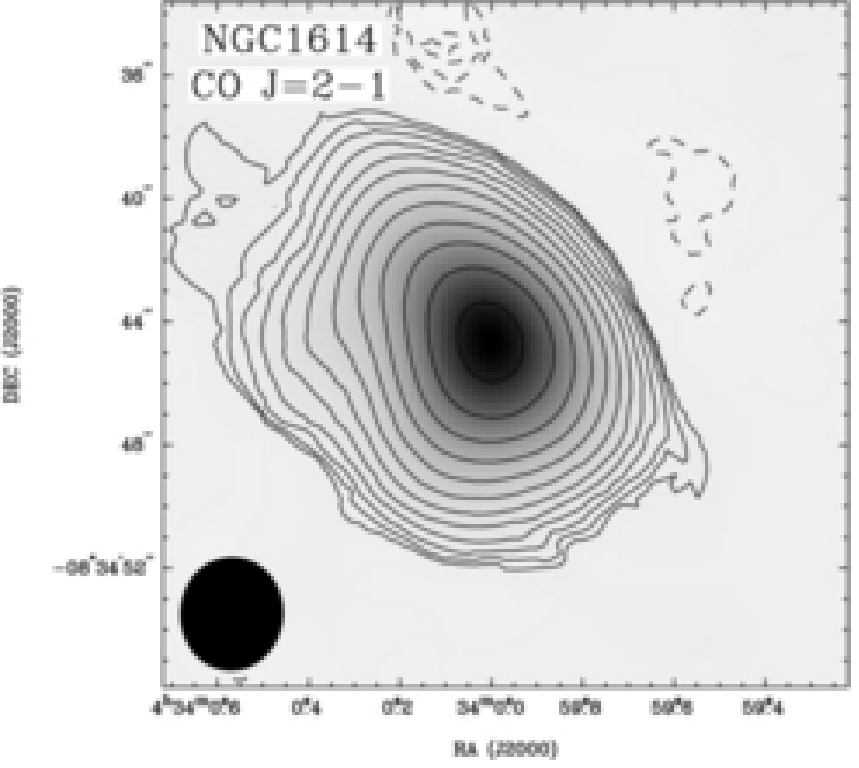}
\includegraphics[angle=-90,scale=.3]{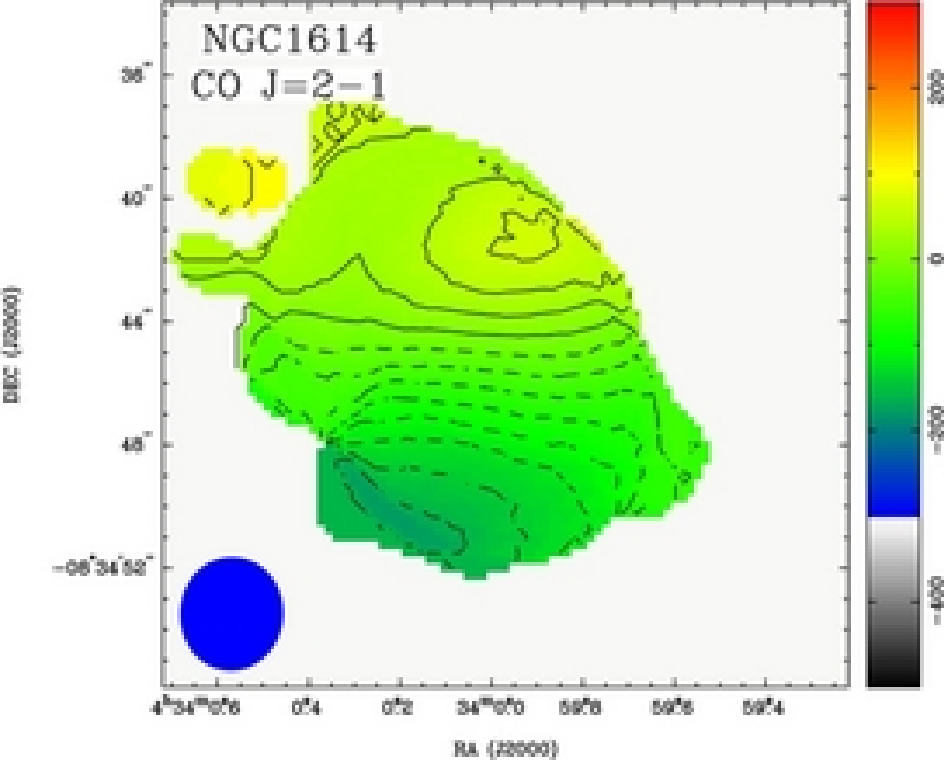}
\includegraphics[angle=-90,scale=.3]{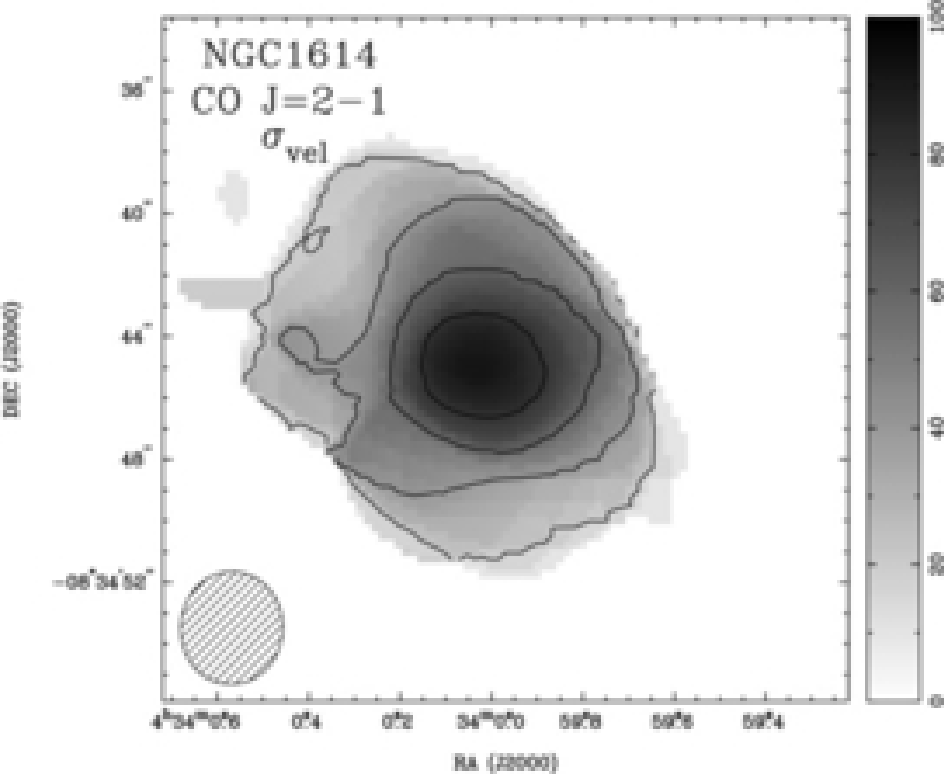}
\includegraphics[angle=-90,scale=.3]{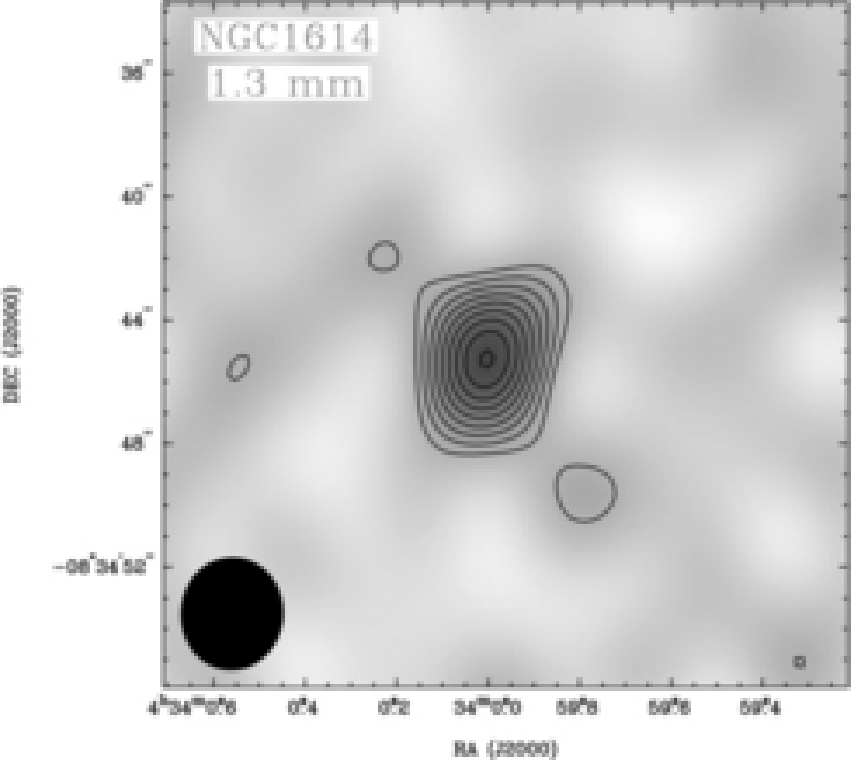}
\includegraphics[angle=-90,scale=.3]{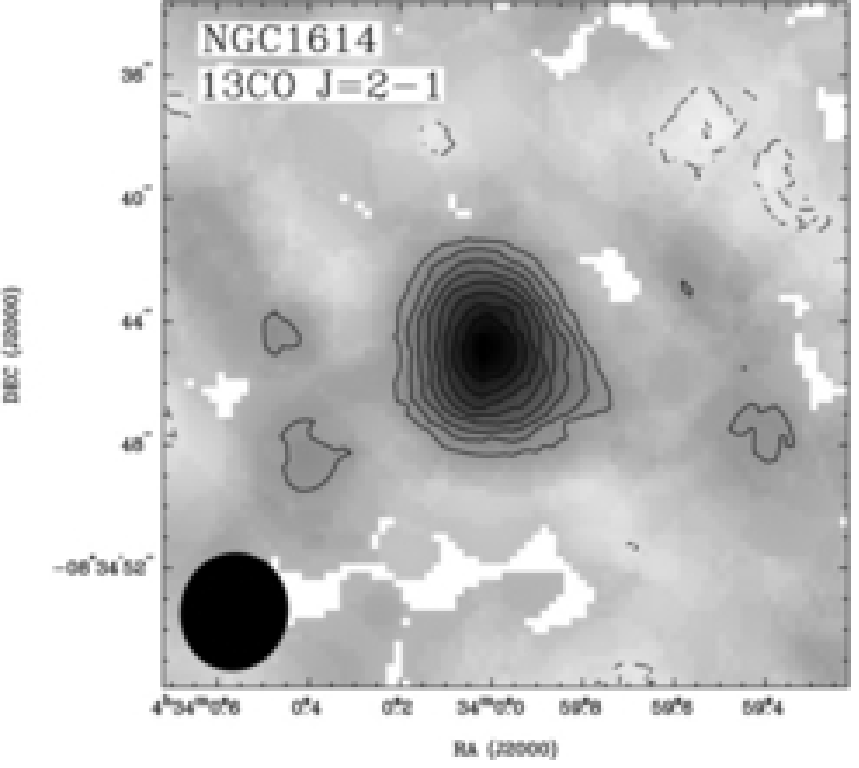}
\caption[NGC1614co21.mom0.eps]{NGC 1614 CO J=2-1 and
1.3 mm continuum maps. Notation as in Figure~\ref{fig-I17208co32}
and~\ref{fig-Arp299co21}.
(a) CO J=2-1 moment 0 map. Lowest
contour is $\pm 2 \sigma = 3.0 $ Jy beam$^{-1}$ km s$^{-1}$ and contours
increase by factors of 1.5. 
(b) CO J=2-1 moment 1 map. Contours are 20 km s$^{-1}$ $\times
(-8,-7,-6,-5,-4,-3,-2,-1,0,1,2,3,4,5)$ 
relative to $cz$. 
(c) CO J=2-1 moment 2 map. Contours are 20 km s$^{-1}$ $\times (1,2,3,4)$.
(d) Uncleaned 1.3 mm map. Lowest contour
is $2 \sigma = 2.6$ mJy and contours increase in steps of $1 \sigma$.
(e) $^{13}$CO J=2-1 moment 0 map. Lowest
contour is $\pm 2 \sigma = 2.5 $ Jy beam$^{-1}$ km s$^{-1}$ and contours
increase in steps of $1 \sigma$. 
\label{fig-NGC1614co21}}
\end{figure}

Figures~\ref{fig-I17208co32}-\ref{fig-NGC1614co21} reveal a diversity of
morphologies and kinematics for the molecular gas in our sample of
galaxies. Nine of the galaxies show a predominantly compact morphology
while the remaining five galaxies range from well-separated pairs of
galaxies to highly complex morphologies. The galaxies with extended CO
emission are all found in the lower luminosity half of our sample,
although some low luminosity galaxies also show a compact
morphology. The source sizes derived from two-dimensional Gaussian
fits to the CO J=3-2 images and the line widths (full width at half
maximum) derived from the 
CO J=3-2 spectra are tabulated in \citet{i08}.
The distributions of the CO J=3-2 and J=2-1 emissions are
qualitatively similar; we will discuss the line ratio
distributions in more detail in a future paper.

The kinematics of the molecular gas are also quite varied. Integrating
over the full emission region in each galaxy
(Figures~\ref{fig-co32_mrk_iras}-\ref{fig-co21_n1614_n5257}), the
line widths at half maximum range from 200 km s$^{-1}$ in Mrk 231 to
650 km s$^{-1}$ in UGC 5101. Some of the line profiles are quite
smooth and gaussian (e.g. Mrk 231), while others have more of a
double-peaked structure (e.g. UGC 5101).  The maps of the velocity
fields (Figures~\ref{fig-I17208co32}-\ref{fig-NGC1614co21})
show many examples of smooth and regular rotation 
among the more compact galaxies, although some of the
compact galaxies, such as Mrk 273 and NGC 6240,
are significantly distorted. Two of
the galaxies with extended CO emission, Arp 299 and VV 114
show very complex and distorted velocity fields. The maps of the
velocity dispersions show that the peak velocity dispersion is
commonly 80-100 km s$^{-1}$. 
However, the peak velocity dispersions for
IRAS 17208-0014, Mrk 273, UGC 5101, NGC6240, and 
NGC 5331S are twice as large as the typical galaxy in our sample.
The lower velocity dispersions seen for NGC 5257/8  are likely due to
the lower signal-to-noise of those maps; similarly, a lower velocity
dispersion 
is seen for NGC 5331S in the CO J=3-2 line than in the CO J=2-1 line.

Our velocity
dispersions are typically about 70\% of the values measured from CO
band-head emission in the near-infrared
\citep{hr06,d06}.
In some cases, the lower
values found in this paper may be related to the somewhat lower
angular resolution of the millimeter-wave CO observations, as
discussed by \citet{hr06}. The near-infrared data may also
overestimate the velocity dispersion if the CO band-head emission is
strongly peaked on 
the nucleus \citep{hr06}. We derive a significantly larger velocity
disperion for IRAS 17208-0014; this may be due to the lower
signal-to-noise ratio of the near-infrared spectrum for this galaxy.


\subsubsection{The relative strength of  the HCO$^+$ J=4-3 emission}

We detect the HCO$^+$ J=4-3 line in seven of the fourteen galaxies in
our sample. Since the weak HCO$^+$ emission is extremely compact, we
compare the line strength to the peak line strength of the CO J=3-2
line. The CO/HCO$^+$ line ratios are 8 and 16 for Mrk 231 and NGC
2623, respectively, and range from $\le$7 to $\le$32 for the other
five galaxies. Treating all the HCO$^+$ lines as detections rather
than upper limits, and
treating the detections of the eastern and central components of VV114
separately, the
average line ratio is 19 with an rms deviation of 9. However, the true
average line ratio for these galaxies is likely somewhat smaller,
since many of the HCO$^+$ detections are lower limits because the line
emission extends to the edge of the spectral window. 

This line ratio
is comparable to the average CO/HCO$^+$ line ratio using the J=1-0 and
J=3-2 lines, respectively, determined for 6 of the galaxies in our
sample by \citet{gc06}, who also find a roughly constant CO/HCO$^+$
line ratio as a function of L$_{\rm FIR}$. High-resolution HCO$^+$
J=1-0 data have been published for four of the galaxies in our sample
by \citet{i06,iman07}. Comparing their data with our new measurements
yields HCO$^+$ J=4-3/J=1-0 line ratios of about 4 in both Mrk231 and
Arp 299, while this same line ratio appears to vary dramatically from
place to place within VV 114. The HCN/HCO$^+$ J=1-0 line ratio
also varies spatially in VV 114 \citep{iman07} and in Arp 299 \citep{i06},
which suggests that there
are interesting physical and/or chemical variations in the gas
properties in these early-stage mergers. Large variations in the
HCN J=4-3/J=1-0 ratio in a sample of four galaxies are evidence for
significant differences in the excitation in the gas phase
\citep{p07a}.


The seven galaxies detected in HCO$^+$ are the galaxies
with strong centrally peaked lines, for which the ratio of the CO peak
flux to the full-width half-maximum of the line is greater than 1
(Table~\ref{tbl-dustmasspeak} and
Figure~\ref{fig-co32_mrk_iras}-\ref{fig-co32_n1614_n5258}). Thus, it
appears that we
have detected HCO$^+$ in all the galaxies in our sample where we have
sufficient sensitivity to detect the peak of the line in a single
channel. The fact that the remaining galaxies are not detected may be
attributed to insufficient sensitivity; while it is possible that some
of these galaxies 
may be relatively weak in HCO$^+$ compared to CO, we have no direct
evidence of any such line ratio variations.


\subsubsection{The relative strength of the $^{13}$CO J=2-1 emission}

Of the nine galaxies observed at 230 GHz with the SMA, we have
detected $^{13}$CO J=2-1 emission in only four galaxies: Arp 299,
VV 114, NGC 5331S, and NGC 1614. All of these galaxies except NGC 5331S
were also detected in HCO$^+$; the only galaxy that was observed at
230 GHz and detected in HCO$^+$ but {\it not} in $^{13}$CO is NGC
2623. This is likely because NGC 2623 is less luminous in CO J=2-1
(compare NGC 2623 with NGC 1614 in Table~\ref{tbl-fluxes}). Given the
average strength of the $^{13}$CO line compared to the CO line in our
four detections, we would not expect to detect any of the remaining
galaxies in our sample at the 3 sigma level or better.

\subsection{Published and Archival Single-Dish Data}

In this section, we summarize published single-dish data for the CO J=3-2
line and 880 $\mu$m continuum from the James Clerk Maxwell Telescope (JCMT). 
To these published results we add unpublished
continuum and spectral line data taken from the JCMT
archive. The sample of single-dish data for the CO J=2-1 line and 1.3
mm continuum is less complete; the available data 
will be discussed in a future paper.

\subsubsection{Single-Dish CO J=3-2 Data}

A search of the JCMT archive reveals that many of the galaxies in our 
sample have been observed in the CO J=3-2 line with the JCMT
with a 15$^{\prime\prime}$ beam.
In addition, we used the JCMT to make small (7x7) Nyquist-sampled 
maps (program M05AC05) or 2$^\prime$ Nyquist-sampled maps with the new
array receiver HARP-B (program M07AC11) of many of our
galaxies in the CO J=3-2 line. Our 
ultimate goal is to combine
these single-dish data with our SMA data to obtain complete maps
without any missing spatial frequencies. 
An example and discussion of
this technique as applied to VV 114 is given in \citet{p08}.
These maps also can be used to estimate the total CO J=3-2
emission from some of our galaxies. By combining our new data with
archival data, we can obtain single-dish CO J=3-2 spectra for all the galaxies
in our sample except NGC 5331.

The quality of the data varies quite significantly from one galaxy to
another. For three galaxies (VV 114, NGC 2623, NGC 1614), we have
maps covering 35$^{\prime\prime}$ on a side that clearly include
all the emission from the galaxy. For Arp 299 and NGC 5258, we have
maps covering 2$^\prime$ on a side but with bad baselines in some of
the outer pixels; in addition, these data needed to be scaled up by a
factor of 1.25 to make 
observations of spectral line standards consistent with
previous results. We also adopt a higher calibration uncertainty
(30\%) for these data sets.
Mrk 273, Mrk 231, IRAS 10565+2448, UGC 5101,
IRAS 17208-0014, NGC 6240, and Arp 193
were all observed in a single co-ordinated program in 1999 (program M99AN19,
PI. Papadopoulos) with many observations of calibrators to check for
calibration consistency. 
Arp 55 has only a single 
spectrum centered roughly on the south-western source that is visible
in our interferometric maps (Figure~\ref{fig-Arp55co32}). This flux
measurement must be considered 
a lower limit since this pointing would detect very little emission from
the stronger northern source. For NGC 5257, there is a 
single spectrum published in
\citet{y03}. We reduced these data again (program M01AC03, PI 
Seaquist), including the
calibration data, and there appear to be some problems with 
the calibration during these observations. The main beam efficiency is 
20\% lower than the standard value and the observations of the
spectral line standards are also lower by about a similar amount.
Although we have corrected for these effects in calculating the
CO flux, the calibration must be considered more
uncertain for this galaxy. 

In converting the single-dish fluxes from their intrinsic units
of K(T$_A^*$) km s$^{-1}$ to Jy km s$^{-1}$, we assume an aperture
efficiency of 0.5, which gives a scaling factor of
31.3 Jy K$^{-1}$. All of the galaxies in our sample that have not been
mapped are sufficiently
compact that we assume a point-source distribution in
converting from K to Jy. The exception is Arp 55, for which
we assume the emission
fills the main beam of the telescope and thus divide by the main
beam efficiency $\eta_{MB} = 0.63$ before multiplying by 31.3 Jy K$^{-1}$.
For the four galaxies with small maps, each spectrum was converted
to Jy km s$^{-1}$ using the point source scaling factor, and then
the spectra were summed and normalized by the number of pixels per 
15$^{\prime\prime}$ beam in the map to obtain the final integrated flux.
The resulting single-dish CO J=3-2 fluxes are given in Table~\ref{tbl-co32}.

\subsubsection{Single-Dish 880 $\mu$m Data}

Published 850 or 800 $\mu$m fluxes from the JCMT 
are available for all the galaxies in our
sample except NGC 5331 and Arp 299.
Data presented by \citet{r96} for Mrk 231, Mrk 273, and UGC 5101 were
obtained with the single-pixel bolometer UKT14 at 800 $\mu$m, while data 
from \citet{k01} for IRAS 17208-0014 and NGC 6240 and \citet{d00} 
for the remaining galaxies were obtained with the bolometer
array SCUBA at 850 $\mu$m. In addition, 
SCUBA data for VV 114, NGC 5331, and Arp 299 are
available in the JCMT archive. These data were processed into
images using an automatic pipeline \citep{dF08} and the fluxes were
measured by us from those 
images.


We then rescaled the fluxes by dividing by a factor of
$1.13 \pm 0.02$ (850 $\mu$m) or $1.40 \pm 0.06$ (800 $\mu$m)
to obtain an estimate of the continuum flux at 880 $\mu$m.
These scaling factors assume a dust emissivity $\beta = 1.5 \pm 0.5$.
The resulting single-dish 880 $\mu$m continuum fluxes are given in
Table~\ref{tbl-cont}. The published uncertainties vary somewhat
in their meaning from author to author. \citet{k01}
simply adopt an uncertainty of 30\%. The errors given by
\citet{r96} and for VV 114, NGC 5331, and Arp 299 are
purely statistical measurement errors with no calibration uncertainty
added in. The errors given by \citet{d00} include a calibration
uncertainty of 10\% (15\% for NGC 5257 and NGC 5258). In comparing
the single-dish and SMA fluxes, we have adopted a 20\% calibration
uncertainty for each telescope and adjusted the published
uncertainties for the single-dish data accordingly.

\subsection{Comparison of interferometric and single-dish
  fluxes\label{missing}}  

A comparison of the CO J=3-2 fluxes recovered by the SMA with
the fluxes seen by the JCMT shows that the interferometric data miss
a significant fraction of the CO J=3-2 emission for eight of the galaxies
in our sample.
In these galaxies,
we are typically missing  50\%  of the total single-dish 
flux. However, only for three galaxies in our
sample, Arp 299, VV 114, and NGC 1614, 
does this missing flux manifest itself as significant
negative bowls in the moment maps 
(Figures~\ref{fig-Arp299co32},
~\ref{fig-VV114co32}, and~\ref{fig-NGC1614co32}).
These are the three galaxies with the largest
absolute amount of missing flux and so it makes sense that they are
the ones in which the missing flux leaves a detectable negative signal in the
maps.

We can also calculate the percentage of continuum
flux missed by the interferometric observations 
by comparing the continuum flux detected with the SMA to the
single-dish continuum flux detected with the JCMT. In doing this 
calculation, we also correct for the contribution of the CO J=3-2
line within the SCUBA filter using the formula given by
\citet{s04}. The results given in
Table~\ref{tbl-cont} show that, while the interferometric observations
detect all of the single-dish continuum flux in four galaxies (and
possibly in a fifth, NGC 5258, if the
off-nuclear continuum source is not time variable),
for the remaining nine galaxies typically  50-80\% of the continuum
flux is missed by the interferometer. 

This result is somewhat surprising, as the few sources studied
previously at millimeter wavelengths at high resolution show quite
compact continuum emission. For example,
\citet{s99} detect all the 1.3 mm continuum emission in Arp 220 
and find it is contained in two compact components associated with
the merging nuclei. For Mrk 231, \citet{ds98} also detect
all of the 1.3 mm continuum emission seen in single-dish
data \citep{c92}, although in this case the emission is thought to
be primarily non-thermal emission from the central AGN, with
thermal emission from dust contributing only 20\% of the total flux.
Mrk 273 and Arp 193 have also been detected at 1.3 mm by \citet{ds98},
but there are no single-dish data for these galaxies that can be used to
determine the total flux.

The large amount of missing continuum
flux in many of our sample galaxies suggests that
a significant fraction of the 880 $\mu$m emission occurs on
moderately large spatial scales 
even in systems with very luminous compact cores.
(The 15$^{\prime\prime}$ beam of the JCMT subtends 3-14 kpc at
the distances of the galaxies in our sample.)
A plot of the percentage of flux missing at 880 $\mu$m versus the
percentage missing in the CO J=3-2 line (not shown) shows reasonable  agreement
for all galaxies, although on average there is a slightly higher
percentage of the total flux missing in the continuum images. 
The good agreement between the percentage of missing flux seen in
the CO J=3-2 line and the 880 $\mu$m continuum suggests that the missing
continuum flux comes from dust emission associated with molecular
gas in the more extended disks of the galaxies. Such gas could
be associated with a disk-wide
starburst (as opposed to a nuclear starburst) or more quiescent star
formation as is seen in less luminous spiral galaxies. 
The difference between
our results and previous results for Arp 220 and Mrk 231
is likely due to the fading contribution from dust emission at 1.3 mm
combined with the importance of non-thermal and possibly free-free 
emission in these two galaxies.

\section{The Gas-to-Dust Mass Ratio in the Nuclear Regions\label{sec-gasdust}}

We can use our new CO J=3-2 and 880 $\mu$m data to estimate the
gas and dust masses, and in particular the 
gas-to-dust mass ratio, in the central regions of these luminous infrared
galaxies. The gas-to-dust ratio is interesting because
 it allows us to probe the physical
properties of the interstellar medium on kiloparsec scales in
regions of galaxies with intense heating from starburst activity
and (perhaps) vigorous accretion. Alternatively, it can be used to
provide an independent check of the gas masses determined
from the CO emission lines. The gas mass in the central region
is a critical quantity for understanding the evolution of these
galaxies, as it determines  the fuel for the current activity, be it
a central starburst or accretion onto an active nucleus.

\subsection{Gas and Dust Mass Calculations\label{calc}}

We have calculated the dust mass from the 880 $\mu$m flux
assuming a dust emissivity at 880 $\mu$m, $\kappa = 0.9$ cm$^2$
g$^{-1}$, appropriate for molecular cloud envelopes \citep{h95,j00}.
With this assumption, the dust mass is given by
\begin{equation}
M_{dust} = 74220  S_{880} D_L^2 (\exp(17/T_D) -1) /\kappa \hskip6pt
{\rm (M_\odot)}
\end{equation}
where $S_{880}$ is the 880 $\mu$m flux in Jy and $D_L$ is the luminosity
distance in Mpc.
We calculated the dust temperature, $T_D$, from 
optically thin modified black-body fits to published global photometric data
between 60 and 800 $\mu$m, following the procedure of \citet{k01}.
Our derived temperatures for
Mrk 231, Mrk 273, UGC 5101, and NGC6240 are between 5 K and 10 K
warmer than the temperature of the
cold component derived by \citet{a07}. However, our temperatures are
generally in good agreement (except for Mrk231, for which we are 8 K
colder) with the values derived by \citet{yp07} in fits including new
data at 350 $\mu$m.

To calculate the molecular hydrogen gas mass from the CO J=3-2 emission
line, we need to know the CO-to-H$_2$ conversion factor, $X_{\rm CO}$. While
this conversion factor is now reasonably well determined for normal
galaxies \citep{strong88,ws90,sm96,d01,r03},
it has been clear for many years that the standard
conversion factor does not apply to starbursting systems like the
galaxies studied here \citep{bs96,ds98}.
In addition, the conversion factor has been determined for the CO J=1-0
line only. To apply the appropriate conversion factor to our CO J=3-2
data also requires us to know or assume an appropriate CO J=3-2/J=1-0
line ratio.

To calculate the H$_2$ gas mass, we adopt the revised conversion factor
advocated by \citet{ds98}, $M_{\rm H_2} = 0.8 L^\prime_{\rm CO}(1-0)$, where
$M_{\rm H_2}$ is the H$_2$ gas mass in $M_\odot$ and $L^\prime_{\rm CO}(1-0)$ is the
luminosity of the CO J=1-0 line in K km s$^{-1}$ pc$^2$. 
[This equation corresponds to a CO-to-H$_2$ conversion factor
of $0.5 \times 10^{20}$ H$_2$ cm$^{-2}$ (K km s$^{-1}$)$^{-1}$.]
To convert our measurements in Jy km s$^{-1}$ to units of K km s$^{-1}$ pc$^2$,
we use the equation
\begin{equation}
{L^\prime \over {\rm K~km~s^{-1}~pc^2}} = 3.2546\times 10^7
({S_{\rm CO} \over {\rm Jy~km~s^{-1}}}) ({D_L \over {\rm Mpc}})^2
({\nu_0 \over {\rm GHz}})^{-2} (1+z)^{-1}\end{equation}
where $S_{\rm CO}$ is the CO integrated intensity,  $\nu_0$ is the
rest frequency of the CO transition, and $z$ is the redshift.
To determine
the CO J=3-2/J=1-0 line ratio, we compared the interferometric and
single-dish CO J=3-2 data discussed here with published 
interferometric CO J=1-0 data of similar angular resolution \citep{ds98},
and single-dish data for Mrk 231 \citep{s97} and Mrk 273
\citep{gs04}. The average single-dish line ratio for the latter two galaxies
is 0.48, while the average interferometric line ratio is 0.56, where
both values are calculated on the temperature scale.
Assuming absolute calibration uncertainties of 20\% for both data sets,
these two values agree well, and so we adopt an average CO J=3-2/J=1-0
line ratio of 0.5 in calculating the H$_2$ gas mass. Thus, the
H$_2$ gas mass is calculated via
\begin{equation} M_{\rm H_2} = 1.6 L^\prime_{\rm CO}(3-2) \end{equation} 
where $L^\prime_{\rm CO}(3-2)$ is the
luminosity of the CO J=3-2 line in K km s$^{-1}$ pc$^2$. 
(Note that to obtain the total mass in molecular gas, $M_{\rm mol}$, 
M$_{\rm H_2}$ would need to be multiplied by a factor of 1.36 to
account for helium.)
Gas and dust masses and the gas-to-dust mass ratio calculated
from the total detected CO J=3-2 and 880 $\mu$m luminosity of
each galaxy or galaxy component are given in 
Table~\ref{tbl-dustmass}, while similar quantities calculated from
the peak CO J=3-2 emission and 880 $\mu$m flux in a single
beam are given in Table~\ref{tbl-dustmasspeak}.

\subsection{Uncertainties in Gas and Dust Masses}

The primary source of uncertainty in the H$_2$ masses we have
calculated is the CO-to-H$_2$ conversion factor. We have
adopted here the value advocated by \citet{ds98} as being the most
appropriate for ultraluminous infrared galaxies. The Milky
Way value for this conversion factor is about 5 times larger
\citep{s97}. Thus, if some of the galaxies in our sample, particularly
the less luminous ones, have an interstellar medium more similar to
that of the Milky Way than that of ultraluminous infrared galaxies,
the H$_2$ masses in Tables~\ref{tbl-dustmass} and~\ref{tbl-dustmasspeak}
could be underestimated by up to a factor of five. Detailed radiative
transfer and dynamical modelling can reduce the uncertainty in this
factor by probing the physical conditions directly \citep{i07} and may
provide an independent determination of the CO-to-H$_2$ conversion
factor, and will be presented in a future paper.

A second source of uncertainty comes in our adopted value of
the CO J=3-2/J=1-0 line ratio. This uncertainty is probably not as
large as the uncertainty caused by the CO-to-H$_2$ conversion factor,
as the line ratio we have adopted as being appropriate for Mrk 231
and Mrk 273 agrees very well with the line ratio seen in star forming
regions in more normal spiral galaxies such as M33 \citep{w97} and in
the more highly inclined galaxies from the sample of
\citet{dumke01}. However, many of the galaxies in the \citet{dumke01} sample
have line ratios of 1-1.4, so if it introduces any bias,
our adopted line ratio would
overestimate the H$_2$ mass by a factor of at most 2-3. Also, any
systematic changes in the line ratio with luminosity or evolutionary
phase would introduce similar systematic effects into the gas-to-dust
mass ratio. A spatially resolved comparison with high resolution CO
J=1-0 images 
of these galaxies, which would reduce the uncertainty due to the line
ratio, will be presented in a future paper
\citep[see also][]{i07}.

One major source of uncertainty in the dust masses we have calculated is
the dust temperature. An alternative to assuming optically thin dust
with emissivity varying as $\nu^\beta$ is to
assume that the 60 and 100 $\mu$m fluxes can be fit with
an optically thick blackbody as in \citet{s97}.
This method gives, on average,  dust temperatures that are 1.5 times larger
and dust masses that are 0.55 times smaller than the values given
here. The net result 
of assuming optically thick emission in the far-infrared would be to
increase the gas-to-dust mass ratio by about a factor of two.
Conversely, if we have overestimated the dust temperature by using
single component fits (\S\ref{calc}), this would have a net effect of
decreasing the gas-to-dust mass ratio.


Another source of uncertainty comes from the value of the
dust emissivity that we have assumed. A value of $\kappa
= 0.9$ cm$^2$ g$^{-1}$ is appropriate for moderately dense gas in
star-forming regions in the Galaxy
\citep{h95,j00}, 
while \citet{d00} assumed
$\kappa = 0.77$ cm$^2$ g$^{-1}$ \citep{dl84,h93} in deriving dust masses for
their survey of nearby galaxies. It is possible that the dust
emissivity is modified drastically in the hot, dense central
regions of these galaxies. However, without sufficient data
at high angular resolution to fit physical dust models (see below),
it is impossible at the moment to determine how large an effect this
might be. In any case, the dust masses determined here are likely
uncertain by a factor of less than 2 due to the uncertain value of $\kappa$.

The ideal procedure for determing dust mass would be to fit a
detailed physical model to the full infrared spectral energy
distribution,  as has been done by \citet{g03} for the
dwarf galaxy NGC 1569 or \citet{r04} for NGC 7331. However, only
near-infrared and ground-based mid-infrared 
data can approach the angular resolution of
our 880 $\mu$m SMA data. 
The critical mid- and far-infrared
spectral regions can be covered by the {\it Spitzer Space Telescope}, for which
the angular resolution ranges from 1.5$^{\prime\prime}$ at 3.6
$\mu$m to 47$^{\prime\prime}$ at 160 $\mu$m. Near-infrared data from
{\it Spitzer} can probe the hot dust continuum 
\citep{h06},
while data for the mid- and far-infrared part of the spectrum 
only allow
a determination of galaxy-wide dust properties and do not isolate
the central regions discussed here. In addition, diffraction-limited 
ground-based
mid-infrared photometry at 12.5 $\mu$m is available for nine galaxies in
our sample \citep{s00,s01}; however, no data beyond 30 $\mu$m with
arcsecond-scale resolution are currently available from any
instrument. We defer a more detailed analysis of the
dust temperature and properties inferred by combining our
submillimeter data with infrared data to a future paper.

\subsection{Total Gas and Dust Masses from SMA Data}

The central H$_2$ gas masses calculated from the total luminosities
detected with the SMA range from
 $1.4 \times 10^{9}$ $M_\odot$ for NGC 5257 to 
$1.2 \times 10^{10}$ $M_\odot$ for NGC 6240
while the central dust masses range from $3\times 10^6$ $M_\odot$
for NGC 1614 to $8\times 10^7$ $M_\odot$ for 
Mrk 231 (Table~\ref{tbl-dustmass}).
The central gas-to-dust ratios also show a wide variation,
from as low as 29 for Arp 55(NE) to as high as 720
for NGC 6240. 
Excluding the three galaxy components with non-detections in the continuum,
the average gas-to-dust ratio is
$215 \pm 53$ (rms deviation 207). This value is about 40\%
larger than the typical Milky Way value of $\sim$150
\citep{d07}, and there is a large
dispersion of the individual values about the average. Our value
is also significantly smaller than the value of 540 (rms deviation
290) determined by \citet{sanders91} using single-dish CO J=1-0 data
and far-infrared luminosities. However, that study adopted a
CO-to-H$_2$ conversion factor that is six times larger than our value, 
which almost certainly results in an
overestimate of the total amount of molecular gas in these
galaxies. Adopting the same conversion factor as used in this paper
would give a gas-to-dust ratio of 90 for the data in
\citet{sanders91}, somewhat smaller than the ratio obtained here.

The extremely low values seen for Arp 55 and, to some
extent, for UGC 5101 and Mrk 231 may be due to a significant contribution
to the 880 $\mu$m flux from non-thermal or radio free-free emission. 
However, Arp 55(NE)
shows no evidence for a central AGN that could be a source
of non-thermal emission (see Appendix). 
There is some
evidence to suggest that UGC 5101 contains a buried AGN (see
Appendix), which could increase the 880 $\mu$m flux and lead
to a small overestimate in the dust mass and an underestimate
of the gas-to-dust ratio. 
Mrk 231 is the one galaxy in our sample for which a dominant
AGN contribution has been identified previously at millimeter wavelengths
\citep{ds98}; it is possible that we have underestimated the contribution
from the AGN at 880 $\mu$m  in our calculations.
Mrk 273, Arp 299, NGC 6240, and NGC 2623 have all been identified as
containing an AGN from X-ray data
\citep{x02,z03,k03,m03};  however, the contribution of
non-thermal emission from the AGN at millimeter wavelengths appears
to be small (see Appendix).

Extremely high values for the gas-to-dust ratio are measured
for NGC 6240, VV 114, and NGC 1614.
For NGC 6240, the very broad CO emission line
reduces the bandwidth available to measure the continuum flux and produces
a large uncertainty in the flux. In addition, the CO J=3-2 emission
may be moderately optically thin \citep{d07}, which could result in
the gas mass being overestimated.
VV 114 and NGC 1614 are some of the nearest galaxies in
our sample and, along with Arp 299, are the only ones for which large 
amounts of missing flux
produce negative bowls in the CO J=3-2 maps. Because these galaxies are
relatively nearby, the SMA has greater sensitivity to low surface
density gas that is likely to be spatially extended than it would have
in some of our more distant
galaxies. Figures~\ref{fig-VV114co32} and \ref{fig-NGC1614co32} 
show that the CO-emitting regions in these
galaxies are quite extended;
however, the dust continuum
emission does not appear to be significantly extended in these three galaxies.
The compact nature of the dust emission suggests that the SMA data are not
sufficiently sensitive to pick up continuum emission from dust outside
the peak central concentration. In addition, the continuum data suffer
slightly more from the missing flux problem than the spectral line
data do (\S~\ref{missing} and Tables~\ref{tbl-co32} and~\ref{tbl-cont}) because the line
emission is intrinsically more compact in an individual spectral channel.

An examination of the relative
strengths of the CO J=3-2 emission and the 880 $\mu$m emission suggests
that this is a reasonable explanation. For example, if the gas-to-dust
mass ratio were 150,  
then, using the equations for dust and gas
masses given above, we would expect the flux in the CO J=3-2 moment map
to be roughly 10$^4$ times brighter (in Jy km s$^{-1}$) than the
flux in the 880 $\mu$m map (in Jy). For NGC  1614 and Arp 299, if we
compare the two maps using contours scaled by this factor of 10$^4$, 
we see a very
similar extent and relative strength in the CO J=3-2 and 880 $\mu$m
maps. For VV 114, the central and western peaks
are similar in extent and relative strength in the CO J=3-2 and 880 $\mu$m
maps. However, the
eastern peak is fainter
in 880 $\mu$m continuum than we would expect by a factor of about two,
which suggests that either the gas-to-dust ratio or some other
property of the gas and dust (such as CO emissivity or temperature) is
different in this region.

\subsection{Gas and Dust Masses in the Central Kiloparsec}

The previous discussion combined with 
the large range in the gas-to-dust mass ratios seen in 
Table~\ref{tbl-dustmass} led us to examine the gas-to-dust ratio
using a single resolution element to probe the 
most central region of each galaxy or galaxy component. For
these calculations, we compared the H$_2$ mass calculated from 
the integrated CO J=3-2 intensity in a single beam at the peak of
the emission with the peak 880 $\mu$m continuum intensity
(Table~\ref{tbl-dustmasspeak}). Note that
since many of the continuum sources are not resolved, the peak 
continuum emission given in Table~\ref{tbl-dustmasspeak} has the
same value as the integrated flux given in Table~\ref{tbl-fluxes}.

The average gas-to-dust ratio including all 15 galaxies or galaxy
components with continuum detections in Table~\ref{tbl-dustmasspeak} is 
$120 \pm 28$ (rms deviation 109). This value is smaller
than the average value obtained from the values in
Table~\ref{tbl-dustmass} (i.e., using the total mass
detected with the SMA data) and agrees somewhat 
better with the typical Milky Way
gas-to-dust ratio. Perhaps more importantly, the dispersion about
the mean value is smaller by about a factor of two when only the
peak values are used, which suggests some of the dispersion seen in
the integrated values can be attributed to the better mass sensitivity of
the CO observations (see previous section). 


Thus our new SMA data suggest that the gas-to-dust mass ratio in
the central kiloparsec of these luminous and ultraluminous infrared
galaxies is very similar to the gas-to-dust ratio measured in 
the Milky Way. This result is somewhat surprising given that the
dust in these nuclear regions is subject to more intense heating
(as well as perhaps processing due to shocks) compared to typical
regions in the Milky Way. Perhaps the main zone of activity is very
compact and so does not dominate the energetics of the ISM in the
entire central kiloparsec; this could be tested by sub-arcsec
resolution observations in a direct star formation
tracer. Alternatively, perhaps most of the gas and dust reside in high
column density clouds or cores that help to shield much of the mass
from external heating. 

\subsection{Possible Variations in Gas and Dust Properties}

The normal gas-to-dust ratio seen in these galaxies provides additional
validation for the CO-to-H$_2$ conversion factor derived by \citet{ds98}
and adopted in this analysis. 
In this section, we check for any
systematic differences in the gas-to-dust ratio among the galaxies in 
our sample that may point to differences in
the interstellar medium properties.
Our overall picture 
is one in which the formation and heating of dust tracks the location of 
molecular gas, regardless of whether that gas is distributed largely in 
discrete virialized clouds (as in the Galaxy) or to a large extent in a 
diffuse intracloud medium (as in ULIRGs, where the CO-to-H2 conversion 
factor is consequently lower).
We might expect there to be some differences in our sample, which
spans a factor of ten in infrared luminosity and
exhibits nuclear separations from less than 1 kpc (if multiple nuclei
are seen at all)  to 20 kpc.
Relatively undisturbed galaxies with a larger nuclear separation are
more likely to have a normal interstellar medium with a more normal
CO-to-H$_2$ conversion factor, which would increase the gas mass
and the gas-to-dust ratio.

Among the fourteen objects in our sample are five systems in which the two 
progenitor  galaxies can still be clearly distinguished
(Arp 299, Arp 55, VV 114, NGC 5331, NGC 5257/8), while the remaining
nine objects show a single nucleus or a pair of nuclei separated by
less than 1 kpc. If we calculate the gas-to-dust ratio in the central
kiloparsec for the extended and the compact galaxies separately, we find
no significant difference between the two samples. If we divide the sample
as a function of luminosity, we find that the average gas-to-dust ratio 
calculated for the lower-luminosity galaxies (160-180 depending on
where we divide the sample) is roughly three times larger
than the gas-to-dust ratio for the more luminous galaxies (50-70).
Thus, there appears to
be a strong dependence of the central gas-to-dust ratio on infrared
luminosity in our
sample, but not on nuclear separation. 

These nominally
discrepant ratios are in fact compatible with a reasonably uniform 
gas-to-dust ratio of $\sim 100-150$ once plausible corrections are
made.
We consider here simply changes to the
CO-to-H$_2$ conversion factor,  how the dust temperature
is derived (optically thick versus optically thin),
and the origin of the submillimeter luminosity. We do not
consider varying the absolute value of the dust emissivity, $\kappa$,
since this is relatively poorly constrained in any case. 
Since the less luminous galaxies show on average good agreement with
the Milky Way's gas-to-dust ratio, we focus on possible changes to our analysis
that could increase the gas-to-dust ratio in the more luminous galaxies.

One possibility is that the dust temperature is underestimated by our
optically thin models and that in fact a higher dust temperature,
such as those  derived
by \citet{s97}, is more appropriate. Using the optically
thick dust temperatures would raise the gas-to-dust ratio in these galaxies
by about a factor of two. A second possibility is that there is significant
non-thermal or free-free emission contaminating the 880 $\mu$m flux. 
This is 
difficult to quantify but would also act to raise the gas-to-dust ratio.
There can  
be a significant contribution to the far-infrared luminosity from a
buried AGN and this 
contribution can increase with increasing $L_{\rm FIR}$ \citep{t01}.
Assuming tha tthe AGN also contributes significant emission at 880 $\mu$m,
this effect has the right sense to explain the trend of the
gas-to-dust ratio with $L_{\rm FIR}$. In contrast, 
adopting a larger value for the CO-to-H$_2$ conversion factor (to bring
it closer to the value in normal spiral galaxies) would also
increase the gas-to-dust ratio, but 
would not be consistent with previous studies of 
these extremely luminous galaxies.



\section{Correlations between gas and dust properties in the sample\label{sec-corr}}

\subsection{Testing for correlations}

We have searched for correlations among the gas and dust properties in
the galaxies in our sample. 
The physical and observational quantities that were considered were: 
far-infrared luminosity; 
dust temperature;
the peak H$_2$ surface density derived from the CO J=3-2 emission; 
CO J=3-2 beam area;
total H$_2$ gas mass derived from the CO J=3-2 emission detected with the SMA;
the ratio of the far-infrared luminosity to the total H$_2$ mass;
the ratio of the far-infrared luminosity to the peak H$_2$ mass;
the peak gas-to-dust mass ratio;
and the CO J=3-2/2-1 line ratio
measured from the total flux detected with the SMA.
This list includes both physical quantities associated with the star
formation and ISM properties as well as properties of the sample (such
as beam area and distance) that may introduce selection effects.
We note that the CO J=3-2/2-1 line ratio could be underestimated if
the SMA maps resolve out a larger fraction of the flux in the higher
frequency line. 
We also note that the peak H$_2$ surface density may be underestimated
in IRAS 17208-0014, which has a bright unresolved core, and also
somewhat underestimated in six galaxies (Mrk 231, Mrk 273, IC 694 (Arp
299), Arp 55 NE, Arp 193, and NGC 2623) for which the CO emission is
resolved along the long axis but only slightly extended relative to
the beam along the short axis of emission.
We calculated the correlations using the method described in
\citet{as96} to correct for any distance biases in our sample.

\begin{figure}
\includegraphics[angle=0,scale=0.7]{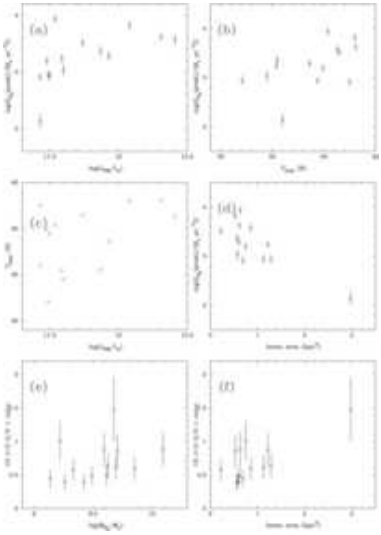}
\caption[multi_panel_good.eps]{Pairs of quantities 
for the galaxies in our sample that showed significant correlations. 
(a) Peak H$_2$ surface density versus far-infrared luminosity. 
(b) Peak H$_2$ surface density versus dust temperature.
(c) Dust temperature versus far-infrared luminosity. 
(d) Peak H$_2$ surface density versus beam area. Note that this
correlation depends heavily on the two galaxies with the most
divergent beam areas.
(e) CO J=3-2/2-1 line ratio versus total H$_2$ mass detected with the
SMA.
(f) CO J=3-2/2-1 line ratio versus beam area.
\label{fig-correlations}}
\end{figure}

\begin{figure}
\includegraphics[angle=0,scale=0.7]{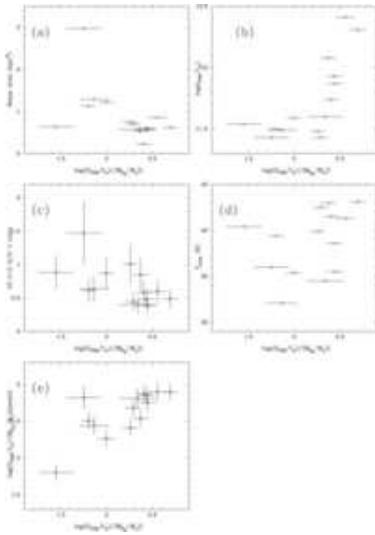}
\caption[multi_panel_bad.eps]{
Correlations of five quantities with
the ratio $L_{\rm
  FIR}/M_{\rm H_2}$, 
where $M_{\rm H_2}$ is the total mass detected with the SMA.  
All these correlations become insignificant if one of NGC 5257 or
Arp 299 is removed from the analysis.
(a) Beam area. 
(b) Far-infrared luminosity $L_{\rm FIR}$.
(c) CO J=3-2/2-1 line ratio.
(d) Dust temperature 
(e) Ratio of far-infrared luminosity to peak H$_2$ mass.
\label{fig-bad_correlations}}
\end{figure}

We identified eleven correlations that are statistically significant
at the 95\% confidence level or better ($ p \le 0.05$,
where $p$ is the probability of a false correlation after
  removing the effect of distance;
Table~\ref{tbl-corr}):
far-infrared luminosity and peak H$_2$ surface density;
far-infrared luminosity and dust temperature;
dust temperature and peak H$_2$ surface density;
beam area and peak H$_2$ surface density;
CO J=3-2/2-1 line ratio and beam area; 
CO J=3-2/2-1 line ratio and total H$_2$ mass detected with the SMA;
and
the ratio of the far-infrared luminosity to the total H$_2$ mass
detected with the SMA
with each of far-infrared luminosity, dust temperature, beam area,
CO J=3-2/2-1 line ratio, and
the ratio of the far-infrared luminosity to the peak H$_2$ mass.
However, the correlation of surface density with beam area becomes
consistent with a null correlation if we use only the twelve galaxies
for which the linear resolution of the beam lies between 0.5 and 1.1
kpc, e.g., excluding NGC 5257 and Arp 299.
In addition, all five of the correlations with
the ratio of the far-infrared luminosity to the total H$_2$ mass
become consistent with a null correlation if we remove NGC 5257
or Arp 299 from the data set. Thus, we conclude that
these five correlations are less robust. 
The remaining five correlations are all significant at the 99\%
confidence 
level except for far-infrared luminosity with dust temperature, which
is significant at the 98\% confidence level; however, this
correlation is well known from previous studies with larger samples
\citep{soifer89}.
Plots of these pairs of quantities are shown in
Figures~\ref{fig-correlations} and~\ref{fig-bad_correlations}.

The correlations of peak H$_2$ surface density with the far-infrared
luminosity and dust temperature suggest that galaxies with higher
gas surface densities are more rapidly producing hot young stars,
which in turn heat the gas and dust more efficiently. 
This correlation is consistent with the observed systematic correlation
of the HCN luminosity (which traces dense gas mass) with $L_{\rm FIR}$
\citep{gs04}.
The far-infrared luminosity can be used to estimate the
star formation rate \citep{k98}, although there can also
be a significant contribution from a buried AGN \citep{t01}.
Thus, this correlation suggests that higher
star formation rates and/or AGN activity 
are associated with higher gas surface densities
inside the central kiloparsec. (Note that this conclusion is somewhat different
from the correlation between 
star formation efficiencies and gas surface densities seen by
\citet{s91} and discussed 
in \S~\ref{sec-scov}.)
Interestingly, the total H$_2$ mass
detected with the SMA does not correlate with the far-infrared
luminosity in our sample, although this may be partly due to the
limited range of $L_{\rm FIR}$ in our sample (see \citet{i08} for a
discussion of this correlation in the context of a larger sample
including high-redshift galaxies.)
This lack of correlation suggests that it is the concentration of gas 
into the central regions that is important for generating the high
star formation rates seen in these merging and merger-remnant
galaxies, rather than the total amount of fuel available 
on somewhat larger scales. 
This result suggests that the increased star formation rate 
and/or AGN activity inferred for
the more luminous infrared galaxies is primarily a result of the
increased availability of fuel for star formation in the central
kiloparsec.

The CO J=3-2/2-1 line ratio has been used as an indicator of temperature
in the molecular gas \citep{w97}. However, for the galaxies in our
sample, there is no obvious correlation between this ratio and
the dust temperature derived from the 60 and 100 $\mu$m IRAS
data. One possible reason for this lack of correlation is that
the derived dust temperatures are based on global measurements while
the CO J=3-2/2-1 line ratio traces material more concentrated to the
center.
The line ratio does correlate with both
the total molecular gas mass and the beam area
(Fig~\ref{fig-correlations}). We can think of no obvious 
physical reason
why the line ratio should correlate with the beam area, and so this
may be produced by better sensitivity (particularly important for the
CO 3-2 line) in the lower resolution
data. A similar effect could produce the correlation with total mass,
although it could also be hinting at a temperature dependence in the
CO-to-H$_2$ conversion factor, which would affect the derived mass.
We do not detect
$^{13}$CO J=2-1 emission from enough of our galaxies to make 
similar correlation plots involving this line. We will present an
analysis of the various CO line ratios, both integrated and spatially
resolved, in combination with radiative transfer models in a future paper.

\subsection{Does the central gas surface density correlate
with $L_{\rm FIR}/M_{\rm H_2}$?\label{sec-scov}}

\begin{figure}
\includegraphics[angle=-90,scale=.65]{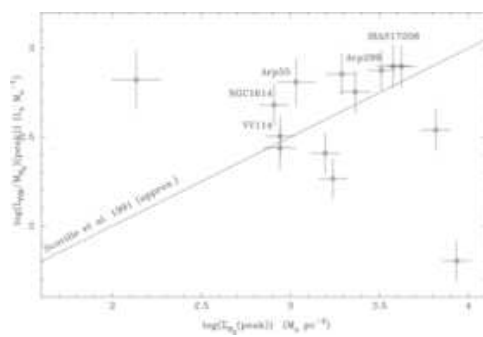}
\caption[sigma_lm.eps]{The ratio of global far-infrared luminosity to
the peak H$_2$ gas mass detected with the SMA versus
peak H$_2$ surface density. Only the highest surface density component
is shown for composite systems.
The error bars shown represent
measurement errors on the CO J=3-2 emission as well as a 20\% calibration
uncertainty. Note that the calibration uncertainty has the same systematic
effect on both quantities, since both are derived from the same data set.
No correlation is seen,  in conflict with
the earlier study by \citet{s91}. The approximate relationship
seen by \citet{s91} (corrected by a factor of
six to account for the different CO-to-H$_2$ conversion factor) is
shown as the straight line. The five galaxies in common between
this study and that of \citet{s91} are labeled on the plot.
\label{fig-sigma_L/M}}
\end{figure}

\citet{s91}  found that the central gas surface density
increased as the ratio of far-infrared luminosity to nuclear gas mass
increased. The far-infrared luminosity is a good tracer of
the star formation rate \citep{k98}; thus, if most of the far-infrared
luminosity occurs in the central regions, then this
ratio traces the star formation efficiency of the central, high
density starburst. \citet{s91} interpreted the observed
trend as indicating that higher star formation efficiencies are
produced by higher gas surface densities. However, our data do not
show any evidence of a similar correlation
(Figure~\ref{fig-sigma_L/M}) and so it is 
worth examining further the possible explanations for the
disagreement between these two results.

\citet{s91} adopted a standard CO-to-H$_2$ conversion factor and
so their gas surface densities can be expected to be a factor
of six higher and their infrared luminosity to H$_2$ mass
ratios a factor of six lower than our values. However, this
would only shift the absolute scale of their correlation and
not change the correlation itself. Indeed, when we estimate their
relationship and correct for the different gas masses, we find that
it agrees well with the location of most of the points in our
analysis (Figure~\ref{fig-sigma_L/M}).
\citet{s91} used the CO J=1-0
transition in their study, which is more sensitive to cooler,
more spatially extended emission than is the CO J=3-2 transition
used here. However, by isolating the nuclear emission from any more
extended emission, \citet{s91} may have mitigated the contribution
of any extended disk gas in their study.

An 
additional issue worth considering is the possibility of different sample 
selection or observational biases.
Our galaxy sample spans a slightly 
smaller range of central gas surface density 
(a factor of 60) than does the sample in \citet{s91},
which had a range of a factor of 100. However, our range of 
surface density drops to only a factor of 10 if the galaxy with the
poorest spatial resolution (NGC 5257) is removed from the sample.
On the other hand, some of the large
range in \citet{s91} may be due to the range of spatial resolution discussed
below.
The sample in \citet{s91} has a factor of 10 larger range in
$L_{\rm FIR}$ and a factor of 5 larger range in the nuclear or peak
gas mass. However,
the net result is that the range of $L_{\rm FIR}/M_{\rm H_2}$ (a factor of 10) is 
similar in the two studies
(although the range in our sample drops to a factor of only 4 if we
remove NGC 6240).
\citet{s91} calculated this ratio using only the gas mass
in the central nuclear source that was also used to calculate the
gas surface density, 
while we have used the peak H$_2$ mass in the central
beam, which would be a difference in the method if many of the sources
in the \citet{s91} sample were spatially resolved. However,
most of the sources in the \citet{s91} sample appear to have been
unresolved in the original data sets, which implies that
the method used to calculate the surface density and
the $L_{\rm FIR}/M_{\rm H_2}$ ratio is very similar in this paper
and in \citet{s91}. 

\citet{s91}
used CO J=1-0 data for a sample of 14 galaxies, including
5 galaxies from our sample, with resolutions ranging from 0.1 to 3.2
kpc. Although their {\it range} of resolution is similar to that of the
sample presented here, the {\it distribution} of resolutions 
is much broader than for our data set, where
all but two galaxies (Arp 299 and NGC 5257/8) 
have resolutions between 0.7 and 1.1 kpc.
This distribution of resolutions might introduce systematic effects into
the analysis of \citet{s91} in the sense that galaxies observed with better 
angular resolutions could have higher gas surface densities.
Indeed, the average resolution for the seven galaxies at the high
end of the correlation is 0.66 kpc, while the average resolution for the
remaining seven galaxies is 1.14 kpc.

To examine the effect of resolution further, we looked at the five galaxies
that are common to both samples  (Arp 299/IC694,
IRAS 17208-0014, Arp 55, VV 114, and NGC 1614). The values of surface 
density and $L_{\rm FIR}/M_{\rm H_2}$ ratio are very similar for Arp 299
and NGC 1614, which are two galaxies for which the angular resolutions of
the CO J=1-0 and the CO J=3-2 data agree within a factor of two. Both
IRAS 17208-0014 and VV 114 move to higher surface densities
and $L_{\rm FIR}/M_{\rm H_2}$ ratios using the CO J=3-2 data, which have
angular resolutions a factor of 2-3 better than the CO J=1-0 data.
The most dramatic change is for Arp 55, for which the CO J=3-2 data have
a factor of 8 better angular resolution: its surface density and
$L_{\rm FIR}/M_{\rm H_2}$ ratio increase dramatically, to the point where it is
larger in both quantities than NGC 1614.


In summary, the better uniformity in angular resolution in our
sample compared to that of \citet{s91} gives us confidence that
the lack of correlation seen between the peak H$_2$ surface density
and the ratio of infrared luminosity to H$_2$ mass seen in our
analysis is a real effect
and not an artifact of our sample selection or observing techniques.
While such a correlation may exist for galaxies with lower 
far-infrared luminosities (below the luminosity cutoff of our
sample), it cannot help us in understanding
the high star formation rates in the luminous systems studied here.
The data presented here imply a star formation rate
(\S~\ref{sec-corr}), not a star 
formation efficiency, that increases with the central gas surface density.
We will explore these interesting correlations further in a future paper.

\section{Conclusions}

In this paper, we have
presented new data obtained with the Submillimeter Array 
for a sample of fourteen luminous and ultraluminous infrared
galaxies selected to have distances $D_L < 200$ Mpc and
far-infrared luminosities $\log L_{\rm FIR} > 11.4$. 
We have obtained data 
in the CO J=3-2, CO J=2-1, $^{13}$CO J=2-1,
and HCO$^+$ J=4-3 lines as well as continuum data at
880 $\mu$m and 1.3 mm with spatial resolutions of order 1 kpc
or better in all but one of the target galaxies. We present
integrated intensity, velocity field, and velocity dispersion maps
for the $^{12}$CO lines, integrated intensity maps for
the continuum, $^{13}$CO, and HCO$^+$ lines, and peak and integrated
spectra for all the detected lines.

We have compared our CO J=3-2 and 880 $\mu$m continuum fluxes detected
with the SMA with published, archival, and new data from the
James Clerk Maxwell Telescope. This comparison shows that
the interferometric data miss a significant fraction 
(typically 50\%) of the
CO J=3-2 emission for eight of the galaxies in our sample
and also a significant fraction (typically 50-80\%) of
the continuum flux for nine of the galaxies.
This large amount of missing continuum
flux suggests that
a significant fraction of the 880 $\mu$m emission 
in these systems occurs on
moderately large spatial scales. 
The good agreement between the percentage of missing flux seen in
the CO J=3-2 line and the 880 $\mu$m continuum suggests that the missing
continuum flux comes from dust emission associated with molecular
gas in the more extended disks of the galaxies.

We have combined our CO and continuum data to determine the
gas-to-dust mass ratio in the central regions of these galaxies.
We adopt the smaller value of the  CO-to-H$_2$ conversion factor from
\citet{ds98} and calculate the dust temperature by fitting a modified
blackbody function as in \citet{k01}.
Because of the lower signal-to-noise ratio in the continuum data,
we find that we obtain more consistent measurements of the gas-to-dust
mass ratio if we use a single beam to probe the central region of each
galaxy or galaxy component. We find an average gas-to-dust mass
ratio of $120 \pm 28$ (rms deviation 109), very similar to the
value of 150 determined for the Milky Way. This similarity between
the gas-to-dust ratio in these luminous systems and that in the Milky
Way is somewhat surprising, given that the dust is subject to more
intense heating from the starburst and possibly accretion activity
compared to typical regions in the Milky Way. 

We have searched for correlations among nine
physical and observational quantities
for the galaxies in our sample. We find five correlations that appear
to be statistically significant as well as robust to small changes in
the exact galaxy sample. The most interesting correlation is
that of peak H$_2$ surface density with the far-infrared
luminosity. Since the far-infrared luminosity
can be used to estimate the star formation rate,
these correlations suggest that galaxies with higher gas surface
densities inside the central kiloparsec have a higher star formation rate.
We do not see a significant correlation of total H$_2$ mass with
the far-infrared luminosity, which suggests that the increase
in star formation rate is due to the increased availability of
molecular gas as fuel for star formation in the central regions,
rather than the total amount of gas available on somewhat larger scales.

Our data do not show any evidence of a significant correlation between
central gas surface density and the ratio of far-infrared luminosity
to nuclear gas mass. This lack of correlation is different from
the results of \citet{s91}, who interpreted their observed correlation as
indicating that higher star formation efficiencies result from higher
gas surface densities. We suggest that the correlation seen by
\citet{s91} was produced by the wider distribution of spatial
resolutions in their data set  and is not an intrinsic property of
these very luminous galaxies. To reiterate, our new data show that
it is star formation {\it rate}, not star formation {\it efficiency},
that increases with the central gas surface density in luminous and
ultraluminous infrared galaxies

There are a number of additional papers in preparation or planning
that will present more detailed analysis of various aspects of the data.
We will compare the results from
this survey with similar observations of high-redshift
submillimeter galaxies to study the gas properties of a wide range of
luminous galaxies using the CO J=3-2 line to trace the molecular gas
content \citep{i08}.
A detailed analysis of the molecular gas properties of NGC 6240 has
already been presented in \citet{i07};
we will  
present similar detailed analyses of the gas, dust, and star formation
properties individual
galaxies such as VV 114 \citep{p08} and Arp 299.
We will 
examine the physical properties of the molecular gas for the entire
sample using
spatially resolved radiative transfer models, similar to what has been
done for NGC6240 \citep{i07}, as well as carry out dynamical analysis
and modeling of the galaxies, both of which can give an independent
estimate of the CO-to-H$_2$ conversion factor, similar to the analysis
of \citet{ds98}.
This physical analysis will also allow us
to place constraints on the origin of the OH megamaser activity
in luminous infrared galaxies \citep{darling07}.
We will combine our high-resolution SMA data with
{\it Spitzer} data to compare the properties of the warm and cold dust
\citep[see also][]{a07,m07}.
Finally, we will make detailed comparisons between
the molecular gas and dust properties of these U/LIRGs and  the
predictions of numerical simulations of merging galaxies
\citep[e.g.,][]{c06,c07}.

\acknowledgments

The Submillimeter Array is a joint project between the Smithsonian 
Astrophysical Observatory and the Academia Sinica Institute of Astronomy 
and Astrophysics and is funded by the Smithsonian Institution and the 
Academia Sinica.
The James Clerk Maxwell Telescope is operated by The Joint Astronomy 
Centre on behalf of the Particle Physics and Astronomy Research Council 
of the United Kingdom, the Netherlands Organisation for Scientific Research, 
and the National Research Council of Canada.
This research has made use of the NASA/IPAC Extragalactic Database (NED) 
which is operated by the Jet Propulsion Laboratory, California Institute 
of Technology, under contract with the National Aeronautics and Space 
Administration.
We are grateful to the many SMA observers who helped to take the
data presented in this paper and to the SMA TAC for giving this
project a high priority.
We thank James di Francesco for
making available to us the pipeline processed archival SCUBA data
for VV 114, NGC 5331, and Arp 299, Padelis Papadopoulos for access to
his single dish CO data, and the anonymous referee for a very useful
and prompt referee report. C.D.W.  and J.G. acknowledge support by  the
Natural Science and Engineering Research Council of Canada (NSERC). 
A.J.B. acknowledges support by National Science 
Foundation grant AST-0708653. M.J. acknowledges support by the Academy of 
Finland grant 124620.
 
{\it Facilities:} \facility{SMA}, \facility{JCMT}.

\appendix

\section{Discussion of individual galaxies}



\subsection{IRAS 17208-0014}

\citet{v95} classify the optical spectrum of IRAS 17208-0014 as HII.
High-resolution near-infrared imaging shows no direct evidence
of an AGN, as the nuclear emission is extended in all bands
\citep{sco00}. The [NeV] line, which is an indicator of an AGN, is not
detected in this galaxy \citep{f07}.
High-resolution radio imaging at 4.8 GHz also shows
no evidence for an AGN \citep{bk06}.
Near-infrared images
show extended emission with a brighter nucleus
\citep{sco00}.  

However, the millimeter continuum fluxes in Table~\ref{tbl-fluxes} suggest 
there may be a significant non-thermal or free-free component present. The 880
$\mu$m and 1.4 mm continuum fluxes are inconsistent with dust with
$\beta$ = 1.5 at greater than the 1.5$\sigma$ level even after
including 20\% calibration uncertainties.
If we assume
that any nonthermal component 
depends inversely on frequency \citep{condon92} while the dust component
varies as $\nu^{3.5}$ (dust emissivity $\beta = 1.5$), then the
non-thermal emission would contribute 40\% of the flux
at 880 $\mu$m.

\subsection{Mrk 231 (UGC 8058, VII Zw 490)}

Mrk 231 is the only object in our sample to show strong emission
from an AGN. \citet{v95} classify the optical spectrum as Seyfert 2,
while \citet{g98} suggest an AGN-like radiation field based on
mid-infrared emission lines and limits. \citet{s98} suggest an AGN
based on their 22 GHz VLA data; \citet{l03} find that VLBI
imaging data are most consistent with an AGN morphology, although
they point out that the AGN may not be responsible for the majority
of the bolometric luminosity of the galaxy. However, \citet{d04} find
that the nuclear starburst contributes 25-40\% of the bolometric luminosity.
Mrk 231 also
contains a hard X-ray source that is variable on timescales of a few
hours \citep{g02}.

The millimeter continuum fluxes in Table~\ref{tbl-fluxes} suggest 
there is a significant non-thermal component present. \citet{ds98}
attributed all but 20\% of their 1.3 mm continuum flux to
non-thermal emission. If we assume that any nonthermal component
depends inversely on frequency \citep{condon92} while the dust component
varies as $\nu^{3.5}$ (dust emissivity $\beta = 1.5$), then the
non-thermal emission contributes perhaps 25\% of the flux
at 880 $\mu$m.

\subsection{Mrk 273  (UGC 8696, VV 851, I Zw 71)}

\citet{v95} classify the optical spectrum of Mrk 273 as LINER.
\citet{g98} suggest that the mid-infrared radiation has equal
contributions from a starburst and an AGN. 
\citet{a07} detect both [NeV] emission and continuum emission from
dust hotter than 300 K, both of which indicate the presence of an AGN.
\citet{x02}
detect hard X-ray emission from the northern nucleus, while \citet{p03}
detect the Fe line in X-rays, both
suggesting the presence of an AGN. However, \citet{s98} find that the
compact 22 GHz core can be fit by a model of clustered supernovae.
Near-infrared images
show extended emission with two bright central sources
\citep{sco00}.  

The relative strength of the 1.3 and 2.6 mm continuum emission
\citep{ds98} suggests that some non-thermal emission is present.
However, the contribution at 880 $\mu$m would be small and well within the
20\% calibration uncertainty of the data.

\subsection{IRAS 10565+2448} 

This galaxy has been relatively poorly studied.
\citet{v95} classify the optical spectrum of IRAS 10565+2448 as
HII. \citet{c91} do not find any compact radio continuum peak
at 1.49 GHz.
Near-infrared images
show extended emission with a brighter nucleus
\citep{sco00}.  
The [NeV] line, which is an indicator of an AGN, is not
detected in this galaxy \citep{f07}.

The millimeter continuum data are consistent with pure dust emission,
although the 1.3 mm continuum point is on the large side of what
is possible, leaving open the possibility of a small non-thermal
contribution at 880 $\mu$m.

\subsection{UGC 5101}

\citet{v95} classify the optical spectrum of UGC 5101 as LINER.
Although \citet{g98} suggest the mid-infrared radiation field is more
similar to that of a starburst than an AGN,
\citet{a04} find that [NeV] emission implies this source contains
a buried AGN. \citet{a07} detect continuum emission from
dust hotter than 300 K, which also indicates the presence of an AGN.
\citet{i03} detect hard X-ray emission, including
the Fe line, which they interpret as evidence for a buried AGN.
\citet{l03} use VLBI imaging to find that
an AGN is responsible for at least 10\% of the total radio flux.
Near-infrared images
show extended emission with a brighter nucleus
\citep{sco00}.

The millimeter continuum data are consistent with pure dust emission,
although the 1.3 mm continuum point is on the large side of what
is possible, leaving open the possibility of a small non-thermal
contribution at 880 $\mu$m.

\subsection{Arp 299 (VV 118, NGC 3690, Mrk 171, IC 694)}

High-resolution radio images from 1.4 to 8.4 GHz reveal five bright
compact sources as well as extended emission \citep{n04}.
The bright compact radio source in the nucleus of IC 694 breaks up
into five candidate radio supernova remnants when observed with
the VLBA; one of the sources may possibly be a low-luminosity
AGN \citep{n04}.
X-ray images reveal a population of compact sources, including sources
 in the nucleus of each component which are likely to be AGN
\citep{z03}. Further evidence for an AGN in NGC 3690 comes from
\citet{gm06}, who identify a conical region with Seyfert-like excitation 
emanating from the B1 region.
Mid-infrared images show three compact sources
associated with the two nuclei as well as the emission region north of 
NGC 3690 \citep{s01}; near-infrared images reveal a fourth compact
source
near NGC3690 as well as extended emission \citep{a00}

The combination of the 2.6 mm continuum flux from \citet{a97} with
the 1.3 mm continuum flux from this paper suggests that some
non-thermal component may be present in the eastern component (IC694).
If we assume the non-thermal emission scales inversely with
frequency and that all of the 2.6 mm continuum emission is
non-thermal, then any non-thermal contribution at 880 $\mu$m would
be less than 10\% of the total.

\subsection{Arp 55 (UGC 4881, VV 155)}

Arp 55 is one of the least studied galaxies in our sample. 
\citet{v95} classify the optical spectrum as HII for both
components. \citet{l93} 
detected Arp 55 using VLBI observations, but find the source
to be consistent with a compact starburst or a group of
clustered supernova.

\subsection{Arp 193 (IC 883, VV 821, UGC 8387,  I Zw 56)}

\citet{v95} classify the optical spectrum as LINER.
8.4 GHz images from the VLA show disk-like extended emission with 
no sign of a compact source \citep{c91}.
\citet{rush96} give an upper limit on the soft X-ray luminosity. 
Near and 
mid-infrared images
also show disk-like extended emission \citep{sco00,s01}.

\subsection{NGC 6240 (IC 4625, UGC 10592, VV 617)}

\citet{v95} classify the optical spectrum as LINER.
However,
hard X-ray imaging has revealed two AGNs separated by $\sim
1^{\prime\prime}$ \citep{k03}.
\citet{a07} detect continuum emission from
dust hotter than 300 K, which also indicates the presence of an AGN.
High-resolution radio images from 2.3 to 8.4 GHz reveal two
compact radio sources with properties similar to Seyfert nuclei
\citep{gb04}.
Near-infrared images also show two nuclei as well as 
bright extended emission
\citep{sco00}.  
For a more complete discussion of the multi-wavelength properties of
NGC 6240, see \citet{i07}.

\subsection{VV 114 (Arp 236, IC 1623)}

\citet{v95} classify the optical spectrum as HII for both components.
8.4 GHz images from the VLA show no sign of a compact source
\citep{c91}. However, \citet{l02} find that the mid-infrared
spectrum shows signs of an AGN in the eastern component.
Near and mid-infrared images
show extended emission in both components with a bright compact source
in the eastern component \citep{sco00,s01}.  

The millimeter continuum data are consistent with pure dust emission,
although the 1.3 mm continuum point for IC694 is on the large side of what
is possible, leaving open the possibility of a small non-thermal
contribution at 880 $\mu$m.

\subsection{NGC 5331 (UGC 8774, VV 253)}


This galaxy has been relatively poorly studied.
\citet{a95} classify both components as starbursts based on optical
spectroscopy and \citet{rush96} give an upper limit on the soft X-ray
luminosity. 
\citet{c90} detect both components at 1.49 GHz, with the 
southern component being about twice as strong as the northern component.

\subsection{NGC 2623 (Arp 243, UGC 4509)}

\citet{h83} classify the optical spectrum as LINER.
8.4 GHz images from the VLA show a strong compact radio source
\citep{c91}. \citet{m03} classified this galaxy as an obscured AGN
using X-ray observations from Chandra. Near and mid-infrared images
show a single compact nucleus \citep{sco00,s01}.

The combination of the 1.3 mm continuum flux from Table~\ref{tbl-fluxes} with
the 880 $\mu$m continuum flux from this paper suggests that some small
non-thermal component may be present.
If we assume the non-thermal emission scales inversely with
frequency and that the dust emissivity goes as $\beta=1.5$,
then any non-thermal contribution at 880 $\mu$m would
be less than 10\% of the total.

\subsection{NGC 5257, NGC 5258 (Arp 240, VV 55, UGC 8641,
UGC 8645)}

\citet{v95} classify the optical spectrum as HII and \citet{rush96}
give an upper limit on the soft X-ray luminosity of NGC5258.
8.4 GHz images from the VLA show emission in both components that
is somewhat extended \citep{c91}. 
\citet{s07} present Spitzer mid-infrared images at 3.6,
8, and 24 $\mu$m, which show that the bright arm seen in CO in NGC5258
is also prominent at 24 $\mu$m.

NGC 5258 is an unusual case in that most of
the CO J=3-2 emission seen by the SMA comes from the bright
southern arm, with only weak emission seen from the nucleus. Surprisingly, this
galaxy contains the strongest continuum source in our entire sample
and is detected at the 6$\sigma$ level despite a relatively high
noise in the SMA image. However, this point source is not located
at the center of the galaxy but rather (-15,8) arcseconds to the 
north-west. Thus, the low gas-to-dust ratio seen in Table~\ref{tbl-dustmass}
is most likely spurious.

\subsection{NGC 1614 (Arp 186, Mrk 617)}

\citet{v95} classify the optical spectrum of NGC 1614 as HII,
although the [NII] lines suggest a LINER classification.
\citet{n90} find no direct evidence for an AGN component in
their 4.6 GHz maps. \citet{f06} fit the infrared
spectral energy distribution with a type II AGN model; however,
these model fits suggest that starburst emission is likely the
dominant component at 880 $\mu$m. High-resolution radio,
near-infrared, and mid-infrared imaging reveal a starforming ring with
diameter 1.2$^{\prime\prime}$ \citep{s01,a01,n90}.

NGC 1614 is the only object in our sample besides Mrk 231 to show
clear evidence of a non-thermal component from the ratio of its
millimeter continuum fluxes. The
peak 880 $\mu$m and 1.3 mm continuum fluxes are almost
identical (Tables~\ref{tbl-fluxes} and ~\ref{tbl-dustmasspeak}).
If we assume that any non-thermal component depends inversely
on frequency, then dust emission represents perhaps 60\% of
the total 880 $\mu$m flux and only 40\% of the peak 880 $\mu$m
flux. However, correcting for this putative non-thermal component
produces the highest peak gas-to-dust ratio of any galaxy in
our sample (Table~\ref{tbl-dustmasspeak}).

\clearpage


\begin{deluxetable}{lcccl}
\tabletypesize{\scriptsize}
\tablecaption{\bf The Nearby Luminous Infrared Galaxy Sample \label{tbl-sample}}
\tablewidth{0pt}
\tablehead{
\colhead{Galaxy}  &     \colhead{$\log L_{\rm FIR}$} &  \colhead{D$_L$} & \colhead{$cz$} \\
 &                \colhead{($L_\odot$)} & \colhead{(Mpc)} &  
}
\startdata
IRAS 17208-0014 & 12.41 & 189 & 12835 \\ 
Mrk 231          & 12.31 &  179 & 12642 \\ 
Mrk 273          & 12.08 &  166 & 11327  \\ 
IRAS 10565+2448      & 11.93 &  191 & 12921  \\ 
UGC 5101         & 11.87 &  174 &  11809 \\ 
Arp 299          & 11.74 &  ~~44 & 3088  \\ 
Arp 55           & 11.60 &  173 &  11900  \\ 
Arp 193    & 11.59 &  102 &  7000 \\ 
NGC 6240 & 11.54 & 107 & 7339 \\ 
VV 114           & 11.50 &  ~~87 & 6010  \\ 
NGC 5331         & 11.49 & 145 &   9907 \\ 
NGC 2623         & 11.48 &  ~~80 &   5535\\ 
NGC 5257/8       & 11.43 &  ~~99 & 6775   \\ 
NGC 1614         & 11.43 &  ~~69 &  4778 \\ 
\enddata
\end{deluxetable}

 

\begin{deluxetable}{lcclccclcc}
\tabletypesize{\scriptsize}
\tablecaption{\bf Observational Properties of the Survey \label{tbl-obs}}
\tablewidth{0pt}
\tablehead{
& \multicolumn{4}{c}{CO J=3-2 Observations} & \multicolumn{4}{c}{CO J=2-1 Observations} \\
\colhead{Galaxy } & \colhead{Date} & \colhead{Number of} 
& \colhead{Sensitivity\tablenotemark{a}} & \colhead{Time} 
& \colhead{Date} & \colhead{Number of} 
& \colhead{Sensitivity\tablenotemark{a}} & \colhead{Time} \\
       & \colhead{Observed} & \colhead{Antennas} & \colhead{(mJy (K))} 
& \colhead{(hr)} & \colhead{Observed} & \colhead{Antennas} 
& \colhead{(mJy (K))} & \colhead{(hr)} 
}
\startdata
IRAS 17208-0014 & 20050818,0516 & 6 & 86 (1.01) & 5.9  & ... & ... & ... & ... \\
Mrk 231          & 20060205 & 8 & 33 (0.55) & 3.3 & ... & ... & ... & ... \\
Mrk 273          & 20060204 & 7 & 45 (0.67) & 3.8  & ... & ... & ... & ... \\
IRAS 10565+2448      & 20060202 & 7 & 25 (0.46) & 3.7 & 20060206 & 7 &
15 (0.36) & 4.8 \\
UGC 5101         & 20060130 & 6 & 32 (0.56) & 5.6 & 20060207 & 7 & 18
(0.39) & 5.0 \\
Arp 299       & 20060410 & 7 & 37 (0.087) & 5.6 & 20070327 & 8 & 20
(0.086) & 6.6 \\
Arp 55           & 20060129 & 6 & 40 (0.74) & 4.3 & 20060211 & 7 & 18
(0.44) & 6.3 \\
Arp 193    & 20060416 & 8 & 38 (0.093) & 3.6  & ... & ... & ... & ... \\
NGC 6240 & 20051009,1014 & 7 & 44 (0.24) & 9.9 & ... & ... & ... & ... \\ 
VV 114           & 051113 & 7 & 31 (0.057) & 4.4 & 20051115,1125 & 7 &
15 (0.029) & 7.0 \\
NGC 5331      & 20070325 & 8 & 50 (0.23) & 6.3 
& 20060422 & 8 & 21 (0.047) & 5.0 \\
NGC 2623         & 20070116 & 8 & 12 (0.028) & 6.7 &
20060212,0213,0313 & 6,8 & 20 (0.40) & 13.0  \\
NGC 5257/8    & ... & ... & ... & ... & 20060318 & 8 & 51 (0.12) & 4.8 \\
NGC 5257    & 20060411,20070502 & 7,6 & 127 (0.12) & 3.0 & ... & ... & ... & ... \\
NGC 5258    & 20060411,20070502 & 7,6 & 165 (0.16) & 3.0 & ... & ... & ... & ... \\
NGC 1614         & 20051114 & 7 & 41 (0.079) & 5.1 & 20051112 & 7 & 14
(0.027) & 5.5 \\
\enddata
\tablenotetext{a}{Noise level measured from the dirty map using 
line-free channels with
20 km s$^{-1}$ resolution.}
\end{deluxetable}

\clearpage



\begin{deluxetable}{lcccccccccccccc}
\rotate
\setlength{\tabcolsep}{.05cm}
\tabletypesize{\tiny}
\tablecaption{\bf Interferometric CO and continuum fluxes\label{tbl-fluxes}}
\tablewidth{0pt}
\tablehead{
\colhead{Galaxy } & \colhead{CO J=3-2 flux} & \colhead{beam } &
\colhead{880 $\mu$m\tablenotemark{a}} & \colhead{HCO$^+$ J=4-3
  flux\tablenotemark{b}} &  
\colhead{CO J=2-1 flux} & \colhead{$^{13}$CO J=2-1 flux\tablenotemark{a}} & \colhead{1.4mm\tablenotemark{a}} & \colhead{beam} & 
\colhead{CO J=1-0 flux} & \colhead{3mm} & \colhead{beam} & \colhead{Refs.\tablenotemark{c}} \\
        & \colhead{(Jy km s$^{-1}$)} & \colhead{($^{\prime\prime}$)} & \colhead{(mJy)} & \colhead{(Jy km s$^{-1}$)} & 
\colhead{(Jy km s$^{-1}$)} & \colhead{(Jy km s$^{-1}$)} & \colhead{(mJy)} & \colhead{($^{\prime\prime}$)} & 
\colhead{(Jy km s$^{-1}$)} & \colhead{(mJy)} & \colhead{($^{\prime\prime}$)} 
}
\startdata
IRAS 17208-0014 & 478 $\pm$ 33 & 1.0x0.9 & 48 $\pm$ 10 & ... & 355 $\pm$ 7  &
... & 37 $\pm 3$ & 1.0x0.7 & 132 & $<$5 & 5.1x1.6 & 1,10 \\
Mrk 231          & 308$\pm$8 & 0.9x0.8 & 80$\pm$4 & 25$\pm$3 & 280 & ... & 36 & 0.7x0.5 & 97 & 63 & 1.3x1.1 & 1 \\
Mrk 273   & 441$\pm$ 14 & 0.9x0.8 & 56$\pm$5 & ... &  231 & ... & 8 & 0.6  & 78 & 11 & 1.4x1.3 & 1 \\
IRAS 10565+2448  & 204$\pm$7 & 0.9x0.7 & 15$\pm$3 & ... & 187 $\pm$ 6 & $<$ 2 & 6 $\pm$ 1 & 1.1x1.0 & 68 & $<2$ & 2.3x1.4 & 1 \\
UGC 5101 & 209$\pm$10 & 1.0x0.7  & 37$\pm$9 & ... & 237 $\pm$ 10 & $<$ 4 & 12 $\pm$ 2 & 1.2x0.9 & 50  & ... & 2.1x1.6 & 2 \\
Arp 299            & 2582$\pm$34 & 2.3x1.9 & 101$\pm$7 & $\ge$67 $\pm$ 5 & 1976 $\pm$ 16 & 38.8 $\pm$ 3.4 & 46 $\pm$ 6 & 3.1x1.8 & 397 & 31$\pm$3 & 2.5x2.2 & 3\\
... IC 694            & $\ge$ 1610$\pm$ 27 & `` & 81$\pm$5 & $\ge$50
$\pm$ 4 & 1168 $\pm$ 13 & 17.5 $\pm$ 2.7 & 40 $\pm$ 6 & `` & 242 & 17$\pm$2  & `` & 3 \\
... NGC3690           & $\ge$ 596$\pm$ 17 & `` & 20$\pm$5 & $\ge$17
$\pm$ 3  & 355 $\pm$ 8 & 3.5 $\pm$ 1.1  & $<$4 & `` & 53 & 5$\pm$2 & `` & 3 \\
Arp 55 & 146$\pm$12 & 0.9x0.8 & 26$\pm$7 & ... & 163 $\pm$ 9 & $<$ 6 &  $<2$ & 1.1x0.9   & 83  & ...  & 9x7 & 4\\
... Arp 55(NE) & 94$\pm$9 & 0.9x0.8 & 26$\pm$7 & ... &110 $\pm$ 8 & $<$ 3 & $<2$ & ``   & ...  & ...  & `` & 4\\
... Arp 55(SW) & 52$\pm$8 & 0.9x0.8 & $<8$ & ... & 53 $\pm$ 4 & $<$ 3 & $<2$ & ``   & ...  & ... & `` & 4\\
Arp 193  & 886 $\pm$ 15 & 2.2x2.0 & 39 $\pm$ 4 & $\ge$21 $\pm$ 5 & 450 & ... & 10 & 0.6x0.4 & 161 & $<$5 & 1.3x0.9 & 1 \\
NGC 6240 & 2428 $\pm$ 52 & 1.6x1.3  & 33 $\pm$ 13 & $\ge$166 $\pm$ 12 & 1220 & ... & 5.9 $\pm  $ 0.3 & 0.9x0.5 & 324 &
... & 4.9x4.1  & 7,9 \\
...NGC 6240(central) & 2318 $\pm$ 48 & 1.6x1.3  & 33 $\pm$ 13 &
$\ge$166 $\pm$ 12 & 1220 & ... & 5.9 $\pm  $ 0.3 & 0.9x0.5 & 324 &
... & 4.9x4.1  & 7,9 \\
...NGC 6240(WC) & 110 $\pm$ 19 & 1.6x1.3  & $<$16 & ... & ... & ... & ... & ... & ... &
... & ... & 7,9 \\
VV 114    & 1530 $\pm$ 16 & 2.8x2.0 & 26 $\pm$ 6 & $\ge$17 $\pm$ 2 & 1109 $\pm$ 8 & 10.6 $\pm$ 2.6$\tablenotemark{d}$ & 11 $\pm$ 2 & 4.1x3.0  & 674 & $<$16 & 4.4x3.1 & 5\\      
NGC 5331 & 478 $\pm$ 26 & 2.3x1.0 & 27 $\pm$ 6 & ... & 333 $\pm$ 11 & 8.8 $\pm$ 2.4 & 8 $\pm$ 2 & 3.5x3.1  & 167 & ...  & 5.4x4.4 & 6\\
... NGC 5331S   & 401 $\pm$ 24 & `` & 27 $\pm$ 6 & ... & 250 $\pm$ 7 & 8.8 $\pm$ 2.4 & 8$\pm$ 2 & ``  & 150 & ...  & `` & 6 \\
... NGC 5331N   & 77 $\pm$ 10 & `` & $<$ 12 & ... & 83 $\pm$ 9 & $<$ 4 & $<$ 4 & ``  & 17 & ...  & `` & 6 \\
NGC 2623  & 607 $\pm$ 5 & 2.2x2.0 & 50 $\pm$ 2 & 27 $\pm$ 2 & 267 $\pm$ 8 & $<$ 3 &  17 $\pm$ 3 & 1.2x1.0 & 153 & ...  & 3.5x2.4  & 7  \\
NGC 5257/8 & 1140 $\pm$ 70 & 3.8x3.0 & 104 $\pm$ 21 & ... & 344 $\pm$ 51 &$<$ 20   & $<$ 16 & 3.5x2.8  & 387 & ... & 6.2x3.7 & 6 \\
... NGC 5257 & 286 $\pm$ 47 & `` & $<$ 26 & ... & 109 $\pm$ 24 & $<$ 10  & $<$ 8 & ``  & 137 & ... & `` & 6 \\
... NGC 5258 & 854 $\pm$ 52 & `` & 104 $\pm$ 21$\tablenotemark{e}$ & ... & 235 $\pm$ 45  & $<$ 10 & $<$ 8 &  `` & 250 & ...  & `` & 6 \\
NGC 1614 & 674 $\pm$ 14 & 2.6x2.1 & 27 $\pm$ 7 & $\ge$14 $\pm$ 3 & 670 $\pm$ 7 & 17.3 $\pm$ 2.7 & 21 $\pm$ 3& 3.7x3.3 & 215 & ...  & 4x6 & 8 \\
\enddata
\tablenotetext{a}{Upper limits are 2$\sigma$ in a single beam (per
galaxy component, if applicable).}
\tablenotetext{b}{Lower limits indicated that the HCO$^+$ J=4-3 flux
  of the galaxy is uncertain 
  because the full width of the line may not be contained within the
  spectrometer; see text.}
\tablenotetext{c}{References are for CO J=1-0 and 3 mm
properties; for Mrk 231, Mrk 273, Arp 193, and NGC 6240, 
reference is also for CO J=2-1 and 1.3 mm properties.
1. \citet{ds98}. 2. \citet{g98}.  3. \citet{a97}. 
4. \citet{s88b}.  5. \citet{y94}.  6. \citet{i05}.
7.\citet{bs99}. 8. \citet{s89}. 9. \citet{t99} 10. L. Tacconi \&
A. Baker, private communication.}
\tablenotetext{d}{Only the eastern peak of VV 114 is detected in the 
$^{13}$CO J=2-1 line.}
\tablenotetext{e}{NGC 5258 contains an off-nuclear continuum source whose
flux is quoted here. The nucleus is undetected at $< 30$ mJy (2$\sigma$).}
\end{deluxetable}
\clearpage


\begin{deluxetable}{lcccccc}
\tabletypesize{\scriptsize}
\tablecaption{\bf CO 3-2 single dish and interferometric fluxes 
\label{tbl-co32}}
\tablewidth{0pt}
\tablehead{

\colhead{Galaxy}  & \colhead{CO3-2(SMA)} 
& \colhead{CO3-2(JCMT)\tablenotemark{a}} 
& \colhead{Missing flux\tablenotemark{b}} & \colhead{Percent} \\
        & \colhead{(Jy km s$^{-1}$)} 
& \colhead{(Jy km s$^{-1}$)} 
& \colhead{(Jy km s$^{-1}$)} & \colhead{missing flux\tablenotemark{b}} 
}
\startdata
IRAS 17208-0014 & 478 $\pm$ 33 
& 985 $\pm$ 128  & 510 $\pm$ 260 & (51 $\pm$ 15)\% \\
Mrk 231         & 308$\pm$8 
& 480$\pm$100  & 170$\pm$150 & (35$\pm$23)\%\\
Mrk 273    & 441$\pm$ 14 
& 410$\pm$90  & 0 & 0 \\
IRAS 10565+2448   & 204$\pm$7 
& 470$\pm$60  & 270$\pm$120 & (57$\pm$14)\% \\
UGC 5101  & 209$\pm$10 
& 490$\pm$80 
 & 280$\pm$130 & (57$\pm$14)\% \\
Arp 299 &  2582 $\pm$ 57 
& 4890$\pm$180\tablenotemark{e}  &  2300 $\pm$ 1600 & (47 $\pm$ 19)\%  \\
Arp 55 & 146 $\pm$9 
& $\ge$ 425 $\pm$ 40\tablenotemark{c} & $\ge$ 280 $\pm$ 100 & ($\ge$ 66 $\pm$ 10)\% \\
Arp 193  & 890 $\pm$ 15 
& 1118 $\pm$ 94  & 230 $\pm$ 300 & (20 $\pm$ 24)\% \\
NGC 6240 & 2428 $\pm$ 52 & 3205 $\pm$ 642 & 780 $\pm$ 1030 & (24 $\pm$ 26)\% \\
VV 114 & 1530 $\pm$ 16 
& 2956 $\pm$ 133\tablenotemark{d} &  1430 $\pm$ 680  & (48 $\pm$ 15)\%     \\
NGC 5331S   &  401 $\pm$ 24 & ...  & ... & ... \\
NGC 5331N   & 77 $\pm$ 10  & ...  & ... & ... \\
NGC 2623  & 607 $\pm$ 5
& 620 $\pm$ 44\tablenotemark{d} &  13 $\pm$ 190 & (2 $\pm$ 29)\% \\
NGC 5257 & 286 $\pm$ 47 
& 437 $\pm$ 67 &  151 $\pm$ 133 & (36 $\pm$ 24)\%  \\
NGC 5258 & 854 $\pm$ 52 
& 504 $\pm$ 26\tablenotemark{e} & 0 & 0 \\
NGC 1614 & 674 $\pm$ 14 
&  1471 $\pm$ 62\tablenotemark{d}  &  800 $\pm$ 330 &  (54 $\pm$ 13)\% \\
\enddata


\tablenotetext{a}{For galaxies that were not mapped, conversions from
  K(T$_A^*$) to Jy were done assuming  
the source is 
point-like
compared to the single dish beam, 
except for Arp 55 for which a main beam efficiency $\eta_{MB}=0.63$ was
used. 
The flux for NGC 6240 is
taken from \citet{grev07}.}
\tablenotetext{b}{Uncertainty in missing flux assumes 20\% calibration 
 uncertainty on each flux except for NGC 5258 and Arp 299, for which a
 30\% calibration uncertainty in the JCMT data is assumed.}
\tablenotetext{c}{Flux is from a single spectrum centered near the south-western
source.}
\tablenotetext{d}{Flux is calculated from a 
7x7 map with 5'' spacing.}
\tablenotetext{e}{Flux is calculated from a map made with HARP-B
with 6'' spacing.}
\end{deluxetable}



\begin{deluxetable}{lccccccc}
\tabletypesize{\scriptsize}
\tablecaption{\bf Interferometric and Single Dish 880 $\mu$m continuum fluxes
\label{tbl-cont}}
\tablewidth{0pt}
\tablehead{
\colhead{Galaxy}  & \colhead{JCMT flux\tablenotemark{a}} 
& \colhead{CO3-2\tablenotemark{b}} & \colhead{SMA flux} 
& \colhead{Missing Flux\tablenotemark{c}} & \colhead{Percent\tablenotemark{c}}\\
  & \colhead{(mJy)}       
& \colhead{(mJy)} & \colhead{(mJy)} 
& \colhead{(mJy)} & \colhead{missing flux} 
}
\startdata
IRAS 17208-0014 & 137$\pm$41 & 18.9 & 48$\pm$10 & 70$\pm$31 & (59$\pm$14)\% \\
Mrk 231 & 69$\pm$13 & 9.2 & 80$\pm$4 & 0 & 0 \\
Mrk 273 & 60$\pm$16 & 8.4 & 56$\pm$5 & 0 & 0 \\
IRAS 10565+2448 & 54$\pm$12 & 9.0 & 15$\pm$3 & 30$\pm$15 & (67$\pm$13)\% \\
UGC 5101 & 102$\pm$18 & 9.4 & 37$\pm$9 & 56$\pm$30 & (60$\pm$17)\% \\
Arp 299 & 376$\pm$10 & 93.7 & 101$\pm$7 & 181$\pm$69 & (64$\pm$10)\% \\
Arp 55 & 58$\pm$12 & $\ge$8.1 & 26$\pm$7 & $\le$23$\pm$18 & ($\le$47$\pm$22)\% \\
Arp 193 & 100$\pm$13 & 21.4 & 39$\pm$4 & 40$\pm$24 & (50$\pm$16)\% \\
NGC 6240 & 133$\pm$40 & 61.4 & 33$\pm$13 & 38$\pm$30 & (54$\pm$22)\% \\
VV 114 & 181$\pm$8 & 56.7 & 26$\pm$6 & 99$\pm$33 & (79$\pm$07)\% \\
NGC 5331 & 64$\pm$6 & 9.2 & 27$\pm$6 & 28$\pm$15 & (50$\pm$18)\% \\
NGC 2623 & 81$\pm$12 & 11.9 & 50$\pm$2 & 19$\pm$21 & (27$\pm$23)\% \\
NGC 5257 & 101$\pm$20 & 8.4 & $<$26 & $>$66$\pm$28 & ($>$72$\pm$17)\% \\
NGC 5258 & 150$\pm$28 & 16.4 & 104$\pm$21 & 29$\pm$46 & (22$\pm$29)\% \\
NGC 1614 & 194$\pm$29 & 28.2 & 27$\pm$7 & 139$\pm$45 & (84$\pm$6)\% \\
\enddata


\tablenotetext{a}{See text for a description of the source of the
uncertainty listed here for each galaxy.}
\tablenotetext{b}{SMA CO J=3-2 fluxes used for Mrk 273, NGC 5258, and NGC 5331. 
Arp 55 flux is likely underestimated since spectrum was centered near the
south-western source and so misses most emission from the stronger 
north-eastern source.}
\tablenotetext{c}{Uncertainty in missing flux assumes 20\% calibration 
uncertainty on each flux.}
\end{deluxetable}



\begin{deluxetable}{lcccccccc}
\tabletypesize{\scriptsize}
\tablecaption{\bf Gas and dust masses from total luminosity\label{tbl-dustmass}}
\tablewidth{0pt}
\tablehead{
\colhead{Galaxy}  & \colhead{D$_L$} 
& \colhead{$L'_{\rm CO}(3-2)$} & \colhead{$T_D$\tablenotemark{a}} 
& \colhead{$M_{dust}$} & \colhead{$M_{\rm H_2}$\tablenotemark{b}} & \colhead{Gas/Dust} \\
        & \colhead{(Mpc)} 
& \colhead{($10^9$ K km s$^{-1}$ pc$^2$)} & \colhead{(K)} 
& \colhead{($10^7$ $M_\odot$)} & \colhead{($10^9$ $M_\odot$)}  & 
}
\startdata
IRAS 17208-0014 & 189 & 4.46 & 41.3 
& 4.32 $\pm$ 0.90\tablenotemark{c} & 7.13 $\pm$ 0.49 & 165 $\pm$  36\tablenotemark{c} \\
Mrk 231 & 179 & 2.58 & 43.1 
& 7.67 $\pm$ 0.38\tablenotemark{d} & 4.12 $\pm$ 0.11 & 54 $\pm$ 3\tablenotemark{d}  \\
Mrk 273 & 166 & 3.19 & 43.0 
& 6.17 $\pm$ 0.55\tablenotemark{e} & 5.10 $\pm$ 0.16 & 83 $\pm$ 8\tablenotemark{e} \\
IRAS 10565+2448 & 191 & 1.94 & 38.6 
& 2.50 $\pm$ 0.50 & 3.11 $\pm$ 0.11 & 124 $\pm$ 25 \\
UGC 5101 & 174 & 1.66 & 35.5 
& 5.67 $\pm$ 1.38\tablenotemark{e} & 2.65 $\pm$ 0.13 & 47 $\pm$ 11\tablenotemark{e} \\
IC 694 (Arp 299) & 44 & 0.840 & 41.5 
& 0.655 $\pm$ 0.040\tablenotemark{e} & 1.34 $\pm$ 0.02 & 205 $\pm$ 13\tablenotemark{e} \\
NGC 3690 (Arp 299) & 44 & 0.311 & 41.5
& 0.162 $\pm$ 0.040  & 0.497 $\pm$ 0.014 & 308 $\pm$ 77 \\
Arp 55(NE) & 173 & 0.737 & 34.5
& 4.09 $\pm$ 1.10 & 1.18 $\pm$ 0.11 & 29 $\pm$ 8 \\
Arp 55(SW)  & 173 & 0.408 & 34.5
& $<$ 1.26 $\pm$ 0.63  & 0.652 $\pm$ 0.100 & $>$ 52 $\pm$ 27 \\
Arp 193=IC883 & 102 & 2.45 & 35.4 
& 2.06 $\pm$ 0.21  & 3.92 $\pm$ 0.07 & 190 $\pm$ 20 \\
NGC 6240 & 107 & 7.39 & 40.4 & 1.63 $\pm$ 0.64  & 11.82 $\pm$ 0.25 & 725
$\pm$ 286 \\
VV 114 & 87 & 3.09 & 39.4 
& 0.876 $\pm$ 0.202  & 4.94 $\pm$ 0.05 & 565 $\pm$ 130 \\
NGC 5331N & 145 & 0.427 & 32.1 & $<$ 1.45 $\pm$ 0.73  & 0.682 $\pm$ 0.089 &
$>$ 47 $\pm$ 24 \\
NGC 5331S & 145 & 2.22 & 32.1 & 3.27 $\pm$ 0.73  & 3.55 $\pm$ 0.21 
& 109 $\pm$ 25 \\
NGC 2623  & 80 & 1.04 & 39.9 & 1.40 $\pm$ 0.06\tablenotemark{e}  & 1.66 $\pm$ 0.01 & 118
$\pm$ 4\tablenotemark{e} \\
NGC 5257 & 99 & 0.746 & 36.0 
& $<$ 1.27 $\pm$ 0.63  & 1.19 $\pm$ 0.20 & $>$ 94 $\pm$ 50 \\
NGC 5258 & 99 
& 2.23 & 36.0 & 5.07 $\pm$ 1.02\tablenotemark{f}  & 3.56 $\pm$ 0.22 & 70 $\pm$ 15\tablenotemark{f} \\
NGC 1614 & 69 & 0.860 & 42.5 
& 0.313 $\pm$ 0.081\tablenotemark{c} & 1.38 $\pm$ 0.03 & 440 $\pm$ 114\tablenotemark{c} \\
\enddata
\tablenotetext{a}{ $T_{e}$ is the dust temperature calculated
from fitting a modified blackbody function as in \citet{k01} to published data
from 60 to 800 $\mu$m; see text.}
\tablenotetext{b}{$M_{\rm H_2}$ calculated assuming CO3-2/CO1-0 = 0.5 and 
$M_{\rm H_2} = 0.8 L'_{\rm CO}(1-0)$ \citep{ds98}.}
\tablenotetext{c}{Dust mass and gas-to-dust ratio have been corrected
by a factor of 0.6 to account for a non-thermal contribution
to the 880 $\mu$m flux. See text.}
\tablenotetext{d}{Dust mass and gas-to-dust ratio have been corrected
by a factor of 0.75 to account for a non-thermal contribution
to the 880 $\mu$m flux. See text.}
\tablenotetext{e}{There may be a non-thermal contribution to the
continuum flux of this source; however, any effect on the dust
mass and the gas-to-dust ratio is likely smaller than the
uncertainties. See text.}
\tablenotetext{f}{NGC 5258 contains an off-nuclear continuum source, whose
flux is used here, while the CO J=3-2 emission comes primarily from
the extended southern arm.
Thus, the dust mass and gas-to-dust ratio given here are likely incorrect.}
\end{deluxetable}



\begin{deluxetable}{lccccccccccc}
\tabletypesize{\scriptsize}
\tablecaption{\bf Gas and dust masses in a single beam  \label{tbl-dustmasspeak}}
\tablewidth{0pt}
\tablehead{
\colhead{Galaxy}  & S$_{\rm CO}(3-2)$\tablenotemark{a} 
& \colhead{$L'_{\rm CO}(3-2)$(peak)\tablenotemark{a} } & \colhead{880 $\mu$m\tablenotemark{a}}  
& \colhead{$M_{dust}$\tablenotemark{b}} & \colhead{$M_{\rm H_2}$\tablenotemark{b}} & \colhead{Gas/Dust} \\
        & \colhead{(Jy beam$^{-1}$ km s$^{-1}$)} 
& \colhead{($10^9$ K km s$^{-1}$ pc$^2$)} & \colhead{(mJy beam$^{-1}$)} 
& \colhead{($10^7$ $M_\odot$)} & \colhead{($10^9$ $M_\odot$)}  & 
}
\startdata
IRAS 17208-0014 & 217 $\pm$ 11 & 2.02 & 48 $\pm$ 10 & 4.32 $\pm$ 0.90\tablenotemark{c} & 3.24 $\pm$ 0.16 & 75 $\pm$ 16\tablenotemark{c} \\
Mrk 231 & 193 $\pm$ 3 & 1.62 & 80 $\pm$ 4 & 7.67 $\pm$ 0.38\tablenotemark{d} & 2.58 $\pm$ 0.04 & 34 $\pm$ 2\tablenotemark{d} \\
Mrk 273 & 300 $\pm$ 6 & 2.17 & 56 $\pm$ 5 & 6.17 $\pm$ 0.55\tablenotemark{e} & 3.47 $\pm$ 0.07 & 56 $\pm$ 5\tablenotemark{e} \\
IRAS 10565+2448 & 78.1 $\pm$ 2.2 & 0.743 & 15 $\pm$ 3 & 2.50 $\pm$ 0.50 & 1.19 $\pm$ 0.03 & 48 $\pm$ 10 \\
UGC 5101 & 103 $\pm$ 4 & 0.817 & 30 $\pm$ 4 & 4.60 $\pm$ 0.61\tablenotemark{e} & 1.31 $\pm$ 0.05 & 28 $\pm$ 4\tablenotemark{e} \\
IC 694 (Arp 299) & 877 $\pm$ 4 & 0.457 & 81 $\pm$ 5 & 0.655 $\pm$ 0.040\tablenotemark{e} & 0.732 $\pm$ 0.003 & 112 $\pm$ 7\tablenotemark{e} \\
NGC 3690 (Arp 299) & 309 $\pm$ 4 & 0.161 & 20 $\pm$ 5 & 0.162 $\pm$ 0.040 & 0.258 $\pm$ 0.003 & 159 $\pm$ 40 \\
Arp 55(NE) & 49.2 $\pm$ 3.6 & 0.386 & 21 $\pm$ 4 & 3.30 $\pm$ 0.63 & 0.617 $\pm$ 0.045 & 19 $\pm$ 4 \\
Arp 55(SW) & 22.5 $\pm$ 2.5 & 0.176 & $<$ 8 $\pm$ 4 & $<$ 1.26 $\pm$ 0.63 & 0.282 $\pm$ 0.031 & $>$ 22 $\pm$ 11 \\
Arp 193 & 477 $\pm$ 4 & 1.32 & 39 $\pm$ 4 & 2.06 $\pm$ 0.21 & 2.11 $\pm$ 0.02 & 102 $\pm$ 11 \\
NGC 6240 & 1120 $\pm$ 7 & 3.41 & 26 $\pm$ 8 & 1.28 $\pm$ 0.40 &
5.45 $\pm$ 0.03 & 424 $\pm$ 131 \\
VV 114(east) & 306 $\pm$ 3 & 0.618 & 13 $\pm$ 4 & 0.438 $\pm$ 0.135 & 0.989 $\pm$ 0.010 & 226 $\pm$ 70 \\
VV 114(center) & 223 $\pm$ 3 & 0.450 & 13 $\pm$ 4 & 0.438 $\pm$ 0.135 & 0.721 $\pm$ 0.010 & 165 $\pm$ 51 \\
VV 114(west) & 145 $\pm$ 3 & 0.293 & $<$ 8 $\pm$ 4 & $<$ 0.269 $\pm$ 0.134 & 0.469 $\pm$ 0.010 & $>$ 174 $\pm$ 87 \\
NGC 5331N & 48 $\pm$ 5 & 0.266 &  $<$ 12 $\pm$ 6 & $<$ 1.45 $\pm$ 
0.73 & 0.425 $\pm$ 0.044 & $>$ 29 $\pm$ 15 \\
NGC 5331S & 127 $\pm$ 6 & 0.704 &  27 $\pm$ 6 & 3.27 $\pm$ 0.73 & 1.13 $\pm$
0.05 & 34 $\pm$ 8 \\
NGC 2623  & 430 $\pm$ 1.3 & 0.735 & 50 $\pm$ 2 & 1.40 $\pm$ 0.06\tablenotemark{e} &
1.177 $\pm$ 0.004 & 84 $\pm$ 3\tablenotemark{e} \\
NGC 5257 & 97 $\pm$ 11 & 0.253 & $<$ 26 $\pm$ 13 & $<$ 1.27 $\pm$
0.63 & 0.405 $\pm$ 0.046 &  $>$ 32 $\pm$ 16 \\
NGC 5258 & 60 $\pm$ 10 & 0.157 & $<$ 30 $\pm$ 15 & $<$ 1.46 $\pm$ 0.73 &
0.250 $\pm$  0.042 & $>$ 17 $\pm$ 9 \\
NGC 1614 & 275 $\pm$ 4 & 0.351 & 21 $\pm$ 4 & 0.243 $\pm$ 0.046\tablenotemark{c} & 0.561 $\pm$ 0.008 & 231 $\pm$ 44\tablenotemark{c} \\
\enddata
\tablenotetext{a}{Values measured at the peak emission except for
  NGC 5258, where it is measured at the peak in the nuclear region. See Table~\ref{tbl-obs}
for the beam size for each galaxy.}
\tablenotetext{b}{$M_{\rm H_2}$ and $M_{dust}$ are calculated with
the same assumptions and dust temperature used in Table~\ref{tbl-dustmass}.}
\tablenotetext{c}{Dust mass and gas-to-dust ratio have been corrected
by a factor of 0.6 to account for a possible non-thermal contribution
to the 880 $\mu$m flux. See text.}
\tablenotetext{d}{Dust mass and gas-to-dust ratio have been corrected
by a factor of 0.75 to account for a non-thermal contribution
to the 880 $\mu$m flux. See text.}
\tablenotetext{e}{There may be a non-thermal contribution to the
continuum flux of this source; however, any effect on the dust
mass and the gas-to-dust ratio is likely smaller than the
uncertaintites. See text.}
\end{deluxetable}


\clearpage


\begin{deluxetable}{lcccccccc}
\tabletypesize{\tiny}
\rotate
\tablecaption{\bf Probability $p$ of a false correlation after
  removing the effect of distance \label{tbl-corr}}
\tablewidth{0pt}
\tablehead{
\colhead{Quantity} &  \colhead{$T_{\rm D}$} &
\colhead{$\Sigma_{\rm H_2}$(peak)} &
\colhead{beam area}  &
\colhead{$M_{\rm H_2}$(total)} &
\colhead{$L_{\rm FIR}/M_{\rm H_2}$(total)} &
\colhead{$L_{\rm FIR}/M_{\rm H_2}$(peak)} &
\colhead{gas-to-dust ratio (peak)} &
\colhead{CO J=3-2/2-1 line ratio} 
}
\startdata
$L_{\rm FIR}$ & {\bf 0.0168} & {\bf 0.0014} & 0.0552 & 0.2244 & {\bf
  0.0250} & 0.2042 & 0.5774 & 0.4554 \\
$T_{\rm D}$ & ... & {\bf 0.0016} & 0.0719 & 0.4902 & {\bf 0.0321} &
   0.0997 & 0.2671 & 0.6234\\
$\Sigma_{\rm H_2}$(peak) & ... & ... &  {\bf 0.0063} & 0.2084 & 0.1219
  &  0.9933 & 0.7035 & 0.7662 \\
beam area  & ... & ... & ... &  0.1202 & {\bf 0.0284} & 0.5258 & 0.5393 &
   {\bf 0.0005}\\
$M_{\rm H_2}$(total) & ... & ... & ... & ... &  0.2107 & 0.6571 &
   0.1044 &  {\bf 0.0074} \\
$L_{\rm FIR}/M_{\rm H_2}$(tot) & ... & ... & ... & ... & ... & {\bf
  0.0140} & 0.0916 & {\bf 0.0092} \\
$L_{\rm FIR}/M_{\rm H_2}$(pk) & ... & ... & ... & ... & ... & ... &
  0.2411 &  0.2213 \\
gas-to-dust ratio (peak) & ... & ... & ... & ... &
... & ... & ... & 0.0578 \\ 
\enddata
\end{deluxetable}

\clearpage

\clearpage
                                                                             
\clearpage

\clearpage

\clearpage

\end{document}